\documentclass[3p,sort&compress]{elsarticle}

\usepackage[T1]{fontenc}
\usepackage{lineno}
\usepackage[hidelinks]{hyperref}
\modulolinenumbers[5]
\usepackage{subcaption}
\usepackage{array, multirow, booktabs, bm, color, todonotes, amsmath, hhline, appendix, amssymb}

\def\ket#1{\left | #1 \right \rangle} 
\def\bra#1{\left \langle #1 \right |} 

\numberwithin{equation}{section}

\makeatletter
\renewcommand\p@subfigure{\thefigure\,}

\DeclareCaptionLabelFormat{mysublabelfmt}{(\alph{sub\@captype})}
\makeatother

\journal{Physics Reports}

\begin{document}

\begin{frontmatter}

\title{Cavity Spintronics: An Early Review of Recent Progress in the Study of Magnon-Photon Level Repulsion}

\author{Michael Harder\corref{mycorrespondingauthor}}
\ead{michael.harder@umanitoba.ca}

\author{Can-Ming Hu\corref{memail}}
\ead{hu@physics.umanitoba.ca}
\address{Department of Physics and Astronomy, University of Manitoba, Winnipeg, Canada R3T 2N2}

\begin{abstract}
Light-matter interactions lie at the heart of condensed matter physics, providing physical insight into material behaviour while enabling the design of new devices.  Perhaps this is most evident in the push to develop quantum information and spintronic technologies.  On the side of quantum information, engineered light-matter interactions offer a powerful means to access and control quantum states, while at the same time new insights into spin-photon manipulation will benefit the development of spintronic technologies.  In this context the recent discovery of hybridization between ferromagnets and cavity photons has ushered in a new era of light-matter exploration at the crossroads of quantum information and spintronics.  The key player in this rapidly developing field of cavity spintronics is a quasiparticle, the cavity-magnon-polariton.  In this early review of recent work the fundamental behaviour of the cavity-magnon-polariton is summarized and related to the development of new spintronic applications.  In the last few years a comprehensive theoretical framework of spin-photon hybridization has been developed.  On an intuitive level many features can be described by a universal model of coupled oscillators, however the true origin of hybridization is only revealed by considering a more comprehensive electrodynamic framework.  Here both approaches are summarized and an outline of a quantum description through the input-output formalism is provided.  Based on this foundation, in depth experimental investigations of the coupled spin-photon system have been performed.  
For example, it has been found that hybridization will influence spin current generated through the spin pumping mechanism, demonstrating a firm link between spin-photon coupling and spintronics.  Furthermore several in-situ coupling control mechanisms, which offer both physical insight and a means to develop cavity-spintronic technologies, have been revealed.  
The many recent developments within this field represent only the first steps in what appears to be a bright future for cavity spintronics.  Hopefully this early review will introduce new explorers to this exciting frontier of condensed matter research, lying at the crossroads of magnetism and cavity quantum electrodynamics.
\end{abstract}

\begin{keyword}
cavity spintronics, cavity-magnon-polariton, hybridization, magnon-photon coupling, level repulsion
\end{keyword}

\end{frontmatter}

\tableofcontents


\section{Introduction}

Emergence, the appearance of macroscopic properties which qualitatively differ from a system's microscopic behaviour, is one of the most intriguing concepts in physics.  From hydrodynamics \cite{Kovtun2012, ForsterBook}, to spacetime \cite{Seiberg2006, El-Showka2012, Padmanabhan2014}, to superconductivity \cite{PooleBook} to novel quantum materials \cite{Tokura2017}, the concept of emergence underlies a wealth of physical phenomena,\footnote{Actually the concept of emergence extends well beyond physics and even science, underlying the key characteristics of many complex systems.  For example, cellular functions in molecular biology \cite{Hartwell1999, Barabasi2004}, self organization in economics \cite{Harper2012}, scaling of large networks \cite{Barabasi1999} and language development \cite{MacWhinney1998} are all emergent phenomena.} even producing entirely new degrees of freedom through collective excitations.  
Most interestingly, emergent behaviour usually cannot be anticipated from the properties of the underlying constituents.  For this reason, despite advanced knowledge of the fundamental laws governing many-body solid-state systems, condensed matter research still contains many unexplored paths to discovery \cite{Anderson1972, LaughlinBook}.  This field of cavity spintronics has emerged by following one such path, through the coupling of low loss magnetic materials and high quality microwave fields.  

The foundation underlying cavity spintronics is a quasiparticle, the cavity-magnon-polariton (CMP), which displays an intriguing dual spin-photon nature.  This hybrid character highlights the CMP's position at the crossroads of cavity-quantum electrodynamics and magnetism \cite{Hu2015}.  Of course the venerable subject of magnetism stands on its own right as one of the pillars of condensed matter physics, inspiring millennia of scientific discovery and technological innovation \cite{CoeyBook}.  Within this long history, one of the important modern discoveries was the observation that ferromagnetic materials could absorb microwave frequency radiation \cite{Arkadyev1912, Griffiths1946}.  This phenomena results from the emergence of collective spin dynamics,\footnote{From the onset of ferromagnetism at oxide interfaces \cite{Suzuki2015} to the formation of magnetic polaritons \cite{AlbuquerqueBook}, examples of emergence abound in magnetism.}  which results in resonant motion of the magnetization \cite{LaxBook, HillebrandsBookVolI}.  Ferromagnetic resonance has proven to be a powerful technique to investigate a material's magnetic properties and is now routinely used to probe, e.g., demagnetization effects, magnetic anisotropies and damping mechanisms \cite{Harder2016a}.  As magnetic resonance techniques matured, including the development of electron paramagnetic resonance and nuclear magnetic resonance, it was realized that the loss of radiative energy due to inductive processes, known as radiation damping, could play a significant role for magnetization dynamics \cite{Bloembergen1954} (such effects have more recently become the topic of intense discussion \cite{Schoen2015, Klingler2017, Rao2017}).  Radiation damping can be viewed as one effect of the magnetization back action onto the driving microwave field.  Heuristically, losing energy through the emission of radiation only requires coupling back to the environment, thereby influencing the damping at first order.  However if the photons which drive magnetic resonance are trapped for a sufficiently long time, the radiated photon will in turn drive magnetization precession.  It is in this context that cavity spintronics (termed spin cavitronics by some authors) has emerged; when the influence of the magnetization induction is appreciable, its effect on the local microwave field must be accounted for, resulting in a back and forth flow of energy between spin and photonic degrees of freedom which produces a hybridized quasiparticle, the cavity-magnon-polariton.  Cavity spintronics is therefore the study and application of strongly coupled spin-photon physics in condensed matter systems.

From a theoretical perspective understanding the CMP requires the simultaneous solution of Maxwell's equations, which describe electrodynamics, and the Landau-Lifshitz-Gilbert (LLG) equation, which describes magnetization dynamics.\footnote{In contrast to the basic approach to magnetization dynamics, where one assumes that a fixed, applied field drives the magnetization, back action is ignored, and the LLG equation alone is solved.}  On the other hand, from an experimental perspective, realizing strong spin-photon coupling requires a high quality microwave cavity, to confine the photons for extended periods of time, and a large sample-to-cavity filling factor, to increase the number of spins in the microwave cavity.  The latter can be realized using high spin density, low loss ferrimagnetic materials, such as yttrium-iron-garnet (YIG) commonly found in spintronic devices.  The former has been realized for several decades in the field of cavity quantum electrodynamics, where the quantum nature of light is brought to the forefront \cite{HarocheBook}. 

The possibility of realizing hybridized spin-photon excitations in ferromagnetic systems was first discussed theoretically by Soykal and Flatt\'{e} in 2010 \cite{Soykal2010, Soykal2010a}, but only realized experimentally in 2013 \cite{Huebl2013a}.  In this first experiment, Huebl et al. employed a YIG sphere coupled to a planar superconducting resonator at ultra low temperatures, using microwave transmission to measure the key coupling signature, an anticrossing in the eigenspectrum.  Although the absorption of electromagnetic fields by ferromagnets in cavities had been studied for many decades before this observation, the increased filling factor, due to large, high quality samples and confined microwave fields, as well as the shift of focus from fixed to swept frequency experiments, enabled this new discovery.  While initial experiments were performed at low temperatures, it was quickly realized that the experimental conditions for CMP formation could be relaxed, allowing observations at room temperature \cite{Zhang2014}, in 3D microwave cavities \cite{Tabuchi2014, Harder2016b} and using split ring resonators \cite{Stenning2013, Bhoi2014, Gregory2014, Kaur2015, Zhang2017}.  Although YIG, due to its ultra low damping and high spin density, continues to be the prototypical magnetic material for CMP experiments, strong coupling has also been observed in gadollinium-iron-garnet (GdIG) \cite{MaierFlaig2017}, lithium ferrite \cite{Goryachev2017a} and the chiral magnetic insulator Cu$_2$OSeO$_3$ \cite{Abdurakhimov2018}.  Furthermore, while initially motivated by the potential for hybrid quantum information systems \cite{Schuster2010, Kubo2010, Chiorescu2010, Abe2011} with the enhanced coupling of exchange-locked ferromagnets \cite{Garraway2011}, the electrical detection of strong spin-photon coupling has further pushed the CMP into the realm of spintronics \cite{Bai2015, MaierFlaig2016, Bai2017}.  This variety of cavity spintronic measurement techniques which have been established since 2013 have led to a wealth of CMP exploration.  For example, the ultrastrong coupling regime has been reached \cite{Goryachev2014, Zhang2014a, Bourhill2015a}, spin wave strong coupling has been observed \cite{Lambert2015, Zhang2016, MaierFlaig2016, Goryachev2017}, multimode cavity \cite{Hyde2016} and spin \cite{Lambert2015b, Zhang2015g, Morin2016, Bai2017} systems have been realized, exceptional points of the CMP eigenspectrum have been found \cite{Harder2017, Zhang2017a}, coherent perfect absorption has been realized \cite{Zhang2017a} and the effect of Kerr nonlinearities has been studied \cite{Wang2016a, Wang2017}.  Furthermore, a transition from microwave to optical frequencies has recently resulted in the exploration of cavity optomagnonics \cite{Osada2015, Zhang2015b, Haigh2015b, Liu2016, Kusminskiy2016, Braggio2016}.  On the more applied side, voltage control of the CMP using on-chip devices \cite{Kaur2016}, the observation of electromagnetically induced transparency \cite{Kaur2015, Tay2018}, novel quantum information architectures employing magnon dark modes \cite{Zhang2015g}, and the realization of microwave to optical frequency conversion \cite{Hisatomi2016} all point to a bright future for CMP based devices.  

While all of the results mentioned above were based on classical spin-photon properties, recently the quantum frontier has also been encountered with the generation of a qubit-CMP interaction \cite{Tabuchi2014, Tabuchi2015}, the observation of triplet and quintuplet states in active resonators \cite{Yao2017} and low power experiments with single photon excitations \cite{Morris2016}.  All of these discoveries have occurred within the last five years.  Here the theoretical and experimental foundation which underlies these discoveries is outlined, and an early roadmap of cavity spintronics is provided for new researchers entering the field.


\section{Theory of Magnon-Photon Level Repulsion} \label{sec:theory}

Condensed matter physics is a science that requires an active interplay between theory and experiment.  This is a necessity born out of a difficult challenge --- in a system comprised of $\sim 10^{23}$ interacting particles, how can one hope to understand the complex and nuanced behaviour key to physical insight and technological development?  The answer is to not consider the minutia of a given system in its entirety, but rather to reveal the key physical elements which drive the relevant emergent behaviour.\footnote{As is often the case, this approach is not unique to condensed matter, but is also fundamental to all fields of physics.  Even in the reductionist realm of high energy physics, effective field theories play an important role \cite{Georgi1993}, and also provide perhaps the most successful description of quantum gravity \cite{ArkaniHamed2003, Burgess2004}.}  From the tunnelling of supercurrents which underly the Josephson effect \cite{Josephson1962} to the phase transition of the Ising model \cite{Onsager1944}, examples of the insights gained in this manner abound in condensed matter systems.  

Most closely related to the development of cavity spintronics is the general emergent nature of hybridized light-matter excitations known as polaritons, which result from the resonant response of material dynamics.  As there are many types of material excitations there are also many types of polaritons, e.g. phonon-polaritons \cite{Mills1974, AlbuquerqueBook}, exciton-polaritons \cite{AlbuquerqueBook, AndreaniBook}, magnon-polaritons \cite{Mills1974, AlbuquerqueBook, CottamBook, Hartstein1973, Camley1982}, plasmon-polaritons \cite{AlbuquerqueBook, Barnes2003, Barnes2006} and cavity-exciton-polaritons \cite{Deng2010, Khitrova1999, TimofeevBook}.  Although the theoretical details of a polariton may be complex, their general treatment is a simple two step process: (1) determine the material response function in the presence of the excitations; and (2) solve Maxwell's equations using this response function.  However, for the purpose of developling physical intuition, perhaps no model is more helpful than the harmonic oscillator.  As the nature of the cavity-magnon-polariton is the coupling of two resonant systems (spin and photon) such a description already captures many of the key CMP properties.  


\subsection{An Intuitive Model: The Harmonic Oscillator} \label{subsec:harmonic}

\begin{figure}[t!]
\centering
\includegraphics[width=12cm]{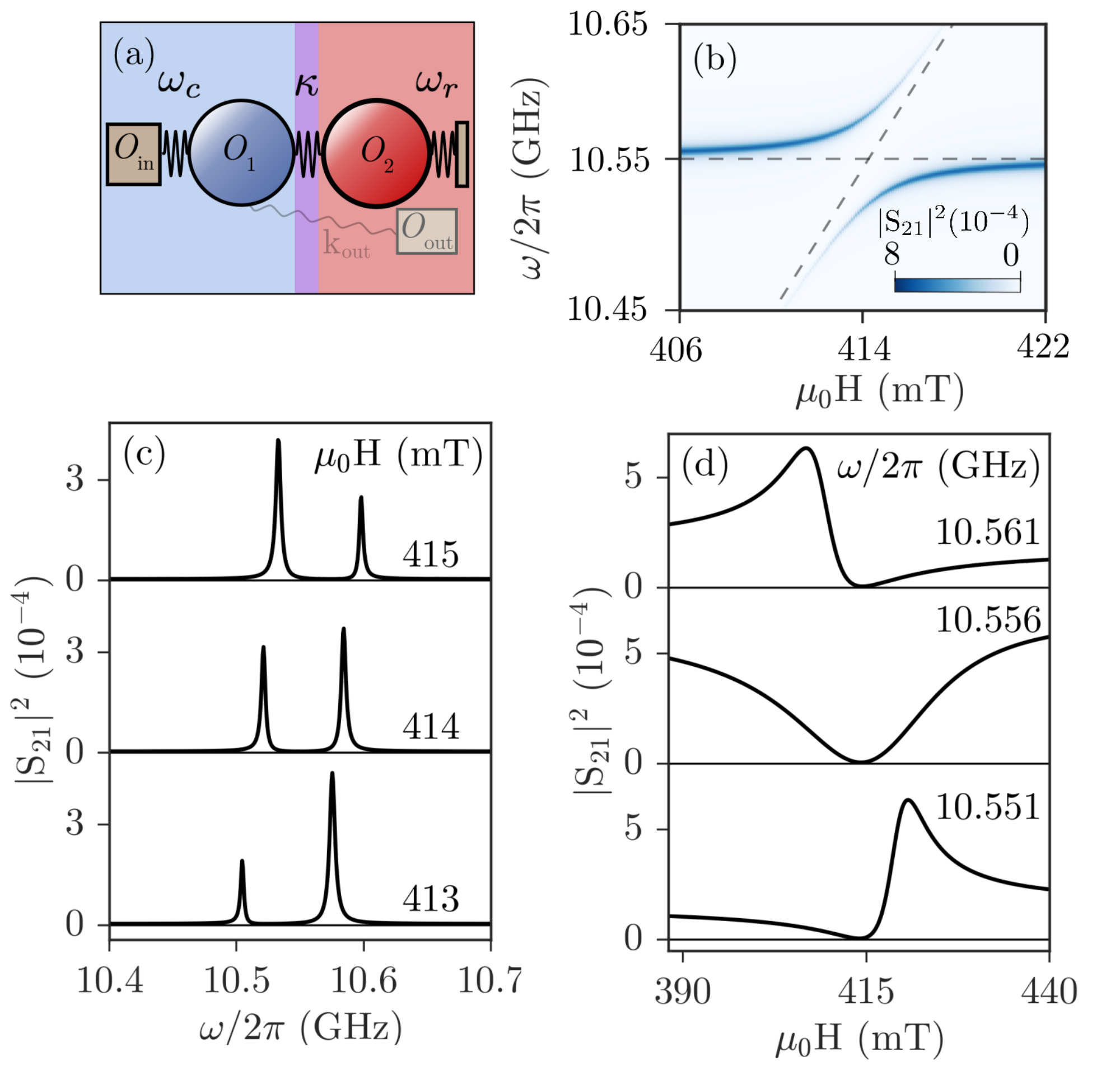}
\caption[Cavity-magnon-polariton transmission spectra in the harmonic model]{CMP transmission spectra in the harmonic coupling model.  (a) Schematic illustration of two coupled oscillators.  $O_1$, representing the cavity, experiences a sinusoidal driving force, and has its motion ``detected" by coupling to $O_\text{out}$ through a stiff spring.  The magnetization resonance is represented by the red $O_2$, while the purple region indicates the coupling between the cavity and spin systems.  (b) The full $\omega-H$ dispersion calculated according to Eq. \eqref{eq:oscT}. (c) Fixed field and (d) frequency cuts made above, at and below the crossing point $\omega_r = \omega_c$ calculated according to Eq. \eqref{eq:oscT}.  A modified version of this figure was originally published in Ref. \cite{Harder2016b}.}
\label{fig:oscFig}
\end{figure}

The basic hybridized system consists of two resonant components, the magnetic material and microwave photon.  Therefore consider a system of two equal mass $(m)$ oscillators, $O_1$ and $O_2$, coupled together via a spring $\kappa$ as shown in Fig. \ref{fig:oscFig} (a) (see e.g. \cite{Harder2016b}).  $O_1$, shown in blue, represents the cavity and is connected to an input plunger, $O_\text{in}$, via a spring with resonance frequency $\omega_c$.  The plunger is driven in constant motion at frequency $\omega/2\pi$ so that $x_\text{in} (t) = x_\text{in} e^{-i\omega t}$.  Physically this plunger represents the constant microwave input to the cavity, providing a driving force, $f(t) = \omega_c^2 x_\text{in}(t)$, to $O_1$.  $O_2$, shown in red, represents the FMR and is attached to a fixed wall with a spring of resonant frequency $\omega_r$.  Damping is introduced via a viscous force with coefficient $\beta$ for $O_1$, which models the intrinsic conductive losses of the cavity, and $\alpha$ for $O_2$, modelling losses due to magnetization damping.  

To extract the hybridized dispersion it is sufficient to set $x_\text{in}  = 0$ and solve the corresponding eigenvalue problem.  This means that the dispersion and line width are determined solely by the properties of the $O_1$, $O_2$ system, the coupling $\kappa$ and the uncoupled resonances and damping, $\omega_c, \omega_r, \beta$ and $\alpha$, and is independent of how the system is driven \cite{Harder2016b}.  The energy absorption of the coupled oscillators can also be calculated in a standard way using a dissipative function for the system \cite{LandauBookMechanics}.  However, the full transmission spectra depends on the coupling of energy into and out of the system and therefore requires an ``output port", $O_\text{out}$, which can be connected to $O_1$ using a spring $k_\text{out}$ as shown in Fig. \ref{fig:oscFig} (a).  The transmission through the system is then $|\text{S}_{21}|^2 = E_\text{out}/E_\text{in}$ where $E_\text{out}$ and $E_\text{in}$ are the kinetic energies of the output and input oscillators respectively.  For high quality cavities, the output coupling $k_\text{out} \ll 1$, and in particular $k_\text{out} \ll \kappa$.  There is also no damping associated with the output.  Therefore the equations of motion for $O_1$, $O_2$ and $O_\text{out}$ are, respectively,
\begin{align}
\ddot{x}_1 + \omega_c^2 x_1 + 2 \beta \omega_c \dot{x}_1 - \kappa^2 \omega_c^2 x_2 &= f e^{-i\omega t}, \label{eq:initialHarmonic1} \\
\ddot{x}_2 +\omega_r^2 x_2 + 2 \alpha \omega_c \dot{x}_2 - \kappa^2 \omega_c^2 x_1 &= 0,  \label{eq:initialHarmonic2} \\
\ddot{x}_\text{out} - k_\text{out}^2 \omega_c^2 x_1 &= 0 \label{eq:initialHarmonicOut}.
\end{align}
Here, by normalizing to $\omega_c$, the damping coefficients $\alpha$ and $\beta$ as well as the couplings $\kappa$ and $k_\text{out}$ are dimensionless.  Since $k_\text{out} \ll 1$ the action of $O_\text{out}$ on $O_1$ does not need to be included in Eq. \eqref{eq:initialHarmonic1} and there is no resonance in Eq. \eqref{eq:initialHarmonicOut}.  Taking $\left(x_1, x_2, x_\text{out}\right) = \left(A_1, A_2, A_\text{out}\right) e^{-i\omega t}$, Eqs. \eqref{eq:initialHarmonic1} and \eqref{eq:initialHarmonic2} can be written in the matrix form $\pmb{\Omega} \textbf{A} = \textbf{f}$ where $\textbf{A} = \left(A_1, A_2\right)$, $\textbf{f} = \left(-f, 0\right)$ and 
\begin{equation}
\pmb{\Omega} = \left(\begin{array}{cc}
\omega^2 - \omega_c^2 + 2 i \beta \omega_c \omega & \kappa^2 \omega_c^2 \\
\kappa^2 \omega_c^2 & \omega^2 - \omega_r^2 + 2 i \alpha \omega_c \omega
\end{array}\right) \label{eq:omegaMatrix}
\end{equation}
while Eq. \eqref{eq:initialHarmonicOut} becomes 
\begin{equation}
A_\text{out} = -\frac{k_\text{out}^2\omega_c^2}{\omega^2} A_1. \label{eq:aout}
\end{equation}
The transmission follows by solving for $A_1$, $\textbf{A} = \pmb{\Omega}^{-1} \textbf{f}$,

\begin{equation}
|\text{S}_{21}\left(\omega, H\right)|^2 = \frac{E_\text{out}}{E_\text{in}} = \frac{m_\text{out} A_\text{out}^2}{m_\text{in} x_\text{in}^2} =  \eta \frac{\omega_c^8}{\omega^4} \frac{|\omega^2 - \omega_r^2 + 2 i \alpha \omega \omega_c|^2}{|\det\left(\pmb{\Omega}\right)|^2} \label{eq:oscT}
\end{equation}
where $x_\text{in} = \omega_c^2 f$ and $\eta = (m_\text{out}/m_\text{in}) k_\text{out}^4$ acts as an impedance matching parameter, depending not only on the output coupling $k_\text{out}$ but also the matching of the input/output ports determined by $m_\text{out}/m_\text{in}$.  The determinant is given by
\begin{equation}
\det\left(\pmb{\Omega}\right) = \left(\omega^2 - \omega_c^2 + 2 i \beta \omega_c\omega\right)\left(\omega^2 - \omega_r^2 + 2 i \alpha \omega_c \omega\right) - \kappa^4 \omega_c^4 \label{eq:oscDet}
\end{equation}
and defines the spectral function of the coupled system.    

The role of the output absorber is apparent from Eqs. \eqref{eq:oscT} and \eqref{eq:oscDet}, it is needed to model the output port and calculate the transmission, since the amplitude of $\text{S}_{21}$ in Eq. \eqref{eq:oscT} is proportional to $k_\text{out}$.  However the output plays no role in the dispersion which is determined by the poles of Eq. \eqref{eq:oscDet}.  These facts mean such a model can be easily extended to multiple cavity/multiple spin wave modes \cite{Harder2016b}.

Due to the general nature of the harmonic model, there are several detailed features which are not accounted for in this description.  For example, the wave vector dependence of either the spin or photon subsystem has not been considered.  Experimentally this is justified since the microwave photon wavelength is much larger than the relevant magnon wavelength and the sample size, and therefore the driving field will be approximately uniform over the sample.  Furthermore, since the hybridization occurs with a cavity photon, the wave vector is fixed and therefore the most relevant feature to consider is the field dependence of the FMR, which is accounted for by allowing $\omega_r$ to vary.  Actually, even in situations where the photon wave vector is controlled through tuning the cavity height, the relevant physical influence is to tune $\omega_c$ and therefore the length scales between the photon and magnon wavelengths do not influence the observed behaviour \cite{Hyde2016}.  Polarization effects of the photons have also not been accounted for in this approach.  This is most relevant in determining the orientation between the local microwave magnetic field and the saturation magnetization, determined by the external bias field and can be included by introducing an orientation dependence to the coupling strength \cite{Bai2016, Bai2017}.  

To illustrate the effects of hybridization within the harmonic model, $|\text{S}_{21} \left(\omega, H\right)|^2$ is calculated  and plotted in Fig. \ref{fig:oscFig} (b) using the experimentally realistic parameters $\alpha = 0.8 \times 10^{-4}$, $\beta = 3 \times 10^{-4}$, $\omega_c/2\pi = 10.556$ GHz, $\eta = 2.3 \times 10^{-10}$ and $|\kappa| = 0.077$ \cite{Harder2016b}.  The bias dependence of $\omega_r$ follows the Kittel formula for a magnetized sphere \cite{Kittel1947, Kittel1948, LaxBook}, $\omega_r = \gamma \left(H+H_A\right)$ and a gyromagnetic ratio $\gamma = 2\pi \times 28 ~\mu_0$GHz/T and shape anisotropy $\mu_0H_A = -37.7$ mT are used.\footnote{Experimentally the value of the coupling strength can be determined from the Rabi gap \cite{Harder2016b}.}  The diagonal dashed line shows the uncoupled FMR dispersion following the Kittel formula and the horizontal dashed line is the uncoupled cavity mode.  The most striking feature of the transmission spectra is an anticrossing in the dispersion, which reflects the coupled nature of the system.  At the crossing point, $\omega_c = \omega_r$, the hybridized modes have the greatest deviation from their uncoupled counterparts.  Also at the crossing point the hybridized modes have their minimum frequency separation, $\omega_\text{gap}$, which is an increasing function of the coupling strength.  Another intriguing feature of the hybridization is the line width broadening/narrowing that can be observed as $H$ is tuned.  Finally, the line cuts at fixed field and frequency are shown in Figs. \ref{fig:oscFig} (c) and (d) respectively.  As may be anticipated for a resonance process, the fixed $H$ cuts are symmetric in $\omega$, appearing as Lorentz-like peaks.  On the other hand, while for fixed $\omega = \omega_c$, $|\text{S}_{21}\left(H\right)|^2$ has a symmetric dip, above and below $\omega_c$ the fixed field line shape has an asymmetry and the polarity of the line shape changes when passing through $\omega_c$.  These features represent the basic signatures of magnon-photon hybridization.


\subsubsection{Simplification of Transmission Spectra} \label{subsubsec:simpFreqOsc}

To compare the harmonic model to other CMP descriptions which have been developed, and to experimental data, it is useful to simplify Eq. \eqref{eq:oscT}.  Since the effect of hybridization is greatest near $\omega_r = \omega_c$ it is appropriate to make a classical rotating wave approximation and neglect higher order terms in $\alpha\omega$ and $\beta\omega$, resulting in the simplified transmission spectra \cite{Harder2016b},\footnote{In this approximation $\omega^2 - \omega_{c,r}^2 \sim \left(\omega - \omega_{c,r} \right) 2 \omega_c$.}
\begin{equation}
|\text{S}_{21}|^2 = \eta \frac{\omega_c^2}{4} \frac{|\omega - \tilde{\omega}_r|^2}{|\left(\omega - \tilde{\omega}_c\right)\left(\omega - \tilde{\omega}_r \right) - \frac{1}{4} \kappa^4 \omega_c^2|^2} \label{eq:oscTSimp}.
\end{equation}
The definitions $\tilde{\omega}_c = \omega_c - i \beta \omega_c$ and $\tilde{\omega}_r = \omega_r - i \alpha \omega_c$ help to simplify the notation and the fixed $H$ eigenmodes, as determined by the poles of $\text{S}_{21}$, are
\begin{equation}
\tilde{\omega}_\pm = \frac{1}{2} \left[\tilde{\omega}_r + \tilde{\omega_c} \pm \sqrt{\left(\tilde{\omega}_r - \tilde{\omega_c}\right)^2 + \kappa^4\omega_c^2}\right]. \label{eq:wpmSimp}
\end{equation} 

\begin{figure}[t!]
\centering
\includegraphics[width=12cm]{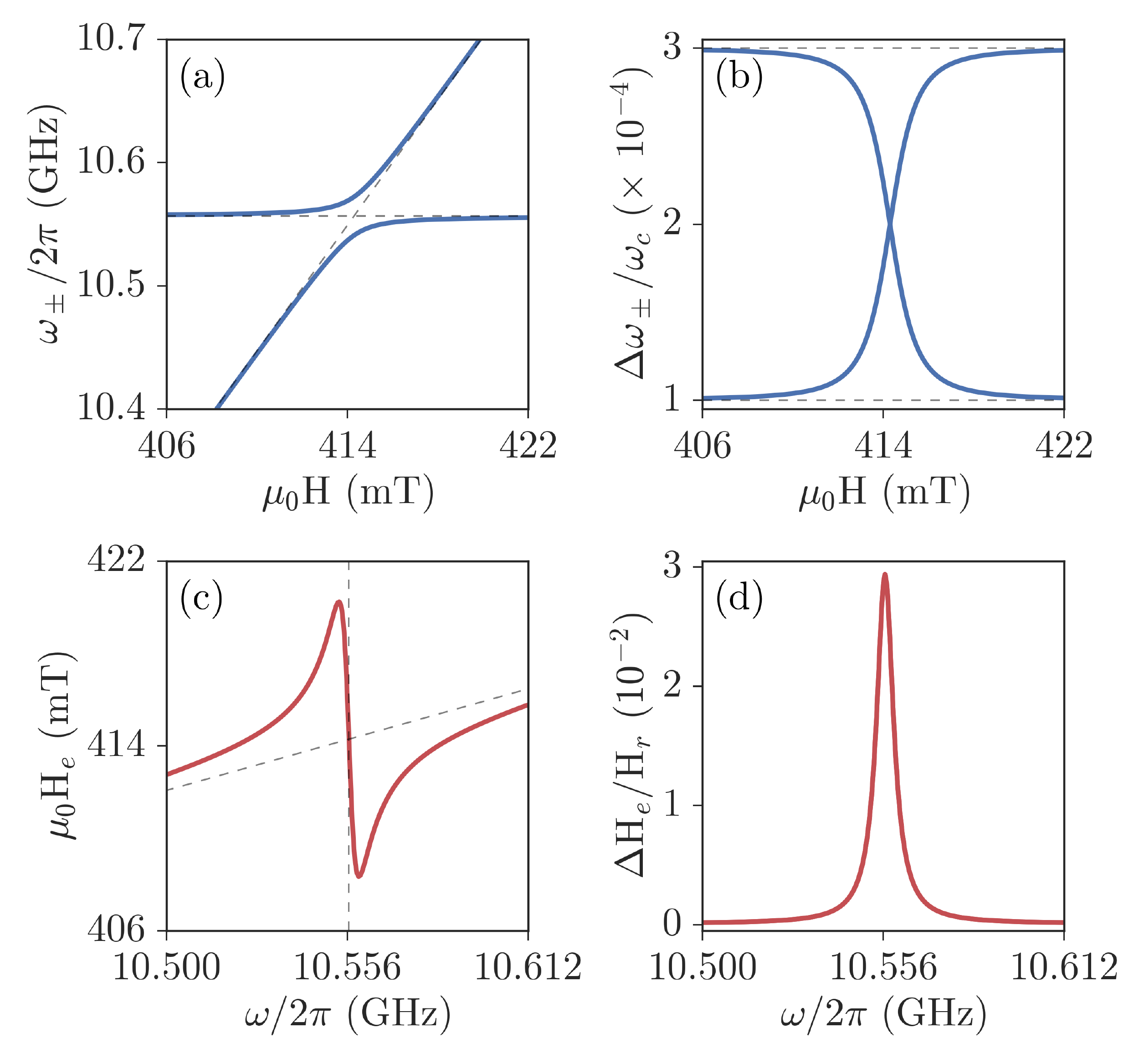}
\caption[Frequency and field dispersion of the cavity-magnon-polariton]{(a) $\omega_\pm$ and (b) $\Delta \omega_\pm$ calculated according to Eq. \eqref{eq:wpmSimp} showing the anticrossing and line width evolution signatures of the spin-photon coupling respectively.  The horizontal and diagonal dashed lines in (a) show the uncoupled cavity and FMR dispersions respectively, while the horizontal dashed lines in (b) indicate the damping limits due to $\alpha$ and $\beta$.  (c) H$_e$ and (d) $\Delta\text{H}_e$ calculated according to Eq. \eqref{eq:fieldDisp}.  The vertical and diagonal dashed lines in (c) indicate the uncoupled cavity and FMR dispersions respectively.  Spin-photon coupling induces an antisymmetric deviation in the frequency swept dispersion at the crossing point and a Lorentz type enhancement of the frequency swept line width.}
\label{fig:freqDisp}
\end{figure}

The real and imaginary contributions of the eigenvalues are plotted in Fig. \ref{fig:freqDisp} (a) and (b) respectively.\footnote{Analytic expressions for the real and imaginary components can be determined from Eq. \eqref{eq:wpmSimp} using de Moivre's theorem as described in Appendix \ref{sec:appOscDispSimp}.}  Panel (a) shows a clear anticrossing, which can already be observed in the full dispersion mapping, with horizontal and vertical dashed lines indicating the uncoupled cavity and FMR dispersions respectively.  The strength of the spin-photon interaction determines the size of the eigenmode separation observed in Fig. \ref{fig:freqDisp} (a), and therefore it is useful to define the Rabi gap,\footnote{The \emph{generalized} Rabi gap is defined as $\omega_+ - \omega_-$ without the constraint $\omega_r = \omega_c$.  Therefore the Rabi gap is the minimum value of the generalized Rabi gap.} 
\begin{equation}
\omega_\text{gap} = \left(\omega_+ - \omega_-\right)|_{\omega_c = \omega_r}, \label{eq:gapDef}
\end{equation}
which can be calculated using Eq. \eqref{eq:wpmSimp} as
\begin{equation}
\omega_\text{gap}= \sqrt{\kappa^4 \omega_c^2 - \omega_c^2 \left(\beta - \alpha\right)^2} \label{eq:gapOsc}.
\end{equation}
Typically $\beta - \alpha \ll \kappa^2$ and therefore $\omega_\text{gap} = \kappa^2 \omega_c$.  This means that in the strongly coupled regime the size of the Rabi gap depends directly on the coupling strength, which can be used to quickly determine the coupling strength experimentally.  However, Eq. \eqref{eq:gapOsc} also indicates that it is possible to have $\omega_\text{gap} = 0$ even when $\kappa \ne 0$.  This special condition actually indicates the presence of an exceptional point in the eigenmode spectrum due to the non-Hermiticity of the spin-photon system \cite{Harder2017}. 

Fig. \ref{fig:freqDisp} (b) shows $\Delta\omega_\pm$, illustrating another feature of coupling, line width evolution.  From Eq. \eqref{eq:wpmSimp} it can be shown that $\Delta \omega_+ + \Delta \omega_- = \omega_c \left(\alpha + \beta\right)$, indicating that although coupling enables energy exchange between the spin and photon subsystems it does not introduce additional dissipation channels, and therefore if $\kappa^4 > 0$ then $\Delta\omega_\pm$ is bounded above and below by $\alpha$ and $\beta$.

The transmission spectra can also be examined for fixed $\omega$.  In this case the pole in the transmission spectra from Eq. \eqref{eq:oscTSimp} is, 
\begin{equation}
\tilde{\omega}_e = \omega - \frac{\kappa^4 \omega_c^2 \left(\omega - \omega_c\right)}{4 \left(\omega - \omega_c\right) + \beta^2 \omega_c^2} + i \left[\alpha \omega_c + \frac{\beta \omega_c^3 \kappa^4}{4 \left[\left(\omega - \omega_c\right)^2 + \beta^2 \omega_c^2\right]}\right]. \label{eq:fieldDisp}
\end{equation} 
This is the field swept analogue of Eq. \eqref{eq:wpmSimp} however in this case there is only one solution.  The real and imaginary components of $\tilde{\omega}_e$ are plotted in Fig. \ref{fig:freqDisp} (c) and (d) respectively using the notation $\omega_e = \text{Re}\left(\tilde{\omega}_e\right)$, $\Delta\omega_e = \text{Im}\left(\tilde{\omega}_e\right)$, $\omega_e = \gamma H_e$, $\Delta \omega_e = \gamma \Delta H_e$ and $\omega_r = \gamma H_r$.  The effects of coupling are evident in the field dispersion plotted in Fig. \ref{fig:freqDisp} (c) from the asymmetric deviation from the dashed line, which indicates the uncoupled FMR behaviour.  Again the deviation due to hybridization is greatest near the crossing point of the uncoupled modes and has a dispersive line shape, with a polarity determined by the sign of the coupling strength $\kappa^4$.  On the other hand the line width, shown in Fig. \ref{fig:freqDisp} (d), has a Lorentz peak at $\omega = \omega_c$.  This resonant enhancement of the FMR line width is in addition to the usual linear $\omega$ dependence and the influence of inhomogeneous broadening and therefore plays an important role in the characterization of FMR \cite{Bai2015}.

Finally, the transmission spectra itself can be simplified which proves beneficial for experimental analysis.  For fixed field measurements $\text{S}_{21}$ can be expanded near $\omega = \omega_\pm$ to find \cite{Harder2016b} (see Appendix \ref{sec:appOscTransSimp}),
\begin{equation}
|\text{S}_{21}|^2_\pm \propto L + \left(\omega_\pm - \omega_r\right)^{-1} D \label{eq:s21LandD}
\end{equation}
where
\begin{equation}
L = \frac{\left(\Delta \omega_\pm\right)^2}{\left(\omega - \omega_\pm\right)^2 + \left(\Delta \omega_\pm\right)^2}, ~~~~~ D = \frac{\Delta \omega_\pm\left(\omega - \omega_\pm\right)}{\left(\omega - \omega_\pm\right)^2 + \left(\Delta \omega_\pm\right)^2}
\end{equation}
are the usual Lorentz and dispersive line shapes.  Therefore the fixed field transmission spectra of the coupled system generally consists of both Lorentz and dispersive contributions, which differs from the Lorentz nature of a single mode.  However generally the term in front of $D$ will suppress the dispersive character unless one of the modes approaches the FMR frequency, where it will be more greatly distorted by the FMR antiresonance \cite{Harder2016}.  Therefore in most cases when analyzing experimental data it is appropriate to fit each peak in the frequency swept CMP transmission spectra independently as a Lorentzian in order to determine the frequency and line width.

On the other hand for the field swept case near $\tilde{\omega}_e$ (see Appendix \ref{sec:appOscTransSimp})
\begin{equation}
|\text{S}_{21}|^2 \propto \frac{\left(q\Delta \omega_e + \omega_r - \omega_e\right)^2}{\left(\omega_r - \omega_e\right)^2 + \Delta \omega_e^2} \label{eq:fano}
\end{equation}
where $q\left(\omega\right) = \left(\omega_e - \omega\right)/\Delta \omega_e$.  The asymmetric transmission of Eq. \ref{eq:fano} is actually a Fano line shape, which typically arises in scattering problems due to the interference between background and resonance processes \cite{Fano1961, ConneradeBook}, with the Fano factor $q$ controlling the degree of asymmetry.  In the case of the CMP $q\left(\omega > \omega_c\right) < 0, q\left(\omega = \omega_c\right) > 0$ and $q\left(\omega < \omega_c\right) >0$, which explains the change in asymmetry of the field swept line shape observed in Fig. \ref{fig:oscFig}.  Despite slight deviations, the Fano function is sufficiently similar to the sum of dispersive and Lorentz line shapes that from a practical perspective the resonance frequencies and line widths of the hybridized modes can be accurately determined by fitting the transmission spectra to a functional form $|\text{S}_{21}\left(\text{H}\right)|^2 \propto L + D$, where $L$ and $D$ are Lorentz and dispersive line shapes with resonance frequency $\omega_e$ and line width $\Delta \omega_e$.  This provides a simple way to analyze experimental results. 


\subsubsection{CMP Eigenvectors and Energy Distribution} \label{subsubsec:eigenvectors}

\begin{figure}[t!]
\centering
\includegraphics[width=9.5cm]{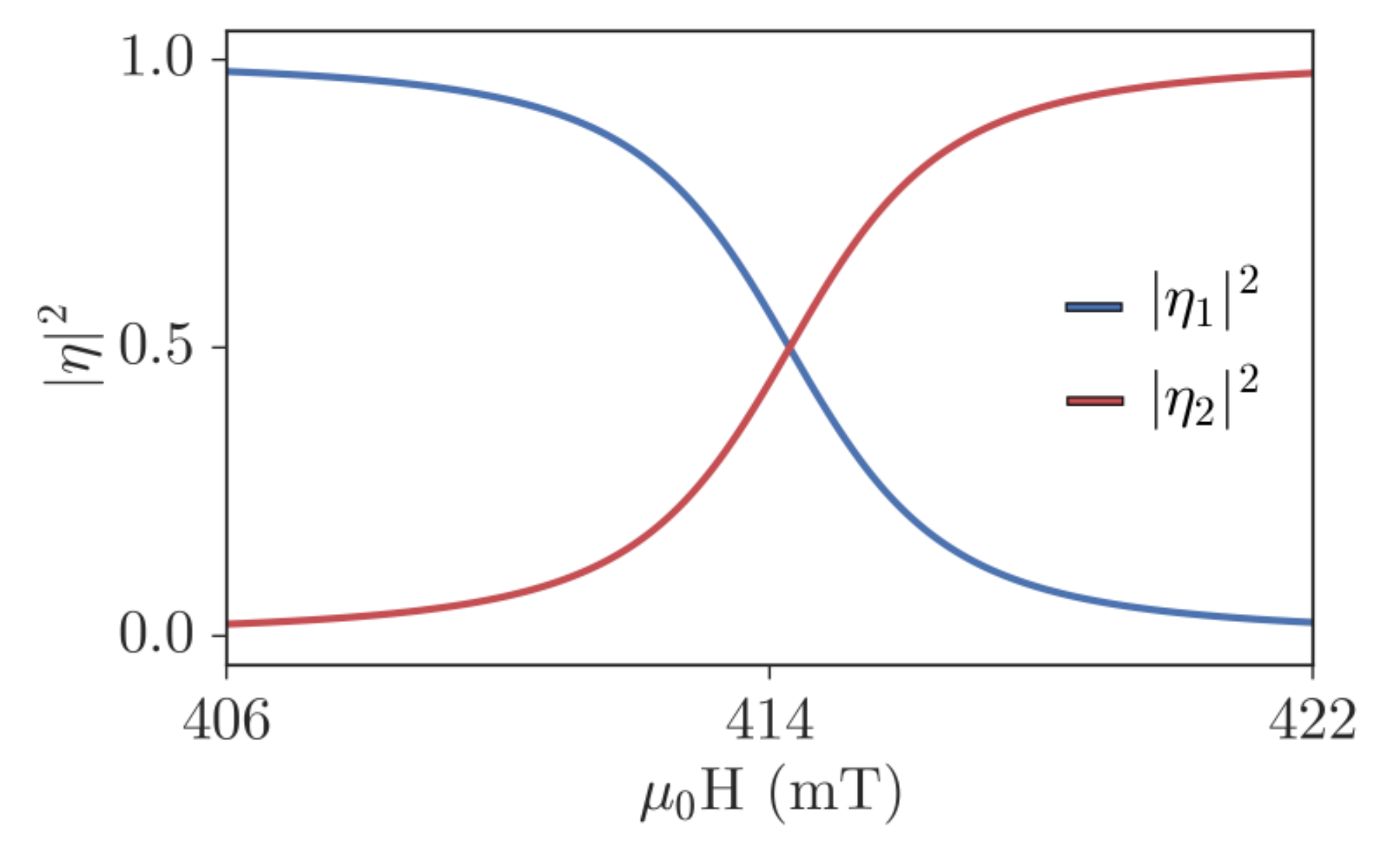}
\caption[Mode composition of the cavity-magnon-polariton]{The Hopfield-like coefficients $\eta_{1,2}$ determine the composition of the polariton branches.  The evolution of $\eta_{1,2}$ with $\mu_0 H$ corresponds to the upper CMP branch transitioning from photon-like to spin-like as the field is increased, with the opposite behaviour for the lower branch.}
\label{fig:hopfield}
\end{figure}

From the dispersion it is clear that the CMP is a hybridization of magnon and photon states.  This fundamental property is also reflected in an important way through the CMP eigenvectors.  The CMP modes, $X_\pm$, are a mixture of spin and photon modes $x_{1,2}$ (see Appendix \ref{sec:appOscNormalModes})
\begin{equation}
\left(\begin{array}{c}
X_+ \\
X_- \\
\end{array}\right) = \left(\begin{array}{cc}
\eta_- & \eta_+ \\
-\eta_- & \eta_+
\end{array}\right) \left(\begin{array}{c}
x_1 \\
x_2 
\end{array}\right),\label{eq:hopfieldTransform}
\end{equation}
where $\eta_\pm = \sqrt{\Omega \pm \Delta}/\sqrt{2\Omega}$ with frequency detuning $\Delta = \tilde{\omega}_r - \tilde{\omega}_c$ and generalized Rabi frequency $\Omega = \sqrt{\left(\tilde{\omega}_r - \tilde{\omega}_c\right)^2 + \kappa^4 \omega_c^2}$.  Eq. \eqref{eq:hopfieldTransform} describes a Hopfield like transformation \cite{TimofeevBook} between the polariton modes ($X_\pm$) and the photon and spin states ($x_{1,2}$), which determines the spin and photon fractions of each CMP branch, as shown in Fig. \ref{fig:hopfield}.  At low fields the upper CMP branch is dominated by the cavity photons, which is why the dispersion approaches that of the uncoupled cavity, whereas at high fields the spin fraction dominates and the dispersion becomes FMR-like.  This situation is reversed for the lower CMP branch.  Since the microwave transmission is sensitive to the photon-like nature of the hybridized modes, the mode composition explains why the strongest signal is observed for the upper (lower) branch when $H \ll H_r$ ($H \gg H_r$).  Meanwhile at the crossing point both upper and lower branches have equal spin and photon contributions, and therefore both branches have equal transmission amplitudes at this point.

At the crossing point, where the fractional composition is equal, and in the absence of damping,
\begin{equation}
X_\pm = \left(\begin{array}{c}
\pm \text{sign} \left(\kappa^2\right) \\
1 \end{array}\right).
\end{equation}
Therefore the coherent nature of the hybridized modes is determined by the sign of the coupling strength. Physically this behaviour is most easily interpreted by considering the Hamiltonian of a coupled oscillator system \cite{KaurThesis2017}
\begin{equation}
2H/m =  \dot{x}_1^2 + \dot{x}_2^2 + \omega_c^2 x_1^2 + \omega_r^2 x_2^2 + \kappa^2 \omega_c^2 x_1 x_2.
\end{equation}
Here the first two terms are the kinetic energies of $O_1$ and $O_2$ respectively, the third and fourth terms are the potential energies of each oscillator and the final term is the interaction energy.  Therefore the coupling strength  determines the sign of the interaction potential and as a result determines the coherent properties of the hybridization.  In the case of the CMP, $\kappa^2 > 0$ and therefore the higher energy mode $X_+ = (1,1)$ has in-phase motion while the lower energy mode $X_- = (-1,1)$ has out-of-phase motion.  

The generality of the oscillator model enables a comparison of the CMP behaviour to the well known case of molecular bonding, where two atoms combine to form bonding and anti-bonding orbitals \cite{KaurThesis2017}.  In that case the bonding potential is attractive, meaning that the analogous $\kappa^2$ would be negative.  Therefore the bonding orbitals have lower energy than either of the two separate atoms and correspond to in-phase oscillations, while the anti-bonding orbitals have out-of-phase motion and a higher energy.  Of course for the CMP the spring which models the spin-photon interaction creates a repulsive potential, leading to the opposite behaviour.  Note that since the Rabi gap is typically determined experimentally, and is proportional to $\kappa^4$, the sign of $\kappa^2$ is experimentally ambiguous.  However the fact that $\kappa^2 > 0$ can be further justified by comparison to the quantum model as described in Sec. \ref{subsec:jaynesCummings} and in detail in Ref. \cite{KaurThesis2017}.

The in-phase and out-of-phase oscillations are a special example of the more general phase relationship between magnon and photon states which create the CMP.  The full eigenmodes can be written as
\begin{equation}
X_\pm \propto \frac{\pm \sqrt{\Omega^2 - \Delta^2}}{\left(\Omega \pm \Delta\right)}x_1 + x_2 = e^{i\phi_\pm} x_1 + x_2
\end{equation}
where 
\begin{equation}
e^{i\phi_\pm} = \frac{\pm \sqrt{\Omega^2 - \Delta^2}}{\left(\Omega \pm \Delta\right)} = \frac{1}{2}\frac{\kappa^2 \omega_c}{\tilde{\omega}_\pm - \tilde{\omega}_c}
\end{equation}
defines the phase $\phi_\pm$.  This relationship reveals an important physical insight into the CMP, which is actually a general property of polaritons: the CMP results from a strict phase correlation between magnon and photon states.  Essentially this reflects the fact that at any given configuration, defined by the magnetic bias field, the hybridized state will have a strictly defined photon/spin composition.  


\subsection{Phase Correlation: An Electrodynamic Approach} \label{subsec:circuit}

Although the harmonic CMP model provides useful intuition, to truly understand the origin of phase correlation it is necessary to carefully examine the material dynamics of the magnetic system.  Since the CMP is formed using a macroscopic spin ensemble, the magnetization dynamics can be described by the Landau-Lifshitz-Gilbert equation \cite{Landau1935, Gilbert2004} which describes the precession of a magnetization $\textbf{M}$ due to a magnetic field $\textbf{H}_f$ 
\begin{equation}
\frac{d\textbf{M}}{dt} = \gamma \textbf{M} \times \textbf{H}_f - \frac{\alpha}{M} \left(\textbf{M} \times \frac{d \textbf{M}}{dt}\right) \label{eq:llg}
\end{equation}
where $\alpha$ is the Gilbert damping parameter, phenomenologically describing the damping of the spin system.  The LLG equation can be solved by splitting the magnetization into a static contribution $\textbf{M}_0$ along the $\widehat{\textbf{z}}$ direction and dynamic components $\textbf{m}(t)$ in the $x$-$y$ plane, $\textbf{M} = \textbf{M}_0 + \textbf{m}(t) = M_0 \widehat{\textbf{z}} + \textbf{m}(t)$.  Similarly $\textbf{H}_f = \textbf{H} + \textbf{h}(t) = H \widehat{\textbf{z}} + \textbf{h}(t)$.  Assuming a harmonic time dependence, $\textbf{m}(t) = \textbf{m} e^{-i\omega t}$ and $\textbf{h}(t) = \textbf{h} e^{-i\omega t}$ and accounting for sample geometry through the demagnetization fields which send, $h_x \to h_x - N_x m_x, h_y \to h_y - N_y m_y$ and $H \to H - N_z M_z$, the linear response is governed by
\begin{equation}
\textbf{m} = \chi \textbf{h}
\end{equation}
where the response function $\chi$ is given by
\begin{equation} 
\chi = A^{-1} \left(\begin{array}{cc}\omega_m \left[\omega_m N_y + \omega_0 - i \alpha \omega\right] & i \omega \omega_m \\
-i\omega \omega_m & \omega_m \left[\omega_m N_x + \omega_0 - i \alpha \omega\right] \end{array}\right)
\end{equation}
with $A = \omega_r^2 - \omega^2 - i \alpha \omega \left(\omega_m \left(N_x + N_y\right) + 2 \omega_0\right)$, $\omega_m = \gamma M_0,\omega_0 = \gamma \left(H_0 - N_z M_0\right)$ and \\ $\omega_r^2 = \gamma^2 \left[H_0 +\left(N_y - N_z\right) M_0\right]\left[H_0 +\left(N_x - N_z\right) M_0\right]$.  Physically $\omega_r$ is the FMR resonance frequency and $\omega_0$ is the resonance frequency in the limit of zero demagnetization fields.  For small damping, so that terms of order $\alpha^2$ may be neglected, the eigenvectors of the response function are the elliptically polarized modes $m^+ = D m_x + i m_y$ and $h^+ = m_x + i D m_y$ with ellipticity $D$,
\begin{equation}
D = \sqrt{\frac{\omega_m N_x + \omega_0 - i \alpha \omega}{\omega_m N_y + \omega_0 - i \alpha \omega}}.
\end{equation}
The magnetization dynamics in terms of the elliptically polarized modes is exceptionally simple,
\begin{equation}
m^+ = \frac{-\omega_m}{\omega - \omega_r + i \alpha \omega}h^+. \label{eq:mdiag}
\end{equation}
Here it is also assumed that $\omega_r \approx \omega_0 + \frac{1}{2}\omega_m\left(N_x + N_y\right)$, which holds well for small demagnetizing fields or in a narrow field range.  The latter is generally true for CMP experiments. 

Eq. \eqref{eq:mdiag} is true for any magnetization dynamics described by the LLG equation.  Typically the microwave fields are taken to be plane waves and the magnetization dynamics then studied.  However for the CMP plane wave solutions are no longer appropriate and in principle it is necessary to self consistently incorporate the cavity boundary conditions into the solutions of Maxwell's equations.  A variety of approaches have been taken to solve this problem in a rigorous way.  For example, Maksymov et al. \cite{Maksymov2015} used numerical solutions of the LLG and Maxwell equations, obtained using a finite-difference time-domain method, to consider a resonator formed by a dielectric-magnetic multilayer, Cao et al. \cite{Cao2014} and Yao et al. \cite{Yao2015} used a scattering approach in a simplified 1D configuration to examine the effects of coupling, with the influence of spin waves explicitly accounted for in Ref. \cite{Cao2014} and finally, the behaviour of the dispersion was examined by Krupka et al. \cite{Krupka2017} and Pacewicz et al. \cite{Pacewicz2018}.  In the former work the authors revisited the role of perturbation theory for sample characterization in traditional cavity based FMR measurements by numerically computing the cavity quality factor in regimes where magnon-photon coupling may play a dominant role.  Indeed they found that proper characterization of the FMR line width should carefully account for the influence of coupling.  Similarly, in Ref. \cite{Pacewicz2018} the authors accounted for the intrinsic permeability of the ferromagnetic material and numerically solved the transcendental equation for the electromagnetic field. 

In all approaches the features of Maxwell's equations which are necessary to describe the CMP are: (i) a resonance (technically due to the cavity or resonator boundary conditions) and (ii) dissipation of the electromagnetic field (typically dominated by conductive losses in the cavity walls).  It is therefore possible to model the properties of the microwave cavity as an RLC circuit, in an approach that is perhaps most familiar to microwave engineers \cite{PozarBook}.\footnote{This approach is also similar to that taken by Bloembergen and Pound in their seminal work on the radiation damping in nuclear magnetic resonance \cite{Bloembergen1954}.}  Therefore consider an artificial RLC circuit with current $j^+(t) = D j_x(t) + i j_y(t)$ \cite{Bai2015, Harder2016b}.  The dynamics of $j^+(t)$ is governed by the RLC equation,
\begin{equation}
R j^+(t) + \frac{1}{C} \int j^+(t) dt + L \frac{dj^+(t)}{dt} = V^+(t)
\end{equation}
where $R$, $L$ and $C$ are the resistance, inductance and capacitance of the model circuit, respectively.  Most importantly this circuit contains a voltage $V^+(t) = i K_c L \frac{dm^+(t)}{dt}$ produced by the precessing magnetization according to Faraday's law.  Here $K_c$ is a coupling parameter which characterizes the back action of the magnetization dynamics onto the cavity field.  The inclusion of $V^+$ is the key step in modelling the CMP \cite{Bai2015, Harder2016b}.  Assuming harmonic time dependence, $j^+(t) = j^+ e^{-i\omega t}$ and $m^+ = m^+ e^{-i\omega t}$ the circuit equation becomes
\begin{equation}
\left(\omega^2 - \omega_c^2 + 2 i \beta_\text{int} \omega_c\right) j^+ = - i \omega^2 K_c m^+ \label{eq:rlcM}
\end{equation}
where $\omega_c = \sqrt{\frac{1}{LC}}$ and $\beta_\text{int} = \frac{R}{2} \sqrt{\frac{C}{L}}$.  Since the field that drives the FMR is generated by the current due to Amp\'ere's law, $j_x$ and $j_y$ will produce fields $h_x$ and $h_y$ according to the phase relation $h_x = K_m j_y$ and $h_y = - K_m j_x$.  Therefore the elliptically polarized fields are $h^+ = -K_m j^+$, where $K_m$ is a coupling constant which characterizes the microwave current induced field, and the coupled equations of motion for the CMP system may be written as,
\begin{equation}
\left(\begin{array}{cc}
\omega^2 - \omega_c^2 + 2 i \beta_\text{int} \omega_c \omega & i \omega^2 K_c \\
-i \omega_m K_m & \omega - \omega_r + i \alpha \omega \end{array}\right) \left(\begin{array}{c}
j^+ \\
m^+
\end{array} \right) = 0. \label{eq:circuitMatrix}
\end{equation}
The determinant of the matrix in Eq. \eqref{eq:circuitMatrix} determines the CMP dispersion.  In the classical rotating wave approximation, and to lowest order in coupling strength and damping near the crossing point, this dispersion is identical to the coupled oscillator system.  Therefore all of the behaviour already discussed in the context of the harmonic model is automatically included in the electrodynamic description.  Importantly however, this more physical approach reveals the origin of the phase correlation in the CMP.  Due to Amp\`ere's law, which locks the phase of the rf current with that of the rf magnetic field, the phase difference between the rf current carried by the cavity mode and the rf magnetization excited in the YIG is fixed to form the hybridized modes of the CMP.

\begin{figure}[b!]
\centering
\includegraphics[width=11cm]{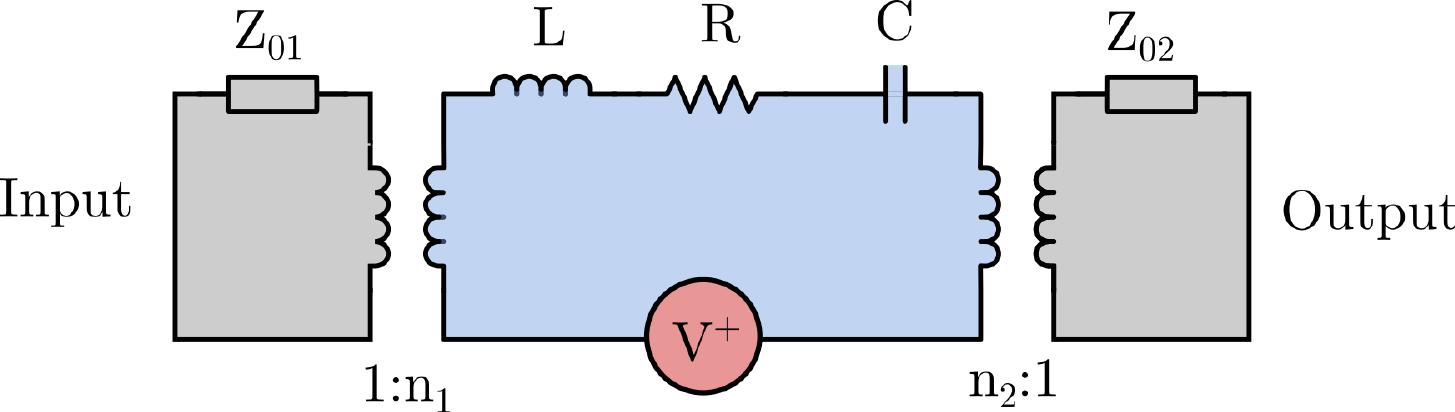}
\caption[RLC circuit for cavity-magnon-polariton]{Circuit model of the cavity-magnon-polariton.  The RLC circuit in blue models the microwave cavity, with a voltage source, shown in red, representing the effect of the magnetization precession. Input and output fields are inductively coupled into the microwave cavity.}
\label{fig:rlcCircuit}
\end{figure}

Microwave circuit theory can now be used to determine the full transmission spectra, as opposed to the dispersion alone.  The artificial circuit used to model the cavity is shown in Fig. \ref{fig:rlcCircuit} and is inductively coupled to input and output ports of impedance and number of turns in the inductor $Z_{01}, n_1$ and $Z_{02}, n_2$ respectively.  The total impedance of the circuit and voltage source is $Z = Z_c + Z_m$ where, from Eq. \ref{eq:mdiag} and the definition of $V^+$,
\begin{equation}
V^+ = \frac{i\omega_m K_c K_m L \omega}{\omega- \omega_r + i \alpha \omega} j^+,
\end{equation}
the inductance due to the voltage source is,
\begin{equation}
Z_m  = \frac{i\omega_c K_c K_m L \omega}{\omega- \omega_r + i \alpha \omega}.
\end{equation}
From Eq. \ref{eq:rlcM} the inductance due the RLC circuit is
\begin{equation}
Z_c = -i \frac{L}{\omega} \left(\omega^2 - \omega_c^2 + 2 i \beta_\text{int} \omega_c \omega\right).
\end{equation}
Therefore the transfer matrix which determines the relationship between the voltages in the input and output circuits is \cite{PozarBook},
\begin{equation}
T = \left(\begin{array}{cc}
A & B \\
C & D \end{array} \right) = \left(\begin{array}{cc}
\frac{n_2}{n_1} & \frac{Z}{n_1 n_2} \\
0 & \frac{n_1}{n_2} \end{array}\right),
\end{equation}
allowing the transmission to be calculated as \cite{Harder2016b}
\begin{align}
\text{S}_{21} &= \frac{2\left(Z_{01} Z_{02}\right)^{1/2}}{AZ_{02} + B + C Z_{01} Z_{02} + D Z_{01}}, \\
\text{S}_{21} & = \frac{2i\omega_c \omega \beta_L \overline{\text{S}}_{21}\left(\omega - \omega_r + i \alpha \omega\right)}{\left(\omega^2 - \omega_c^2 + i \beta_L \omega \omega_c\right)\left(\omega - \omega_r + i \alpha \omega\right) - \omega^2 \omega_m K^2}. \label{eq:s21Circ}
\end{align}
Here the input and output couplings, $2\omega_c \beta^\text{ext}_i = Z_{0i}n_i^2/L$, characterize how microwaves are input and output from the circuit and the loaded damping, $\beta_L = \beta_\text{int} + \beta^\text{ext}_1 + \beta^\text{ext}_2$, and coupling strength, $K^2 = K_c K_m$, have been defined.  The transmission amplitude in the zero coupling limit, $\overline{\text{S}}_{21} = \sqrt{\beta_1^\text{ext} \beta_2^\text{ext}}/\beta_L = \text{S}_{21} \left(\omega = \omega_c, K^2 = 0\right)$, physically defines the power input into the microwave cavity.  The microwave reflection, $\text{S}_{11}$ can also be found directly from the transfer matrix.  For reciprocal systems $n_1 = n_2$ and $Z_{01} = Z_{02}$ and therefore $\text{S}_{11} = -1 + \text{S}_{21}$.  

\begin{figure}[t!]
\centering
\includegraphics[width=12cm]{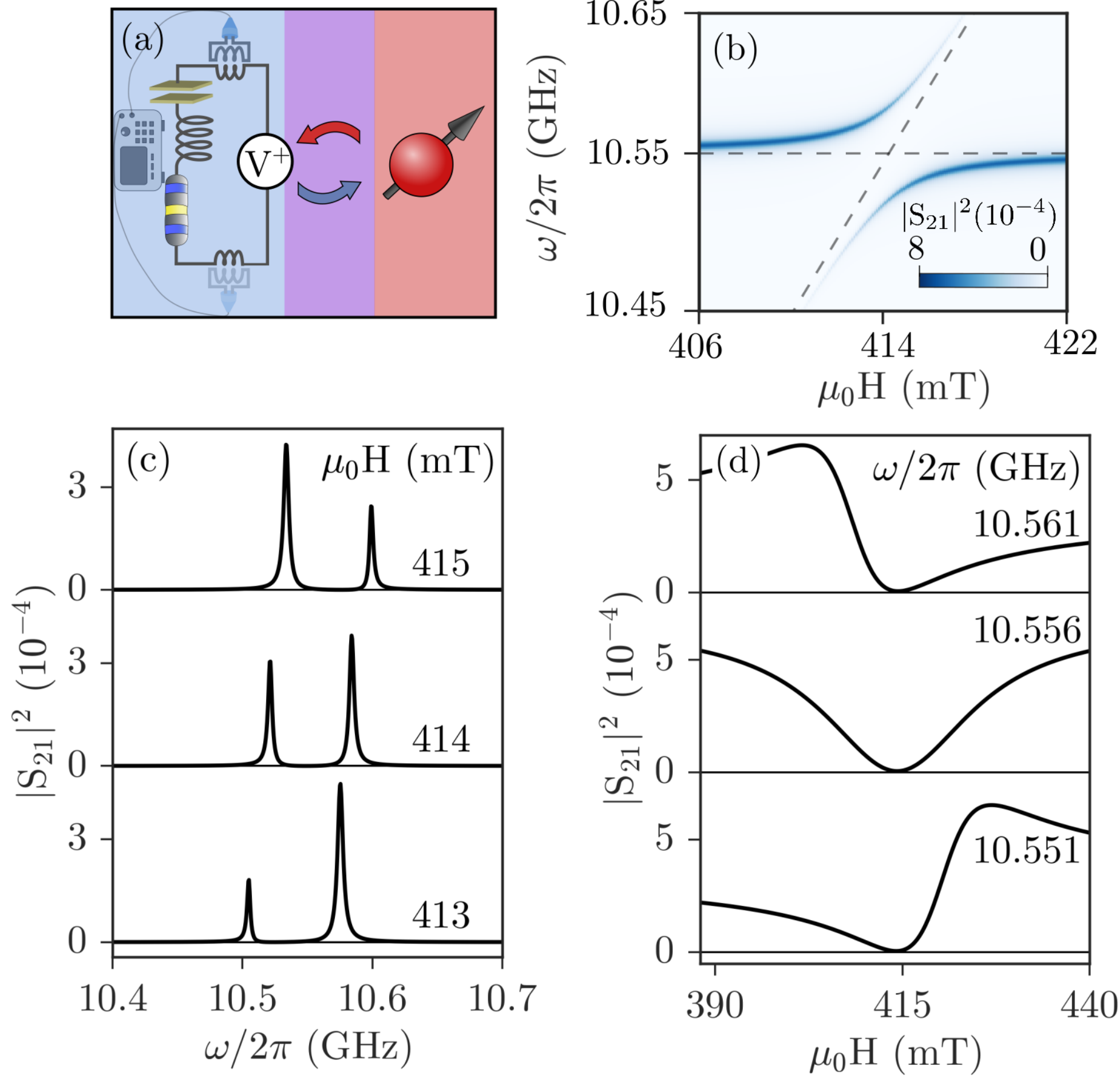}
\caption[Cavity-magnon-polariton transmission spectra in the electrodynamic phase correlation model]{(a) Schematic illustration of the electrodynamic phase correlation which produces the CMP.  The current in an RLC circuit produces a microwave field which drives magnetization precession.  The precessing magnetization induces a voltage via Faraday's law, coupling back to the magnetic field.  (b) The full transmission spectra calculated using Eq. \ref{eq:s21Circ} and experimentally relevant parameters.  (c) Frequency and (d) field line cuts of the transmission spectra.  A modified version of this figure was originally published in Ref. \cite{Harder2016b}.}
\label{fig:rlcFig}
\end{figure}

The transmission properties of Eq. \ref{eq:s21Circ} are plotted in Fig. \ref{fig:rlcFig} using the same experimental parameters summarized previously.  All key signatures of the CMP are consistent with the oscillator model.  In fact by making the rotating wave approximation and neglecting higher order contributions so that $\alpha \omega \sim \alpha \omega_c$, $\text{S}_{21}$ becomes
\begin{equation}
\text{S}_{21} = \frac{i \omega \beta_L \overline{\text{S}}_{21} \left(\omega - \omega_r + i \alpha \omega_c\right)}{\left(\omega - \omega_c + i \beta_\text{int} \omega_c\right)\left(\omega - \omega_r + i \alpha \omega_c\right) - \frac{\omega_c \omega_m K^2}{2}}, \label{eq:circTSimp}
\end{equation}     
which has the same form as the oscillator simplification of Eq. \eqref{eq:oscTSimp}.  Therefore all of the properties, such as line shape, Rabi gap and mode composition, discussed in the context of the harmonic model apply here as well.  

In addition to revealing the nature of hybridization, the phase correlation model also provides a direct interpretation of the two coupled oscillators in the harmonic model: One oscillator represents the precessing magnetization of the spin system and the other represents the microwave current of the cavity (or equivalently the microwave field in the cavity).  Therefore it is then natural to define the spectral functions of the coupled system in a physically meaningful way.  In the limit that $K_c \to 0$, the microwave cavity mode will be described by a single Lorentz peak at the cavity frequency $\omega_c$.  This behaviour can be described by the spectral function 
\begin{equation}
\mathbb{S}_c = \frac{1}{\omega^2 - \omega_c^2 + 2 i \beta_\text{int} \omega_c \omega} \label{eq:spectralCavity}
\end{equation}
in which case the transmission spectra of Eq. \eqref{eq:s21Circ} can be written as
\begin{equation}
\text{S}_{21} = 2 i \beta_L \omega_c \omega \overline{\text{S}}_{21} \left[\text{Re}\left(\mathbb{S}_c\right) + i \text{Im}\left(\mathbb{S}_c\right)\right] \label{eq:s21Cavity}
\end{equation}
and therefore
\begin{equation}
|\text{S}_{21}| = \overline{\text{S}}_{21} = \sqrt{2 \beta_L \omega \omega_c |\text{Im}\left(\mathbb{S}_c\right)|}. \label{eq:s21CavityMag}
\end{equation}
The full spectral function for the cavity, when coupled to the spin system, is defined analogously,
\begin{equation}
\mathbb{S}^\text{coup}_c = \frac{\omega - \omega_r + i \alpha \omega}{\left(\omega^2 - \omega_c^2 + 2 i \beta_L \omega \omega_c\right)\left(\omega - \omega_r + i \alpha \omega \right) - \omega^2 \omega_m K^2}.
\end{equation}
The transmission of the coupled system can then be described by Eqs. \eqref{eq:s21Cavity} and \eqref{eq:s21CavityMag} by replacing $\mathbb{S}_c \to \mathbb{S}_c^\text{coup}$.  This approach is especially relevant to understanding the behaviour of hybridization in bilayer devices, where the spin current is affected \cite{Bai2015, MaierFlaig2016, Bai2017}.  In this case the spectral function of the magnetic system, which would determine the spin pumping voltage is
\begin{equation}
\mathbb{S}_m^\text{coup} = \frac{-i \omega_m K_m}{\left(\omega^2 - \omega_c^2 + 2 i \beta_L \omega \omega_c\right)\left(\omega - \omega_r + i \alpha \omega \right) - \omega^2 \omega_m K^2}.
\end{equation}
The spin pumping voltage (see e.g. \cite{Harder2016a}) would then be $V_\text{SP} \propto |m|^2 = |\mathbb{S}_m|^2$ \cite{Bai2013}.  Taking the plane wave limit, $\omega_c \to 0$, and removing the magnetization back action by setting $K_c \to 0$, the standard result of $V_\text{SP} \propto \left(\Delta H\right)^2/\left[\left(H - H_r\right)^2 + \left(\Delta H\right)^2\right]$ is recovered \cite{Harder2016a}.  In the coupled case this expression will naturally be modified due to the modification of the spectral function.


\subsection{A Quantum Approach} \label{subsec:quantum}

The classical description of the CMP has been used to successfully described important experimental results such as non-local spin current manipulation \cite{Bai2017} and on-chip CMP control \cite{Kaur2016a}. However properly engineered experimental systems are also pushing the quantum frontiers of the CMP through the use of cooperative polariton dynamics \cite{Yao2017} or the introduction of inherently quantum systems \cite{Tabuchi2015}.  The theoretical groundwork for these and a number of additional phenomena, e.g. bistability \cite{Wang2018}, coherent perfect absorption \cite{Zhang2017a} and magnon dark modes \cite{Zhang2015g}, is the Jaynes-Cummings model and input-output formalism, which are summarize here to serve as an introduction to the many developments which are beyond the focus of this early review.

\subsubsection{The Quantum Hamiltonian} \label{subsec:jaynesCummings}
The canonical Hamiltonian of quantum optics is the Jaynes-Cummings model, originally proposed to describe spontaneous emission \cite{Jaynes1963, Garraway2011}.\footnote{Technically it is the Tavis-Cummings model, the $N_s$-spin generalization of the single spin Jaynes-Cummings model \cite{Tavis1968}, which is relevant for a description of the CMP.  However when considering the collection of spins as a single macrospin this distinction is not always made.}  Consider an ensemble of $N_s$ spins coupled to a single mode of a resonant cavity, $\omega_c$.  Assuming $N_s \gg 1$, the Holstein-Primakoff transformation \cite{Holstein1940, MajlisBook} applies and the excitations of the spin system (magnons) can be treated as bosonic particles.  It is also appropriate to consider only the lowest spin excitation at $\omega_r$ (the FMR mode).  The simplest Hamiltonian which describes such a spin-photon interaction should contain a kinetic term for each degree of freedom, plus all possible quadratic interactions, and may therefore be written as,
\begin{equation}
H_\text{CMP} = \hbar \omega_c a^\dagger a + \hbar \omega_r b^\dagger b + \hbar g\left(a+ a^\dagger \right) \left(b + b^\dagger\right).
\label{eq:JCInitial}
\end{equation}
Here $a^\dagger$ ($a$) and $b^\dagger$ ($b$) are the creation (annihilation) operators for the photon and magnon, respectively, and $g$ is the coupling strength.  In the interaction picture, the time evolution of $a$ and $b$ is governed by the non-interacting Hamiltonian, $H_0 = \hbar \omega_c a^\dagger a + \hbar \omega_r b^\dagger b$,
\begin{equation}
\dot{a} = -\frac{i}{\hbar} \left[a, H_0\right], ~~~~~~ \dot{b} = -\frac{i}{\hbar}\left[b, H_0\right],
\label{eq:JCTE}
\end{equation}
and therefore
\begin{equation}
a\left(t\right) = e^{-i \omega_c t} a(0), ~~~~~ b\left(t\right) = e^{-i \omega_r t} b(0).
\label{eq:JCTES}
\end{equation}
As a result, the spin-photon interactions take the form $a b \sim e^{-i\left(\omega_c + \omega_r\right) t}$ and $a b^\dagger \sim e^{-i\left(\omega_c - \omega_r\right)}$ plus Hermitian conjugates.  Since $\omega_c \sim \omega_r$, $e^{-i\left(\omega_c + \omega_r\right) t}$ is rapidly oscillating and will have a small time averaged effect.  This term can therefore be neglected through the rotating wave approximation, which amounts to keeping the component of the interaction whose time evolution (closely) co-rotates with the interaction picture eigenstates.  To physically understand this approximation, imagine a classical precessing dipole which represents transitions in a two-level system.  A linearly oscillating driving field can be decomposed into two counterrotating fields.  One of these fields will co-rotate with the dipole moment, applying a constant torque over many periods.  However the torque due to the counterrotating field component will reverse (with respect to the dipole precession) every period, and therefore will have little average effect \cite{SilvermanBook}.  Neglecting these terms leaves,
\begin{equation}
H_\text{CMP} = \hbar \omega_c a^\dagger a + \hbar \omega_r b^\dagger b + \hbar g\left(a b^\dagger+a^\dagger b\right).
\label{eq:JCFinal}
\end{equation}
This Hamiltonian governs the basic physics of the CMP and can also be derived in an electrodynamic approach using the second quantization of a Heisenberg ferromagnet coupled to light via the Zeeman interaction \cite{Harder2016b}. In the latter method it is also directly revealed that the coupling strength is proportional to $\sqrt{N_s}$ \cite{Harder2016b}\footnote{Though this scaling was already known before Ref. \cite{Harder2016b} and was the initial motivation for pursuing CMP physics \cite{Soykal2010}.} which enables the coupling strength enhancement in ferri or ferromagnetic materials.  The Hamiltonian of Eq. \eqref{eq:JCFinal} can now be used to determine the microwave transmission properties of the CMP.    


\subsubsection{Input-Output Formalism} \label{subsec:inputOutput}

In a CMP experiment external photons are coupled from a microwave feedline into the cavity.  Therefore to determine the transmission through the cavity/spin system described by Eq. \eqref{eq:JCFinal} it is necessary to couple cavity photons to the external photon bath which exists in the microwave feedlines.  This is the realm of the input-ouput formalism of quantum optics, which is summarized here in order to provide a self contained discussion of the CMP.\footnote{An alternative Green's function approach to the scattering of incident photons off of the cavity system has also been used to calculate the transmission spectra of the CMP \cite{Harder2016b}.}  The approach described below follows closely Refs. \cite{WallsBook, Clerk2010}.  

The input-ouput Hamiltonian takes the form
\begin{equation}
H = H_\text{sys} + H_\text{bath} + H_\text{int}^\text{bath},
\end{equation}
where $H_\text{sys}$ describes the cavity and its contents (e.g. cavity photons and magnons in the case of the CMP), $H_\text{bath}$ describes the bath photons and $H_\text{int}^\text{bath}$ is the interaction between the bath photons and the cavity photons.  
\begin{figure}[t!]
\centering
\includegraphics[width=12cm]{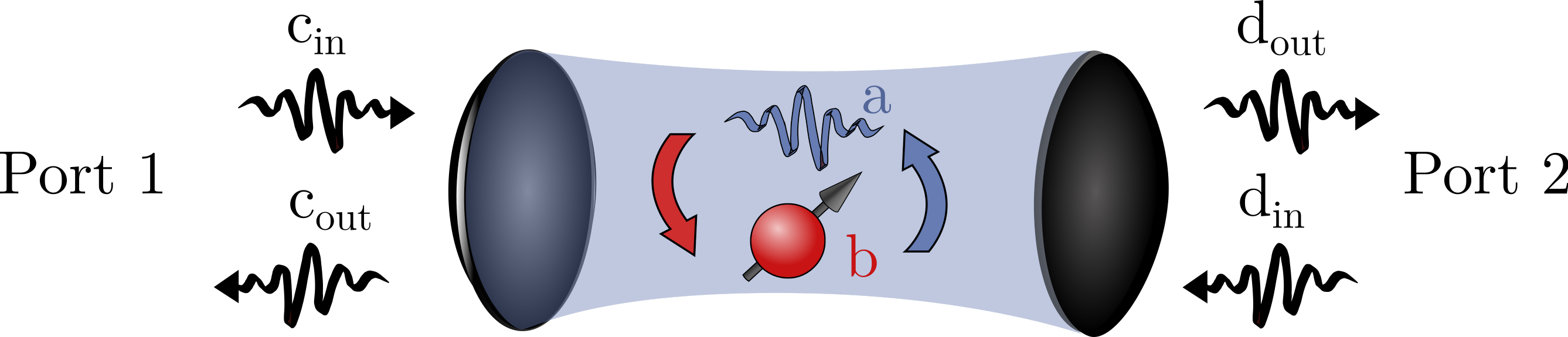}
\caption[Input-output fields]{In the input-output formalism the bath photons at ports 1 and 2 are coupled to the cavity photons, enabling calculation of the cavity transmission properties.}
\label{fig:inoutFields}
\end{figure}
As shown in Fig. \ref{fig:inoutFields}, for a two-sided cavity there are two sets of bath photons, labelled by $c$ and $d$.  Treating the bath photons as harmonic oscillators the bath Hamiltonian can be written as
\begin{equation}
H_\text{bath} = H_\text{bath}^c + H_\text{bath}^d = \sum_q \hbar \omega_q c_q^\dagger c_q + \sum_q \hbar \omega_q d_q^\dagger d_q.
\end{equation}
In the rotating wave approximation the bath-cavity interaction is,
\begin{equation}
H_\text{int}^\text{bath} = - i \hbar \lambda_c \sum_q \left(a^\dagger c_q - c_q^\dagger a\right) - i \hbar \lambda_d \sum_q\left(a^\dagger d_q - d_q^\dagger a\right)
\end{equation}
where $\lambda_{c,d}$ are the bath-cavity coupling rates for ports 1 and 2 respectively, which are assumed here to be mode independent.  Note the difference from the interaction in the Hamiltonian of Sec. \ref{subsec:jaynesCummings}.  This may be surprising since both situations describe the coupling between harmonic oscillators. However the baths act as an additional source of dissipation for the cavity photons, which means that the overall coupling strength should be imaginary.  Therefore here $\lambda_{c,d}$ are chosen to be real, an explicit imaginary factor is added and, to ensure that $H_\text{int}$ is Hermitian, there is an additional negative sign between terms.  Using the entire Hamiltonian the cavity equation of motion is
\begin{equation}
\dot{a} = -\frac{i}{\hbar} \left[a, H_\text{sys}\right] - \lambda_c \sum_q c_q - \lambda_d \sum_q d_q \label{eq:aEOM}.
\end{equation}
The last terms in Eq. \eqref{eq:aEOM} can be evaluated by solving the bath equations of motion in terms of suitably defined incoming and outgoing wave packets of the baths,
\begin{align}
\lambda_c \sum_q c_q\left(t\right) &= -\sqrt{2\kappa_c} c_\text{in}\left(t\right) + \kappa_c a\left(t\right), \label{eq:bathSumsIn}\\
\lambda_c \sum_q c_q\left(t\right) &= \sqrt{2\kappa_c} c_\text{out}\left(t\right) - \kappa_c a\left(t\right), \label{eq:bathSumsOut}
\end{align}
where $\kappa_c$ is the external coupling rate of $c$-bath photons into the cavity.  Similar equations exist for $d_q$.  Eqs. \eqref{eq:bathSumsIn} and \eqref{eq:bathSumsOut} have a simple physical purpose: they link the bath modes to the coupling of a wave packet into ($c_\text{in}$) or out of ($c_\text{out}$) the cavity system.  This can most easily by understood by noticing that Eqs. \eqref{eq:bathSumsIn} and \eqref{eq:bathSumsOut} imply
\begin{align}
c_\text{in}\left(t\right) + c_\text{out}\left(t\right) &= \sqrt{2\kappa_c}a\left(t\right), \nonumber \\
d_\text{in}\left(t\right) + d_\text{out}\left(t\right) &= \sqrt{2\kappa_d}a\left(t\right), \label{eq:cincout}
\end{align}
which simply states that a wave packet of the bath incident on a given port will either be reflected at that port (becoming an outgoing wave) or couple into the cavity.  Eqs. \eqref{eq:bathSumsIn} and \eqref{eq:bathSumsOut} are rigorously derived in Appendix \ref{sec:inoutDetails}.  With the sums over bath photons determined, the cavity equation of motion becomes
\begin{equation}
\dot{a} = -\frac{i}{\hbar}\left[a,H_\text{sys}\right] + \sqrt{2 \kappa_c} c_\text{in} + \sqrt{2\kappa_d}d_\text{in} - \left(\kappa_c + \kappa_d\right) a\left(t\right).
\end{equation}
Taking $H_\text{sys} = H_\text{CMP}$ and including the intrinsic damping of the cavity photon and magnon systems by replacing $\omega_c \to \omega_c - i \kappa_a$ and $\omega_r \to \omega_r - i \kappa_b$ gives  
\begin{equation}
\left[a,H_\text{sys}\right] = \hbar \left(\omega_c - i \kappa_a\right)a + g \hbar b
\end{equation}
and the cavity equation of motion therefore becomes 
\begin{equation}
\dot{a} = - i \omega_c a - \kappa_L a - i g b + \sqrt{2\kappa_c} c_\text{in} + \sqrt{2\kappa_d} d_\text{in}
\end{equation}
where the loaded damping $\kappa_L = \kappa_c + \kappa_d + \kappa_a$ includes losses due to both bath couplings as well as the intrinsic losses of the cavity.  

\begin{figure}[t!]
\centering
\includegraphics[width=12cm]{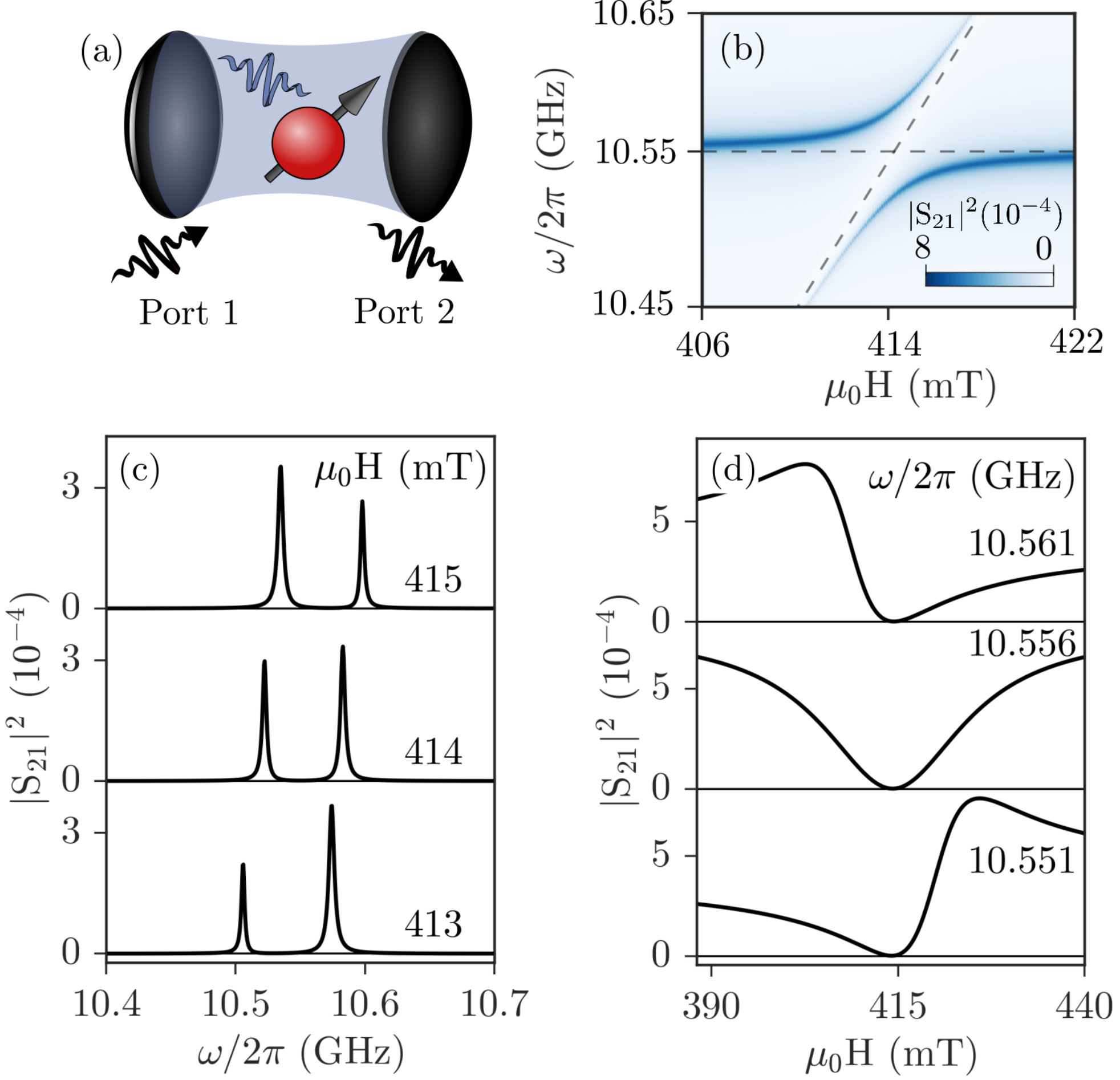}
\caption[Cavity-magnon-polariton transmission spectra in the input-output formalism]{Transmission spectra in the input-output formalism.  (a) A bath photon at port 1 ($c_\text{in}$) is scattered off of the spin-photon system into a final bath photon ($d_\text{out}$).  The input-output fields relevant for transmission are shown.  (b) The full $\omega-H$ dispersion calculated according to Eq. \eqref{eq:inOutS21} using experimentally realistic parameters.  (c) Fixed field and (d) frequency cuts made above, at and below the crossing point $\omega_r = \omega_c$ calculated according to Eq. \eqref{eq:inOutS21}.}
\label{fig:inoutFig}
\end{figure}

Since magnons do not couple directly to the baths, no additional input and output magnons are necessary when considering the magnon equation of motion,
\begin{equation}
\dot{b} = -\frac{i}{\hbar} \left[b, H_\text{sys}\right] = - i \left(\omega_r - i \kappa_b\right)b - i g a,
\end{equation}
which can be solved in terms of $a$ as
\begin{equation}
b = \frac{g a}{\omega - \omega_r + i \kappa_b}.
\end{equation}
Taking the Fourier transform of the cavity photon equation of motion,
\begin{equation}
a = -i A^{-1} \left(\sqrt{2\kappa_c} c_\text{in} + \sqrt{2\kappa_d} d_\text{in}\right),
\end{equation}
where,
\begin{equation}
A = \omega - \omega_c + i \kappa_L - \frac{g^2}{\omega - \omega_r + i \kappa_b}. \label{eq:aFinal}
\end{equation}
The transmission parameters can now be defined in a physically intuitive way.  The transmission $\text{S}_{21}$ is determined by measuring the output photons at port 2 when there is input at port 1 but no input to port 2 and $\text{S}_{11}$ is determined by measuring the output photons at port 1 when there is input at port 1 but no input to port 2.  Analogous definitions hold for the transmission from port 2 to 1, $\text{S}_{12}$ and the reflection at port 2, $\text{S}_{22}$.  Mathematically this means that,
\begin{align}
\text{S}_{21} &= \frac{d_\text{out}}{c_\text{in}}\bigg|_{d_\text{in} = 0},~~~~~ \text{S}_{11} = \frac{c_\text{out}}{c_\text{in}}\bigg|_{d_\text{in} = 0} \nonumber \\
\text{S}_{12} &= \frac{c_\text{out}}{d_\text{in}}\bigg|_{c_\text{in} = 0},~~~~~ \text{S}_{22} = \frac{d_\text{out}}{d_\text{in}}\bigg|_{c_\text{in} = 0}.
\end{align}
Therefore using Eqs. \eqref{eq:cincout} and \eqref{eq:aFinal},
\begin{align}
\text{S}_{21} &= \text{S}_{12} = -\frac{2 \sqrt{\kappa_c \kappa_d}}{i \left(\omega - \omega_c\right) - \kappa_L + \frac{g^2}{i \left(\omega - \omega_r\right) - \kappa_b}}, \label{eq:inOutS21} \\
\text{S}_{11} &= -\left[1+\frac{2 \kappa_c }{i \left(\omega - \omega_c\right) - \kappa_L + \frac{g^2}{i \left(\omega - \omega_r\right)  - \kappa_b}}\right], \label{eq:inOutS11} \\ 
\text{S}_{22} &= -\left[1+\frac{2 \kappa_d}{i \left(\omega - \omega_c\right) - \kappa_L + \frac{g^2}{i \left(\omega - \omega_r\right)  - \kappa_b}}\right].
\end{align}

The properties of the transmission spectra calculated according to Eq. \eqref{eq:inOutS21} are plotted in Fig. \ref{fig:inoutFig} using the same experimentally relevant parameters as the oscillator and phase correlation calculations.  The anticrossing, line width evolution and both the field and frequency cuts agree well with the classical models.  The fact that all three models agree again confirms that the basic features of the CMP arise due to linear harmonic coupling.

\subsection{Summary and Comparison of Models} \label{sec:comparison}

With several models to describe the CMP, a natural question arises:  Which model should be used by an explorer setting out in this new field of cavity spintronics?  This question is even more relevant as new discoveries require model extensions to incorporate new experimental features.  To answer this question it is useful to first realize that in general CMP models are split into two categories: the problem is either solved classically, which means solving the coupled LLG and Maxwell's equations, or approached from a quantum perspective, which means defining a Hamiltonian and determining the eigenfrequencies and transmission properties.  The models presented here have described the most fundamental way to take either approach.  In other words, on the classical level the model using the lowest order magnon mode coupled to a single photon mode was presented.  Meanwhile the quantum approach described here contained only the lowest order linear interactions, again for a single magnon and photon mode.  If an experimental system contains an obvious extension of the classical approach, such as multiple cavity or spin wave modes \cite{Hyde2016, Bai2015}, then it is natural to use an analogous classical model.  On the other hand when an inherently quantum phenomena is investigated, it is natural to extend the quantum mechanical approach by modifying the Hamiltonian, for example through the inclusion of additional interaction terms \cite{Wang2018}.    

On the fundamental level, regardless of the approach the models summarized here demonstrate that CMP hybridization results from the key physics of phase correlation \cite{Bai2015}, and, using the rotating wave approximation near the crossing point all models reduce to a set of coupled equations which have the form,
\begin{equation}
\left(\begin{array}{cc}
\omega -\tilde{\omega}_c & g \\
g & \omega - \tilde{\omega}_r \end{array}\right) \left(\begin{array}{c}
h \\
m \end{array} \right) = \left(\begin{array}{c}
\omega_c h_0 \\
0 \end{array}\right). \label{eq:genMatrix}
\end{equation}
Here $\tilde{\omega}_c = \omega_c - i \beta \omega_c$ and $\tilde{\omega}_r = \omega_r - i \alpha \omega_c$ are the two complex resonance frequencies, $h_0$ is a driving microwave field at frequency $\omega$ and $g$ is the coupling rate.  The top and bottom equations in Eq. \eqref{eq:genMatrix} describe the behaviour of the cavity ($h$) and magnetization ($m$) respectively.  The dispersion of the hybridized modes is then determined by the pole of the matrix determinant from Eq. \eqref{eq:genMatrix} by solving
\begin{equation}
\left(\omega - \tilde{\omega}_c\right)\left(\omega - \tilde{\omega}_r\right) - g^2 = 0, \label{eq:simpDisp}
\end{equation}
and the microwave transmission is determined to be   
\begin{equation}
\text{S}_{21} \propto h = \frac{\omega_c h_0 \left(\omega - \tilde{\omega}_r\right)}{\left(\omega - \tilde{\omega}_c\right)\left(\omega - \tilde{\omega}_r\right) - g^2}.
\label{eq:simpTrans}
\end{equation}
Note that in accordance with Eq. \eqref{eq:inOutS11} the simplified microwave reflection spectra is determined by
\begin{equation}
\text{S}_{11} = -1 + \text{S}_{21}. \label{eq:s21s11Relation}
\end{equation}

Even though all models reproduce the same general behaviour at this fundamental level, such as mode anticrossing, line width evolution, a Lorentz line shape for frequency swept spectra and a Fano line shape for field swept spectra, it is still worth discussing what is gained by each approach.  At this level the real advantage to describing spin-photon hybridization using different methods is that each approach views the spin-photon system from a different angle, revealing new insights which, when put together, offer a more a complete picture of spin-photon coupling.  The insights offered by these different models are both physical and technical in nature.  Physically, the harmonic oscillator model provides a very general basis for understanding CMP physics.  Since the basic features of spin-photon hybridization can be described by coupled oscillators, the main physics is universal and can be observed in many other systems, for example in cavity optomechanics and atomic systems.  This also means that features observed in other systems, or more interestingly, features not observed due to experimental limitations, may be realized through spin-photon coupling.  

While the generality of the harmonic oscillator model is a strength, it is also a weakness in that it does not reveal the origin of the phase correlation which produces the CMP.  Attempting to answer this question led to the development of the dynamic phase correlation model of spin-photon coupling, which clearly reveals the electrodynamic origin of CMP phase correlation.  This approach also unveils the nature of the two oscillators, one represents the cavity mode while the other describes the magnetization dynamics.  Furthermore by establishing this relationship a connection to spin pumping is also revealed, as the voltage is proportional to the spectral function of the magnetization.        

Turning to the quantum description provides further insight into the nature of the coupling, for example this approach reveals the microscopic origin of the coupling strength \cite{Soykal2010, Harder2016b}, revealing new avenues for coupling control through the $N_s$ dependence.  The quantum formalism can also be extended to explore quantum effects of hybridization.  For example, the Green's function calculation \cite{Harder2016b} explicitly demonstrates that the transmission can be calculated by the spin-photon system alone, without considering the properties of the bath photons, which allows straight forward extensions to calculate higher order interactions by accounting for higher order terms in the Dyson expansion. 

Additionally there are a variety of minor advantages to one approach or another.  For example, in order to decouple the two systems in the model the coupling strengths must be controlled independently (at least theoretically).  This is because even without the magnetization back action ($K_c \to 0$) the cavity field still drives the magnetization.  This functionality is provided most naturally by the electrodynamic phase correlation model.  On the other hand, by using the RLC description of the circuit more complex planar resonators can easily be described \cite{Kaur2015, Kaur2016a}.  Alternatively, to describe the reflection spectra the input-output formalism naturally provides the relation $\text{S}_{11} = -1 + \text{S}_{21}$, which can then be adopted in other approaches.  Input-output theory can also be easily generalization to other resonator systems, such as a waveguide.  Finally the simplified model of Eq. \eqref{eq:genMatrix} has a very simple structure, which may easily be extended to multimode systems where several cavity or spin wave modes are present.  Solutions of the hybridized dispersions are also simpler in this approach.  

Therefore the combined understanding of all theoretical approaches described here provides a more complete picture of spin-photon coupling.  Keeping this complete view of the CMP in mind, in the experimental studies that are described below the general formalism defined by Eqs. \eqref{eq:genMatrix} and \eqref{eq:simpTrans} will be used whenever possible.

\section{Experimental Observation of Magnon-Photon Level Repulsion} \label{sec:experiment}


\subsection{Measurement Techniques} \label{subsec:techniques}

\subsubsection{Microwave Transmission and Reflection} \label{sec:transmission}
\begin{figure}[t!]
\centering
\includegraphics[width=5cm]{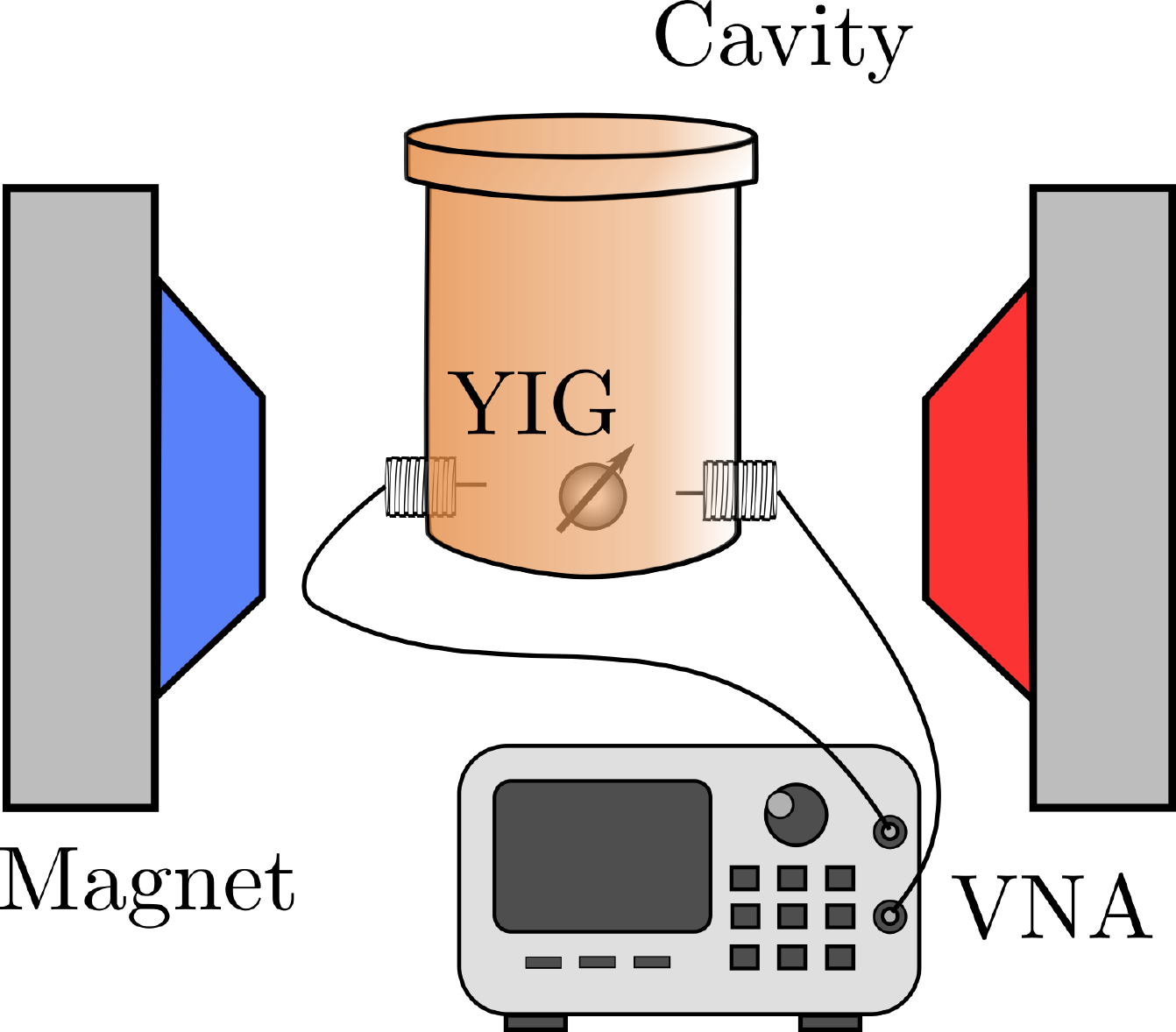}
\caption[Schematic diagram of the microwave transmission measurement system]{Schematic diagram of an experimental configuration used to probe the CMP through microwave transmission.  A VNA injects (and measures) a microwave signal into the cavity which contains the magnetic sample.  An external bias field provided by an electromagnet controls the spin resonance properties.}
\label{fig:transmissionSystem}
\end{figure}
The properties of the CMP were observed for the first time by measuring the microwave transmission spectra of a planar YIG sample in a superconducting resonator at ultra low temperatures \cite{Huebl2013a}.  This technique of microwave transmission has proven to be a useful probe of the CMP that continues to be widely used both at cryogenic and room temperatures.  It is worth noting that to observe the key features of coupling room temperatures are sufficient due to the robust harmonic nature of CMP characteristics.  

A typical experimental setup used to perform microwave transmission or reflection measurements of the CMP is shown in Fig. \ref{fig:transmissionSystem}.  This setup is schematic and many variations on the same theme are possible.  The key requirements to perform such measurements are: (i) a ferrimagnetic material, typically yttrium-iron-garnet (YIG) due to its low Gilbert damping.  In principle the physics of hybridization would be the same for any spin ensemble, except that the observation of strong coupling effects, such as an anticrossing,\footnote{A note on terminology: anticrossing and level repulsion have the same meaning and are used interchangeably here.  With the recent discovery of level attraction in the CMP system, level repulsion may be a better contrasting term.  Hybridization refers to the coupling of the photon and magnon modes which can take the form of either level repulsion or level attraction.} requires that the cooperativity, $C = g_\eta^2/\left(\alpha\beta\right) > 1$ where $g_\eta = g/\omega_c$ is a dimensionless coupling strength normalized to the cavity resonance frequency.\footnote{Actually the strong coupling condition is better defined by the location of the exceptional point \cite{Harder2017}, but this distinction is a subtle point and practically it is simpler to correlate strong coupling with $C > 1$.} (ii) A high quality cavity or resonator.  Again this requirement is flexible and a variety of cavities/resonators have been used, such as a variety of 3D geometries \cite{Tabuchi2014, Zhang2014, Bai2015, Harder2016b, MaierFlaig2016, Souris2017}, planar devices \cite{Huebl2013a, Bhoi2014, Kaur2015}, waveguide resonators \cite{Yao2015} and other special cavity configurations aimed at enhancing local field strengths \cite{Kostylev2016}.  In this context high quality means that $C > 1$ can be achieved.  This has been done with a wide range of quality factors from $Q \sim 1000$ \cite{Harder2016b} to $Q \sim 50$ \cite{Yao2015}.  Active feedback resonators which achieve $Q \sim 10^6 $ have recently been utilized to realize cooperative polariton dynamics \cite{Yao2017}.  (iii) A magnetic bias field use to tune the magnon resonance and cavity resonance into coincidence where the influence of coupling is most notable.  (iv) A microwave source and detector.  This is most easily done with a vector network analyzer (VNA), however depending on the needs of a given experiment other configurations, for example a microwave generator and spectrum analyzer, have also been used.  On a technical note, when performing VNA measurements microwave transmission often yields cleaner data than microwave reflection due to the influence of standing waves in the microwave cables.   Also, although the sample geometry and placement will be experiment dependent, in order to maximize the coupling strength the magnetic sample is often placed at a location of maximum rf magnetic field intensity with the local rf field perpendicular to the static magnetic field.   
\subsubsection{Electrical Detection} \label{sec:electricalDetection}
\begin{figure}[t!]
\centering
\includegraphics[width=12cm]{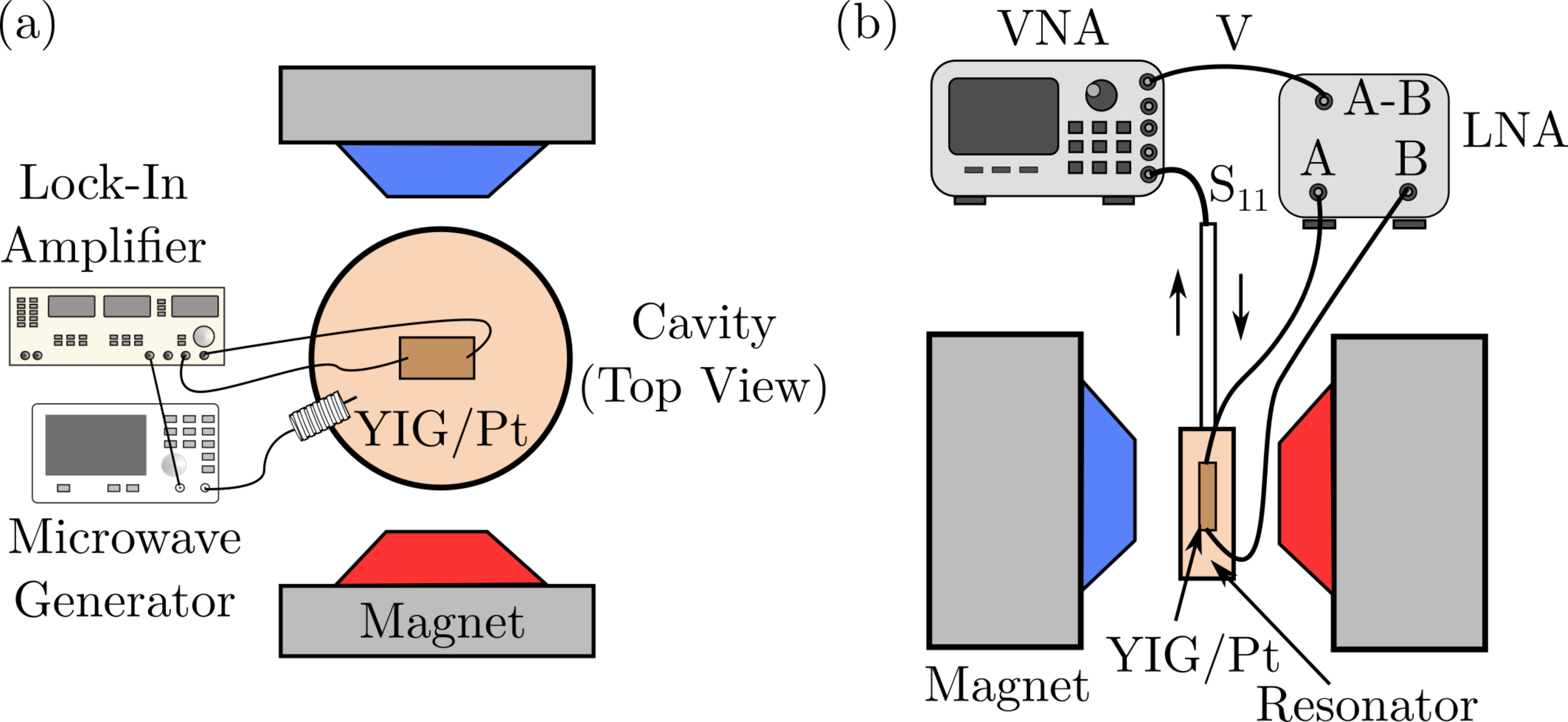}
\caption[Schematic diagram of the voltage detection systems]{(a) Schematic diagram of an electrical detection system using the lock-in technique.  A microwave generator injects a signal into the cavity, which undergoes a low-frequency sinusoidal modulation applied by the lock-in amplifier.  The lock-in amplifier measures the voltage generated through spin-pumping/inverse spin Hall effect.  (b) Schematic diagram of an electrical detection system using low noise amplification.  In this example the microwave transmission can be measured simultaneously.  A VNA injects a microwave signal into the cavity.  The DC voltage generated in the YIG/Pt bilayer is amplified using a low noise differential voltage amplifier (LNA) and recorded through the auxiliary port of the VNA.}
\label{fig:voltageSetup}
\end{figure}

An important early result in cavity spintronics was the demonstration of electrically detected strong coupling \cite{Bai2015}.  Using a YIG/Pt bilayer it was shown that the voltage generated due to spin pumping and the inverse spin Hall effect (ISHE) \cite{Harder2016a} could also detect magnon-photon hybridization.  The importance of this finding is to indicate that hybridization influences the generation of spin current and can therefore be used for spintronic development, for example by providing new control mechanisms \cite{Bai2017}.  It also provides another avenue to probe magnon-photon hybridization.  Along these lines two experimental configurations have been employed, as shown in Fig. \ref{fig:voltageSetup}.  Most key requirements are the same as for microwave transmission measurements.  Again a ferrimagnetic material and high quality cavity are needed to form the CMP.  However since the hybridization will be detected via the spin current, a ferrimagnetic (FM)/normal metal (NM) bilayer is needed as in typical spin pumping measurements.  A microwave generator provides a field to drive the magnetization precession and is sinusoidally modulated by a low frequency signal of a lock-in amplifier, which then measures the voltage generated through the combined effort of spin pumping and the ISHE.  

A slight modification of this setup was used in Ref. \cite{MaierFlaig2016}.  Although the objective of measuring hybridization via electrical detection was the same, there the authors use a VNA to provide the microwave field, which allows simultaneous measurement of the microwave reflection properties.  Since a lock-in amplifier cannot be used in this configuration, a low noise amplifier is required in order to measure the small voltage signal of sever $\mu$V which is typically generated in such bilayer devices.     

\subsubsection{Brillouin Light Scattering} \label{sec:BLS}

In addition to microwave spectroscopy and electrical detection techniques, microfocused Brillouin light scattering (BLS) has also been used to study magnon-photon coupling in a system consisting of a split-ring microwave resonator and an YIG thin film \cite{Klingler2016}. The use of a split-ring resonator loaded with a YIG film enables both microwave and BLS spectra of the hybrid system to be simultaneously recorded. Strong coupling of the magnon and microwave resonator modes is consistently measured by both spectroscopic methods.  By varying the applied magnetic field and microwave frequency, the combined BLS and microwave methods enables the continuous transition of the hybridized modes, from a purely magnonic to a purely photonic mode, to be studied. 


\subsection{Signatures of Level Repulsion} \label{subsec:signatures}

\subsubsection{Transmission and Voltage Spectra} \label{subsec:tandvspectra}

\begin{figure}[t!]
\centering
\includegraphics[width=11cm]{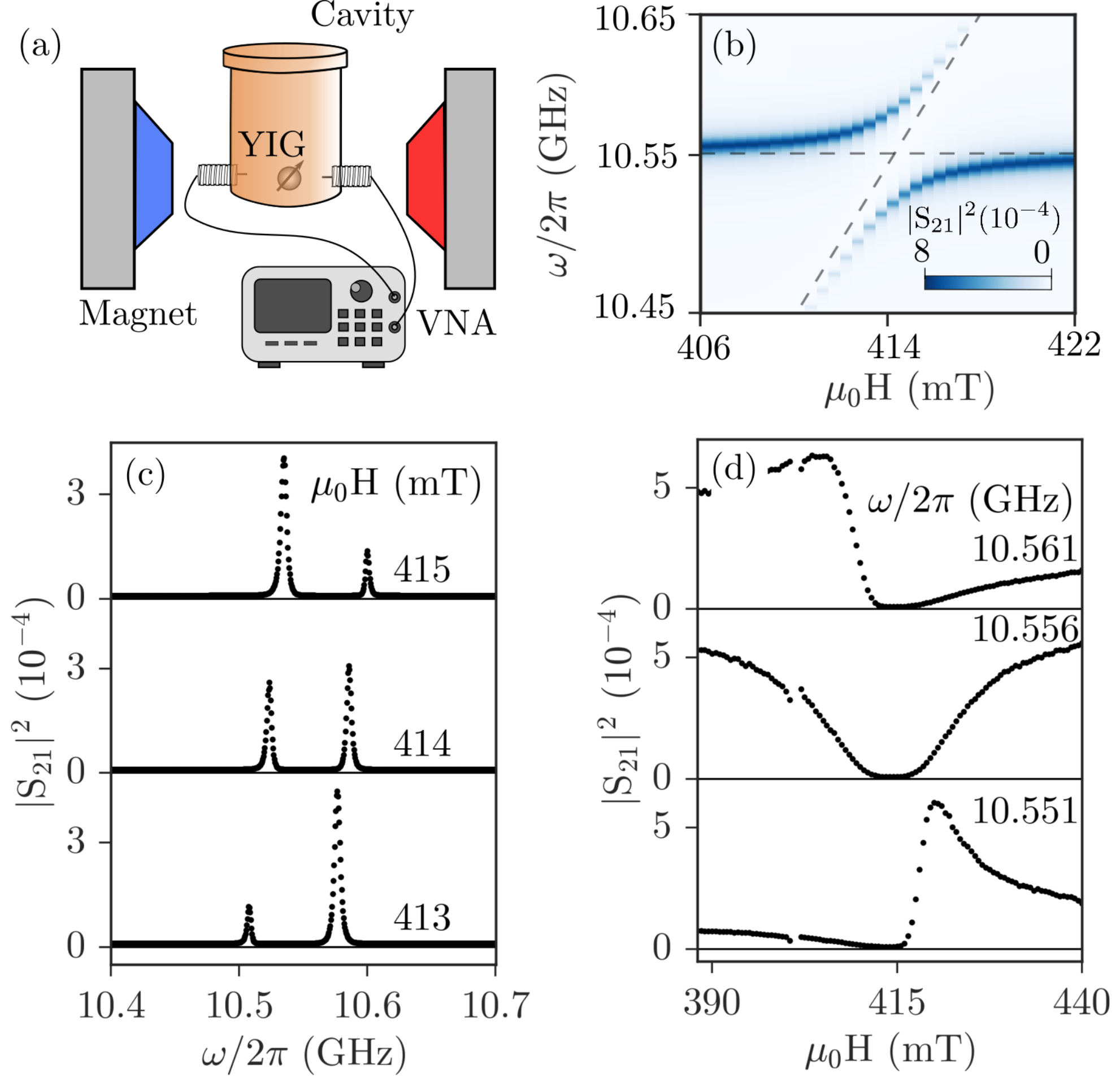}
\caption[Transmission spectra of the cavity-magnon-polariton]{(a) Example of experimental CMP setup for microwave transmission measurements.  (b) $\omega-H$ transmission mapping of the CMP.  Diagonal and horizontal dashed lines indicate the uncoupled FMR and cavity dispersion respectively.  (c) Fixed field and (d) frequency cuts made at, above and below the crossing point $\omega_c = \omega_r$.  A modified version of this figure was originally published in Ref. \cite{Harder2016b}.}
\label{fig:lineShapeTransmission}
\end{figure}

One of the rewording features of magnon-photon hybridization is that the basic physical signatures are revealed directly in the raw experimental data.  An example of such data in a microwave transmission experiment, taken from Ref. \cite{Harder2016b}, is shown in Fig. \ref{fig:lineShapeTransmission}.  In this example the $\omega_c/2\pi = 10.556$, GHz TM$_{011}$ mode a 3D cylindrical copper cavity with diameter of 25 mm and height of 29 mm was coupled to a 1-mm diameter YIG sphere.  A clear Rabi gap is revealed in the eigenspectrum at $\omega_r = \omega_c$, removing the degeneracy at the crossing point and signalling the presence of strong spin-photon coupling.  In this plot the uncoupled dispersions of the cavity and FMR are shown as horizontal and diagonal dashed lines respectively, highlighting the significant deviation of the hybridized modes from their uncoupled behaviour.  The observation of such an anticrossing is the ``smoking gun" of strong spin-photon coupling and shows the dramatic influence of hybridization on the eigenspectrum.  In this example the Rabi gap was $\omega_\text{gap} = 63$ MHz, corresponding to a coupling strength of $g = 31.5$ MHz.  Surprisingly this result agrees remarkably well with first principles calculations using the known spin density of YIG, despite necessary model simplifications to perform the calculation \cite{Harder2016b}.   Coupling rates on the order of several MHz are typical for such experiments, although ultra strong coupling rates in excess of several GHz have also been achieved \cite{Zhang2014a, Bourhill2015, Kostylev2016}, typically through the use of higher frequency whispering gallery modes.  

In the frequency line cuts examined in Fig. \ref{fig:lineShapeTransmission} (c) both CMP branches can be directly seen.  The separation between the two branches at the crossing point ($\mu_0 H_c = 414.5$ mT in this example) defines $\omega_\text{gap}$.  This spectra nicely illustrates the influence of hybridization --- without coupling the two modes observed at $\omega_r = \omega_c$ would overlap and their line width would be very different.  However due to the spin-photon interaction two well separated modes of equal amplitude and equal line width are instead observed.  Above (below) the crossing point the amplitude of the upper (lower) branch decreases sharply as the cavity mode moves away from the FMR frequency and can no longer effectively drive precession \cite{Harder2016b}.  Both branches have a clear Lorentz line shape, consistent with such a resonant process.   The reflection symmetry about the uncoupled dispersions which is evident in Fig. \ref{fig:lineShapeTransmission} (b) can be directly seen in the CMP model by defining the frequency and field detunings, $\delta \omega = \omega - \omega_r$ and $\delta H = H - H_c$, in which case the field swept spectra satisfies $|\text{S}_{21} \left(\omega_r + \delta\omega, H_c + \delta H\right)|^2 = |\text{S}_{21}\left(\omega_r - \delta \omega, H_c - \delta H\right)|^2$.  

Since CMP measurements are typically made as mappings in both $\omega$ and $H$, one data set also contains information about the fixed frequency behaviour, as highlighted in Fig. \ref{fig:lineShapeTransmission} (d).  In this case at $\omega = \omega_c$ a broad symmetric dip is observed, with the transmission line shape becoming asymmetric when $\omega \ne \omega_c$.  A striking feature is that this symmetry change occurs immediately away from the cavity frequency and is opposite above and below $\omega_c$.  In this case a similar reflection symmetry exists about the uncoupled dispersions, $|\text{S}_{21} \left(\omega_c + \delta \omega, H_c + \delta H\right)|^2 = |\text{S}_{21} \left(\omega_c - \delta \omega, H_c - \delta H\right)|^2$.  However a key difference between this line shape symmetry and the analogue for fixed fields is that $\omega_c$ is field independent, whereas $\omega_r$ depends on the field.  Interestingly, in this particular data set, $\text{S}_{21}\left(H\right)$ reveals a new feature that is not easily seen in either the full mapping, $\text{S}_{21}\left(\omega, H\right),$ or the frequency spectra, $\text{S}_{21}\left(\omega\right)$.  Examining the spectra near $\mu_0 H = 404$ mT an additional resonance structure is observed.  This feature is actually the observation of strong coupling with a spin wave, rather than the fundamental FMR mode, which has been experimentally investigated in detail by several authors \cite{Bai2015, MaierFlaig2016, Zhang2016}.   

\begin{figure}[t!]
\centering
\includegraphics[width=11cm]{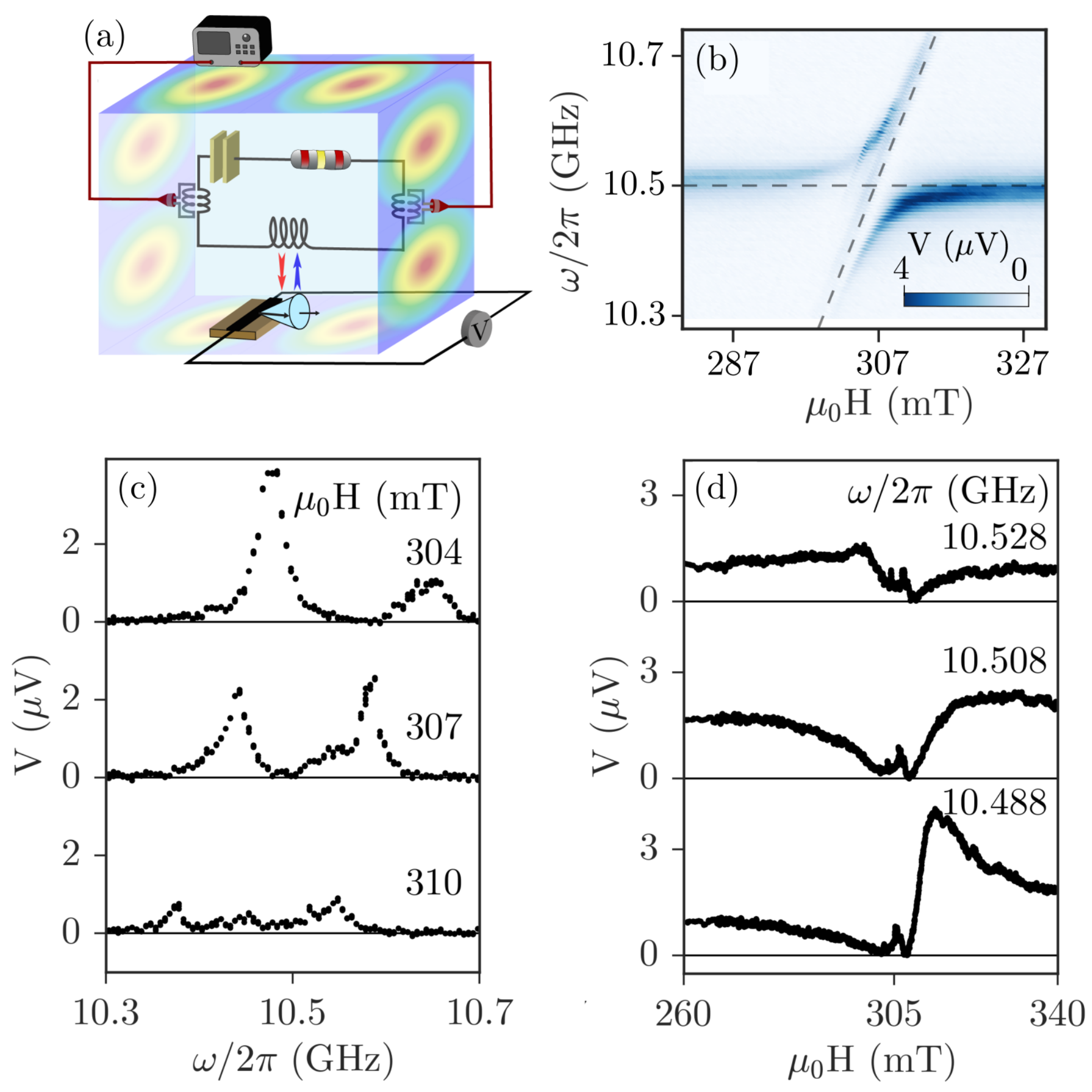}
\caption[Voltage mapping of the cavity-magnon-polariton]{(a) Schematic illustration of the CMP coupling mechanism via the phase correlation model.  (b) The $\omega$-H voltage mapping of the CMP.  Diagonal and horizontal dashed lines indicate the uncoupled FMR and cavity dispersion respectively.  (c) Fixed field and (d) frequency cuts made at, above and below the crossing point $\omega_c = \omega_r$.}
\label{fig:lineShapeVoltage}
\end{figure}

These same CMP features can also be revealed by electrical detection experiments, as highlighted in Fig. \ref{fig:lineShapeVoltage} using data from Ref. \cite{Bai2015}.  In that case a YIG(2.6 $\mu$m)/Pt(10 nm) bilayer with lateral YIG dimensions of 5 mm $\times$ 5 mm, and a Pt layer patterned into a strip of dimension $50~ \mu$m $\times$ 5 mm was coupled to a aluminum cavity of $\omega_c/2\pi = 10.506$ GHz and a loaded quality of $Q = 278$.  The Rabi gap reveals a coupling strength of $g/2\pi$ = 84 MHz. 

Since the key experimental signatures observed through microwave transmission and electrical detection measurements appear to be the same, all analysis procedures suitable for the microwave transmission spectra (e.g. Lorentz line shape fitting for fixed field measurements and Lorentz plus dispersive fitting for fixed frequency measurements) also apply to the electrical detection spectra \cite{Harder2016b}.  Yet despite the similarities there are some advantages and disadvantages to each experimental approach when investigating strong spin-photon coupling.  From the typical results of Figs. \ref{fig:lineShapeTransmission} and \ref{fig:lineShapeVoltage}, it is seen that transmission measurements typically have a better signal to noise ratio and this reduces the uncertainty in the line width measurements extracted from microwave transmission.  On the other hand, higher order spin wave modes can more easily be observed using electrical detection, which is less sensitive to the dominant cavity mode.  However perhaps the most important distinction between the two techniques is that the electrical detection method locally probes the spin system, whereas the microwave transmission technique measures the global properties of the entire system \cite{Bai2017}.  Therefore spin pumping actually provides a method to truly probe the coupling influence on a specific spin device, which is useful for utilizing spin-photon coupling to create new device applications \cite{Bai2017}.

\subsubsection{Avoided Crossing and Line Width Evolution} \label{subsec:avoidedCrossing}
\begin{figure}[t!]
\centering
\includegraphics[width=11.6cm]{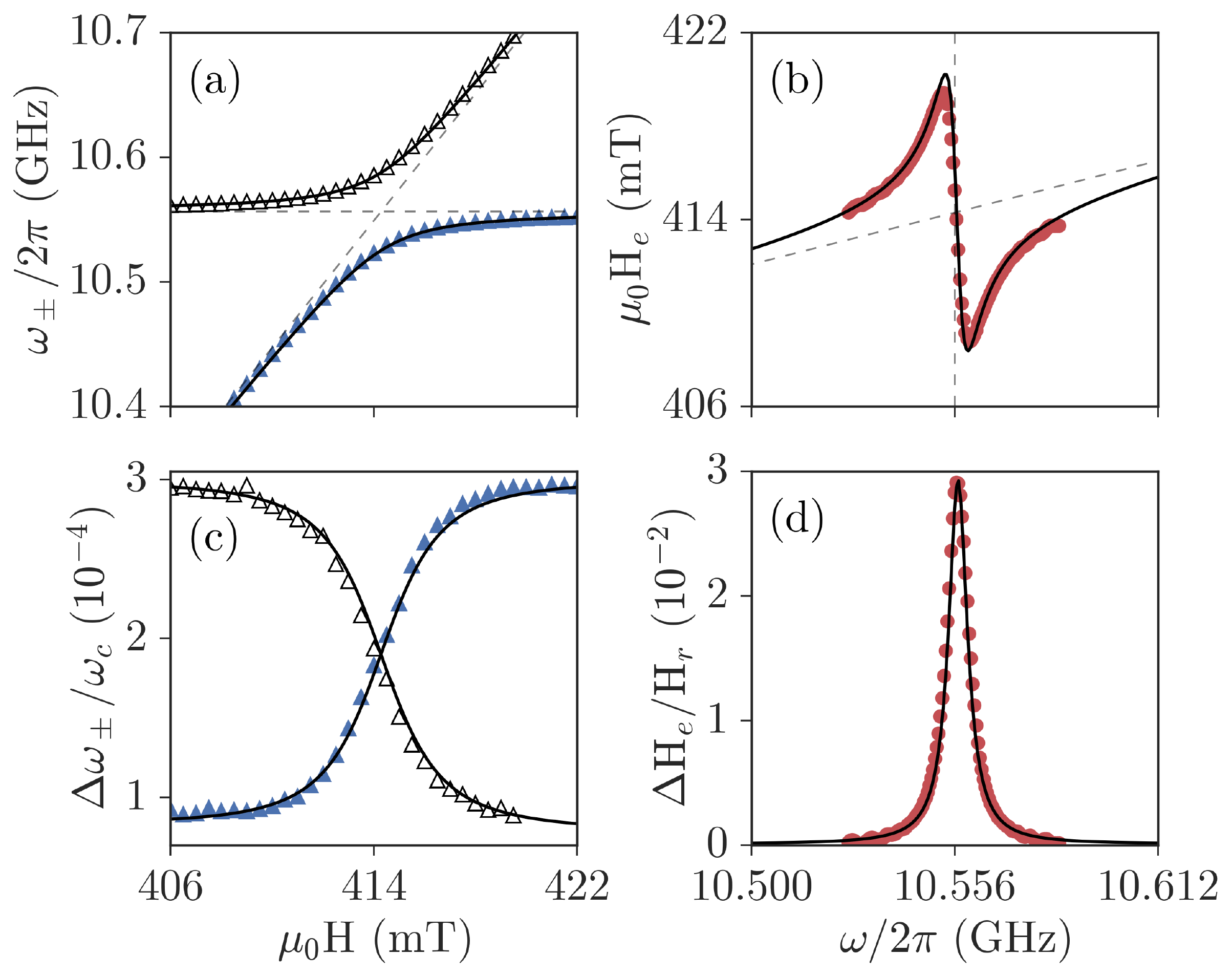}
\caption[Field and frequency swept transmission dispersion and line width]{(a) Frequency swept transmission dispersion and (c) line width.  (b) Field swept transmission dispersion and (d) line width.  Symbols are typical experimental results taken from Ref. \cite{Harder2016b}, with hollow and solid triangles corresponding to the upper and lower CMP branches respectively.  Solid curves are calculations according to Eq. \eqref{eq:wpmSimp} and Eq. \eqref{eq:fieldDisp} for the field and frequency swept spectra respectively.  A modified version of this figure was originally published in Ref. \cite{Harder2016b}.}
\label{fig:transmissionDispersion}
\end{figure}

Both real and imaginary components of the CMP dispersion can be determined by fitting frequency and field mapping data.  For the case of microwave transmission spectra in the strongly coupled regime this can be done by fitting a Lorentz function to each of the two CMP modes and extracting the relevant resonance frequency and line width.\footnote{When the system becomes weakly coupled the two CMP branches merge together and fitting each peak individually is no longer appropriate.  Instead it is either necessary to perform a 2D fit to the data using the full transmission spectra of Eq. \eqref{eq:wpmSimp} or to carefully analyze the line width enhancement of, what appears to be, the single mode (but is actually a combination of the two, narrowly separated modes) \cite{MaierFlaig2016, MaierFlaig2017}.}  This procedure will be able to capture both modes in a window near the crossing point, where both branches have sufficiently large amplitudes.  An example of the dispersion extracted in this way is shown in Fig. \ref{fig:transmissionDispersion} (a) and (c).  Here the open and solid triangles are the upper and lower polariton branches respectively and the solid curves are calculated according to Eq. \eqref{eq:wpmSimp}.  The horizontal and vertical dashed lines in panel (a) indicate the uncoupled cavity and FMR dispersions respectively.  The dispersion again highlights the mode hybridization, which is strongest near the crossing point $\omega_r = \omega_c$ in agreement with the mode composition described by Fig. \ref{fig:hopfield}.  Viewing the CMP from this perspective highlights the damping evolution which is only subtly  present in the full transmission mapping.  An important general feature of the CMP damping evolution is that the line widths of both modes are bounded by $\alpha < \Delta\omega_\pm/\omega_c < \beta$, and are equal at the crossing point.\footnote{Recent investigations of CMP level attraction indicate that this bound may be broken, indicating important extensions of the coupling mechanism.  However in standard CMP hybridization due to level repulsion, these bounds are a strict result of the coupling.}  Actually in general CMP models predict that $\left(\Delta\omega_+ + \Delta\omega_-\right)/\omega_c = \alpha + \beta$.  This relationship is independent of coupling strength, field or frequency and physically indicates that, although the spin-photon coupling enables energy exchange between the two subsystems, coupling does not introduce any additional dissipation channels.  Line width measurements, such as Fig. \ref{fig:transmissionDispersion} (c), have experimentally confirmed this relationship.  

\begin{figure}[t!]
\centering
\includegraphics[width=11.6cm]{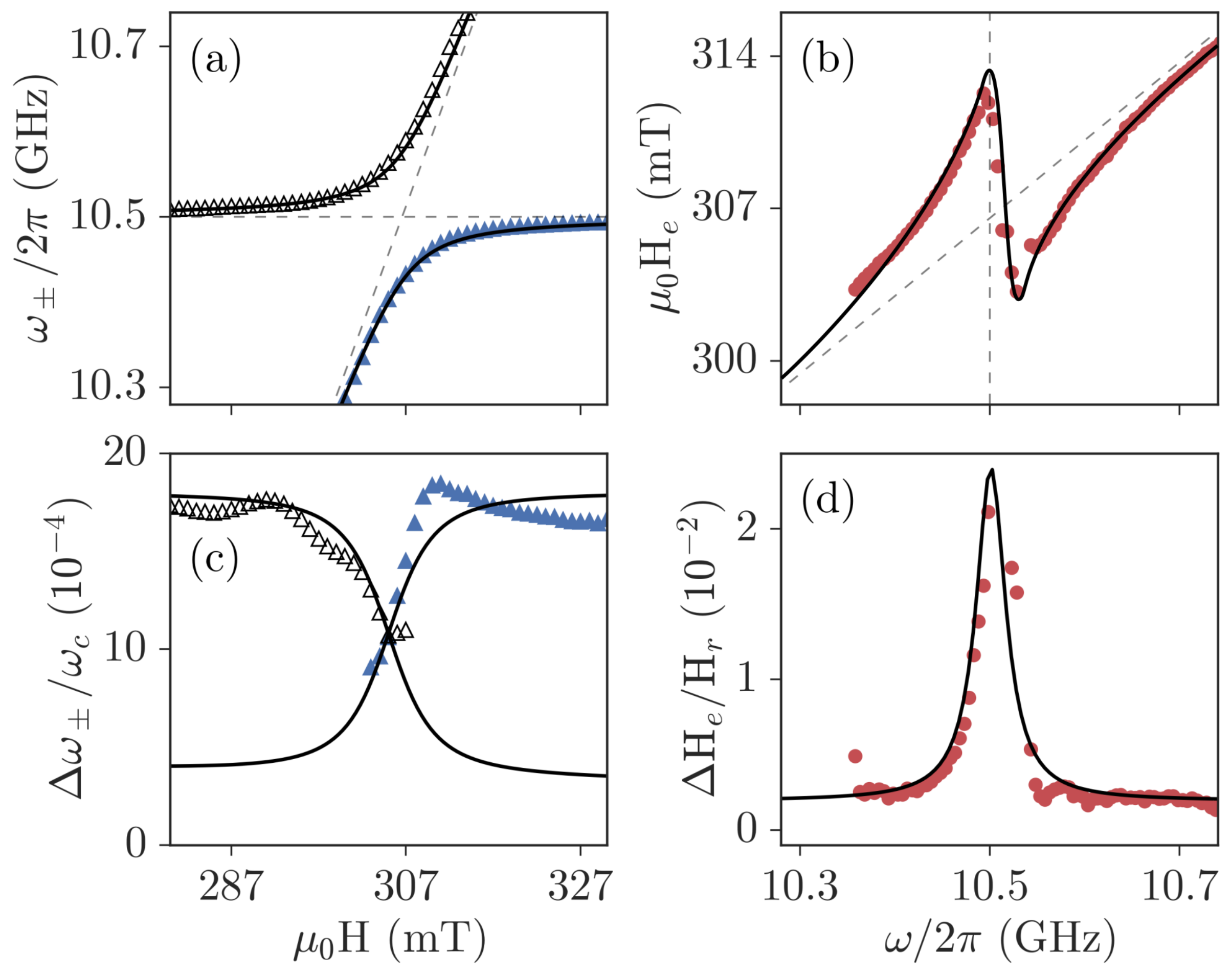}
\caption[Field and frequency swept voltage dispersion and line width]{(a) Frequency swept voltage dispersion and (c) line width.  (b) Field swept voltage dispersion and (d) line width.  Symbols are experimental results, with hollow and solid triangles corresponding to the upper and lower CMP branches respectively.  Solid curves are calculations according to Eqs. \eqref{eq:wpmSimp} and \eqref{eq:fieldDisp} for the field and frequency swept spectra respectively.}
\label{fig:voltageDispersion}
\end{figure}

Similarly, typical frequency swept dispersions and line width are illustrated in Fig. \ref{fig:transmissionDispersion} (b) and (d) respectively with symbols representing experimental data.  In this case the resonance frequency and line width must be extracted by fitting to a Fano line shape (or to a combination of Lorentz and dispersive line shapes).  The solid black lines shown in Fig. \ref{fig:transmissionDispersion} (b) and (d) are calculations according to Eq. \eqref{eq:fieldDisp} while the vertical and diagonal dashed lines in panel (b) indicate the uncoupled cavity and FMR dispersions respectively. The frequency swept dispersion is purely dispersive, while the line width is purely Lorentzian, with a maximum at $\omega = \omega_c$.  Fixed field measurements are a more typical way to perform electrically detected FMR experiments, and therefore reveal another layer of insight into the behaviour of magnon-photon coupling.  Importantly such measurements show that the FMR line width will be drastically influenced by the resonant coupling process.  Despite the fact that $\alpha$ is not changed by the coupling, $\Delta H$ increases drastically at the crossing point where resonant coupling is observed \cite{Bai2015}.  This increase occurs independently of the damping magnitudes, in contrast to the frequency swept spectra, where one mode increases and another decreases depending on the relative size of $\alpha$ and $\beta$.  Normally the FMR line width is described by the relation $\Delta H = \Delta H_0 + \alpha \omega/\gamma$ \cite{SparksBook}, where $\Delta H_0$ is the inhomogeneous broadening resulting from disorder, for example due to inhomogeneities in the crystal structure which produce fluctuations in the anisotropy and magnetization and can result in two magnon scattering which couples the FMR mode to higher order degenerate spin waves \cite{SparksBook}.  However the observation of line width enhancement due to coupling showed that this relation is no longer valid for hybridized systems.  To characterize this behaviour the dispersion of Eq. \eqref{eq:fieldDisp} can be rewritten using the uncoupled photon spectra function, $\mathbb{S}_c = 1/\left(\omega^2 - \omega_c^2 + 2i \beta \omega_c\right)$,
\begin{equation}
\Delta H \left(\omega\right) = \Delta H_0 + \frac{\alpha \omega}{\gamma} + \frac{\omega^2 \omega_m}{\gamma} K^2 \text{Im}\left(\mathbb{S}_c\right). \label{eq:dHNonResonant}
\end{equation}
Since Eq. \eqref{eq:dHNonResonant} uses the full electrodynamic dispersion, rather than its simplification, this expression applies even away from resonance, where deviations from the traditional FMR line width are encountered \cite{Bai2015}.  

Similar dispersion studies have also been performed using electrical detection measurements, an example of which are shown in Fig. \ref{fig:voltageDispersion}.  The results are analogous to those found in microwave transmission.  However from Fig. \ref{fig:voltageDispersion} the effects of decreased signal to noise in this technique are evident, resulting in more scattered fitting results, especially for the line width measurements.  

\subsubsection{Exceptional Points and Hybridization} \label{subsec:EP}

The observable signatures of hybridization, level repulsion and line width evolution, are actually related to the presence of an exceptional point (EP) in the eigenspectrum --- a branch point where the eigenmodes and eigenvectors coalesce \cite{Rotter2010}.  EPs plays an important role in many physical systems \cite{Philipp2000, Dembowski2003a, Dietz2007, Peng2014, Zhen2015, Khanbekyan2015, Hahn2016, Xu2016, Bernier2017} and, the mathematical origin of the magnon-photon EP can be easily understood by examining the hybridized dispersion.  If the damping and coupling strength parameters are allowed to vary it is possible that the square root argument will become zero, and therefore the eigenmodes will become degenerate.  Physically, if the root in the hybridized dispersion is real, there is a gap in the eigenspectrum and strong coupling is observed, while if the root is imaginary there is no gap in the dispersion (which is determined from the real part of the hybridized mode) and instead the coupling influence will be observed in the damping evolution.  If the root is exactly zero, there will be no gap in either the dispersion or the line width evolution at the crossing point and the eigenvalues and eigenvectors will coalesce at this point.  Therefore physically the EP marks the transition between strong coupling, where a gap in the dispersion can be observed, and weak coupling, where no gap can be seen.  

While the coalescence of eigenvalues also occurs near diabolic points, which are associated with Berry's phase \cite{Heiss2000, Heiss1999}, an important distinction is that at the EP the eigenvectors coalesce, as can be seen from Eq. \eqref{eq:eigenvectors}.  This merging of eigenvectors is an interesting consequence of the non-Hermitian nature of the coupled magnon-photon system, which can be made explicit by considering the effective Hamiltonian $H_\text{eff}$,
\begin{equation}
H_\text{eff} = \left(
\begin{array}{cc} 
\tilde{\omega}_c & g \\
g & \tilde{\omega}_r
\end{array}\right),
\end{equation}
which allows the general coupled equations of motion in Eq. \eqref{eq:genMatrix} to be written as,
\begin{equation}
H_\text{eff} \ket{X} = - i \frac{d\ket{X}}{dt}
\end{equation}
where $\bra{X} = (h, m)$.  This effective Hamiltonian is clearly non-Hermitian due to the cavity and spin dissipation rates, which appear in $\tilde{\omega}_c$ and $\tilde{\omega}_r$ respectively. The eigenvalues of $H_\text{eff}$ are given by,
\begin{equation}
\tilde{\omega}_\pm = \frac{1}{2} \left[\tilde{\omega}_c + \tilde{\omega}_r \pm \sqrt{\left(\tilde{\omega}_c - \tilde{\omega}_r\right)^2 + 4 g^2}\right] \label{eq:epGen}
\end{equation}
and therefore by writing the hybridized modes in the form $\tilde{\omega}_\pm = \omega_\pm + i \Delta \omega_\pm$,
\begin{equation}
\left(\tilde{\omega}_+ - \tilde{\omega}_-\right)^2 = \left(\tilde{\omega}_c - \tilde{\omega}_r\right)^2 + 4 g^2.
\end{equation}
At the crossing point, where $\omega_r = \omega_c$, the real and imaginary parts of Eq. \eqref{eq:epGen} yield two conditions which must be satisfied by the hybridized eigenmodes,
\begin{gather}
\left(\omega_+-\omega_-\right)\left(\Delta \omega_+ - \Delta\omega_-\right) = 0, \label{ep:cond1}\\
\left(\omega_+ - \omega_-\right)^2 = \left(\Delta\omega_+ - \Delta\omega_-\right)^2 - \omega_c^2 \left(\beta - \alpha\right)^2 + 4 g^2. \label{eq:epcond2}
\end{gather}
Eq. \eqref{eq:epcond2} will be satisfied if:
\begin{align*}
&\text{(i)}~ \Delta\omega_+ = \Delta \omega_-, \\
&\text{(ii)} ~\omega_+ = \omega_-, 
\intertext{or}
&\text{(iii)} ~\omega_+ = \omega_- ~\text{and}~\Delta\omega_+ = \Delta\omega_-.
\end{align*}
These conditions correspond to the following physical behaviour:
\begin{align*}
&\text{(i) resonance anticrossing, line width crossing}, \\
&\text{(ii) resonance crossing, line width anticrossing}, \\
&\text{(iii) resonance crossing, line width crossing}.
\end{align*}
\begin{table}[b!]
\caption[Hybridized dispersion properties in different coupling regimes]{Behaviour of the hybridized dispersion at the crossing point, $\omega_r = \omega_c$, for different coupling regimes.}
\def\arraystretch{1.5}
\centering
\begin{tabular}{>{\centering\arraybackslash}m{3.5cm}m{2.5cm}m{7.5cm}}
  \toprule                       
Dispersion Constraint & Condition & Physical Description \\ \bottomrule
  $\Delta\omega_+ = \Delta\omega_-$ & $g > \frac{\omega_c|\beta - \alpha|}{2}$ & resonance anticrossing, line width crossing \\
  $\omega_+ = \omega_-$ & $g < \frac{\omega_c|\beta - \alpha|}{2}$ & resonance crossing, line width anticrossing \\
  $\omega_+ = \omega_-$ and $\Delta\omega_+ = \Delta \omega_-$ & $g = g_\text{EP} = \frac{\omega_c|\beta - \alpha|}{2}$ & resonance crossing, line width crossing \\
  \bottomrule
\end{tabular}
\label{table:ep}
\end{table}

Which condition is satisfied will depend on the coupling strength according to Eq. \eqref{eq:epcond2}.  If $g > \omega_c|\beta - \alpha|/2$ then $\omega_+ > \omega_-$ at the crossing point and condition (i) must be satisfied, which is simply the traditional observation of strong magnon-photon coupling.  On the other hand, if $g < \omega_c|\beta - \alpha|/2$, since $\left(\omega_+ - \omega_-\right)^2 \ge 0$ and $\left(\Delta\omega_+ - \Delta\omega_-\right)^2 \ge 0$, at the crossing point $\omega_+ = \omega_-$ and $\Delta\omega_+ \ne \Delta\omega_-$.  The transition between condition (i) and (ii) occurs when $g = g_\text{EP} = \omega_c|\beta - \alpha|/2$.  At this point both the resonance frequency and line width will merge, and therefore the hybridized dispersion will completely coalesce.  This special value of the coupling strength defines the EP, marking the transition between an anticrossing and a crossing with $g \ne 0$.     

\begin{figure}[t!]
\centering
\includegraphics[width=15cm]{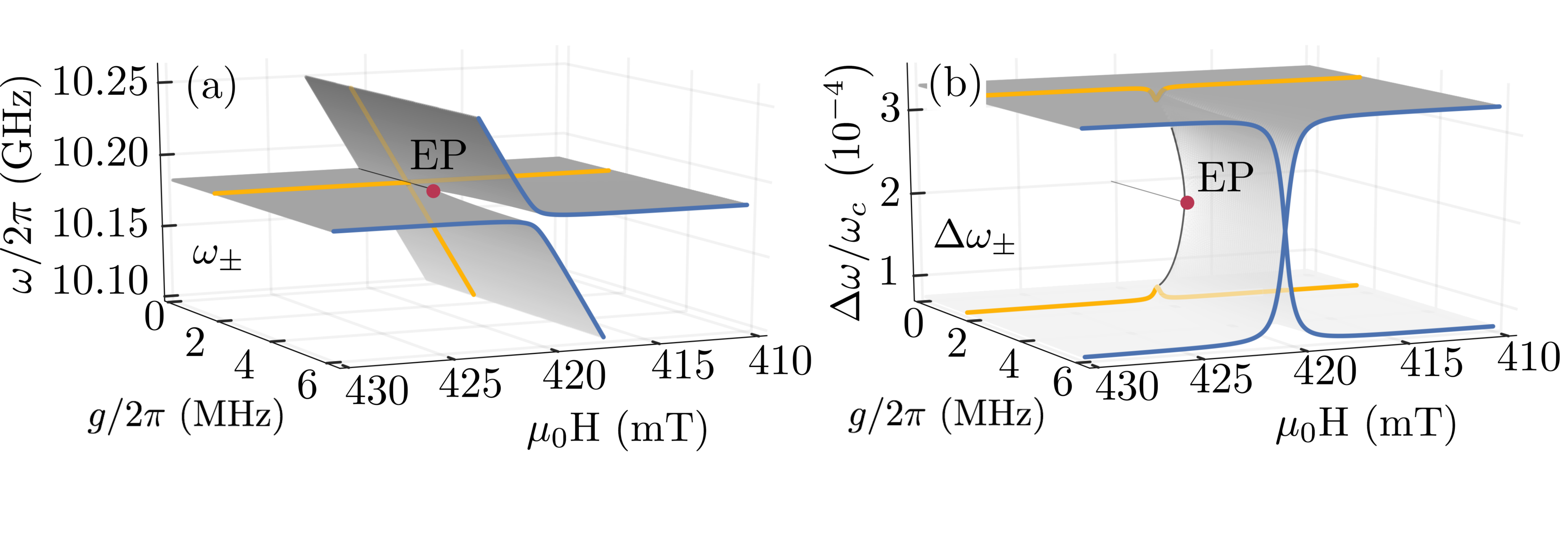}
\caption[Intersecting Riemann sheets of the cavity-magnon-polariton]{The eigenvalue structure of the hybridized magnon-photon system calculated according to Eq. \eqref{eq:simpDisp}.  In both panels the exceptional point (EP) is denoted by a solid red circle.  (a) The eigenvalue dispersion.  For large coupling strength an anticrossing is observed (blue curve).  However for smaller, but non-zero, coupling a dispersion crossing may be seen (yellow curve).  (b) The line width evolution.  In the strong coupling regime a line width crossing is seen (blue curve), whereas when the coupling is reduced a line width gap is found (yellow curve).  A modified version of this figure was originally published in Ref. \cite{Harder2017}.}
\label{fig:riemannSheet}
\end{figure}

Fig. \ref{fig:riemannSheet} displays the geometric influence of the EP by calculating the magnon-photon dispersion for varying coupling strength according to Eq. \eqref{eq:epGen}.  In panel (a) $\omega_\pm$ is shown.  The global structure is that of two intersecting Riemann sheets, which meet at the EP, indicated by a red circle.  At a large fixed coupling strength an anticrossing is observed, as indicated by the blue curve, while at a small fixed coupling strength a crossing is observed, shown by the yellow curve.  The corresponding line width, $\Delta\omega_\pm$, is shown in Fig. \ref{fig:riemannSheet} (b).  Here the eigenspectrum takes the form of two sheets which fold into each other.  In this case the EP marks the position where the sheets separate.  As expected the opposite behaviour is observed.  At large coupling strength the blue curve indicates a line width crossing, while at small coupling strength an anticrossing is observed, as shown by the yellow curve.  

\begin{figure}[t!]
\centering
\includegraphics[width=12cm]{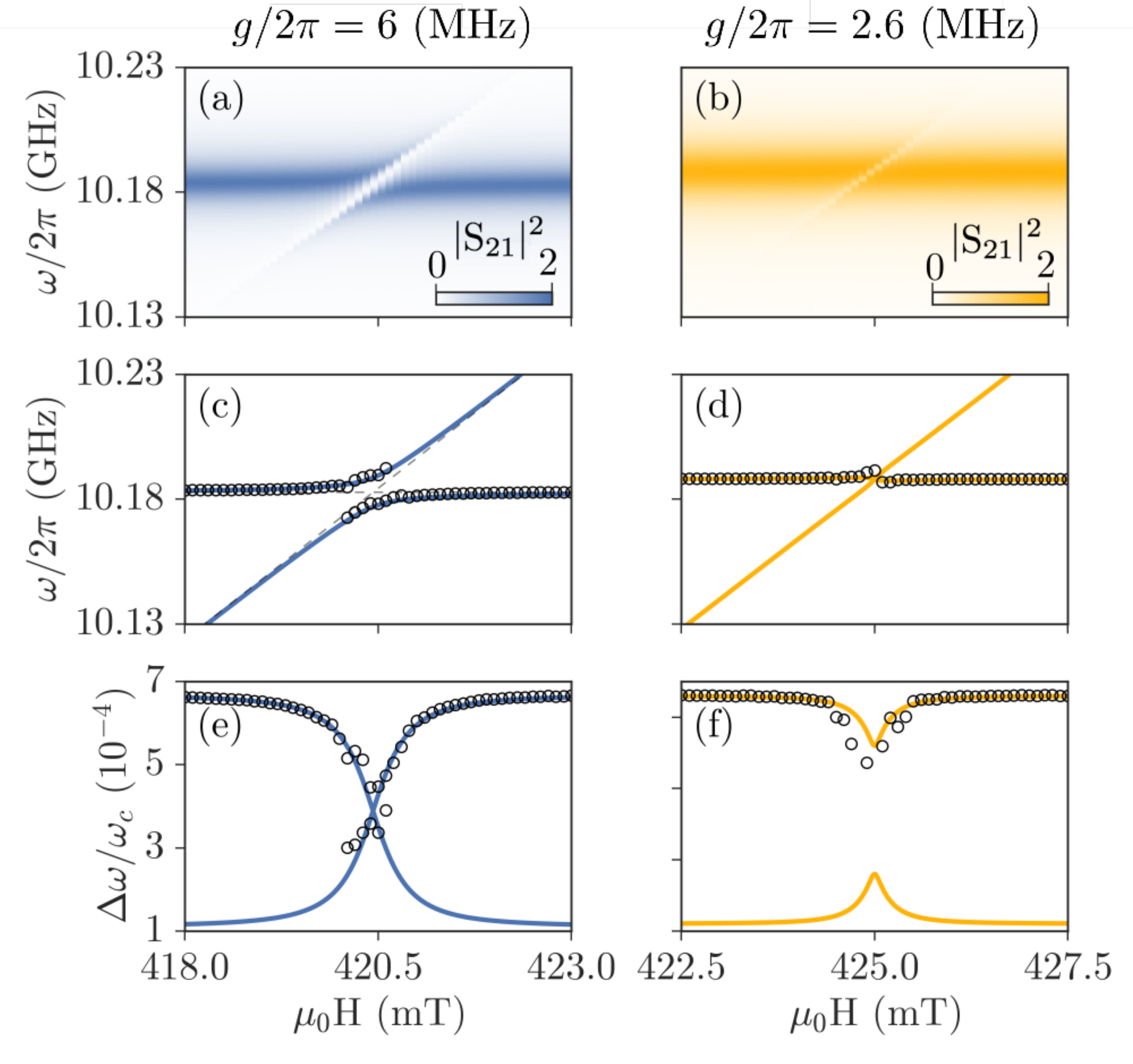}
\caption[Influence of the exceptional point on the cavity-magnon-polariton dispersion]{The influence of the exceptional point on the hybridized magnon-photon dispersion.  The microwave transmission mapping measured for a coupled YIG-cavity system with (a) $g = 6$ MHz and (b) $g = 2.6$ MHz.  Dispersion fitting results for (c) $g = 6$ MHz and (d) $g = 2.6$ MHz respectively.  The line width fits are shown in (e) for $g = 6$ MHz and (f) for $g = 2.6$ MHz.  The former shows the expected crossing while the latter shows a gap.  The solid curves are calculated according to Eq. \eqref{eq:simpDisp}.  The data shown in this figure was originally published in Ref. \cite{Harder2016}.}
\label{fig:EP}
\end{figure}

Experimental observations confirm the strong to weak transition as dictated by the EP.  The transmission spectra measured at $g = 6$ MHz and $g = 2.6$ MHz using a 0.3 mm diameter YIG sphere coupled to a cylindrical microwave cavity are shown in Fig. \ref{fig:EP} \cite{Harder2017}.\footnote{The coupling strength in this experiment was controlled using the technique described in Sec. \ref{sec:angleControl}.}  In this experiment $g_\text{EP} = 3.1$ MHz.  For $g > g_\text{EP}$ a dispersion anticrossing and line width crossing were observed, while for $g < g_\text{EP}$ a dispersion crossing and line width gap were found.  This experiment allows the relationship between the Rabi gap and the cooperativity to be closely examined.  In general, when $\Delta \omega_+ = \Delta \omega_-$ the Rabi gap is written as $\omega_\text{gap} = \sqrt{4g^2 - \omega_c^2 \left(\beta - \alpha\right)^2}$.  If $g \gg \alpha, \beta$, i.e. the system is very strongly coupled, the effect of damping is negligible and $\omega_\text{gap} \sim 2 g$.  This equation provides a useful way to determine the coupling strength without any fitting requirement.  However as the EP is approached this approximation is no longer valid and the full expression for $\omega_\text{gap}$, which includes the effects of damping, must be used.  At the EP, where $g = \omega_c|\beta - \alpha|/2$, the cooperativity is, $C_\text{EP} = g^2/\alpha\beta\omega_c^2 \sim 1/4\left(\alpha/\beta + \beta/\alpha - 2\right)$ \cite{Harder2017}.  As $C_\text{EP}$ is determined solely by the damping, if $\alpha/\beta \gtrsim 6$ (or $\beta/\alpha \gtrsim 6)$, $C_\text{EP} > 1$, which means that a dispersion crossing could be observed even though the cooperativity may exceed one.  Indeed, for the experimental data in Fig. \ref{fig:EP} (d), the crossing occurs at $C = 1.3$.  This leads to an important point regarding the definition of strong coupling.  Typically the following two statements are taken to be synonymous:  (i) The cooperativity exceeds one; and (ii) There is an anticrossing in the dispersion.  However it is possible to observe a dispersion crossing even when the cooperativity exceeds one, and therefore strictly speaking, to maintain the physical picture that strong coupling corresponds to an anticrossing, the EP location, rather than the cooperativity, must be used to define the strong to weak transition.  However it must be noted that this distinction is only relevant in a narrow regime near the EP and in general having a cooperativity greater than one is still a good rule of thumb for the observation of an anticrossing.

\begin{figure}[b!]
\centering
\includegraphics[width=9.6cm]{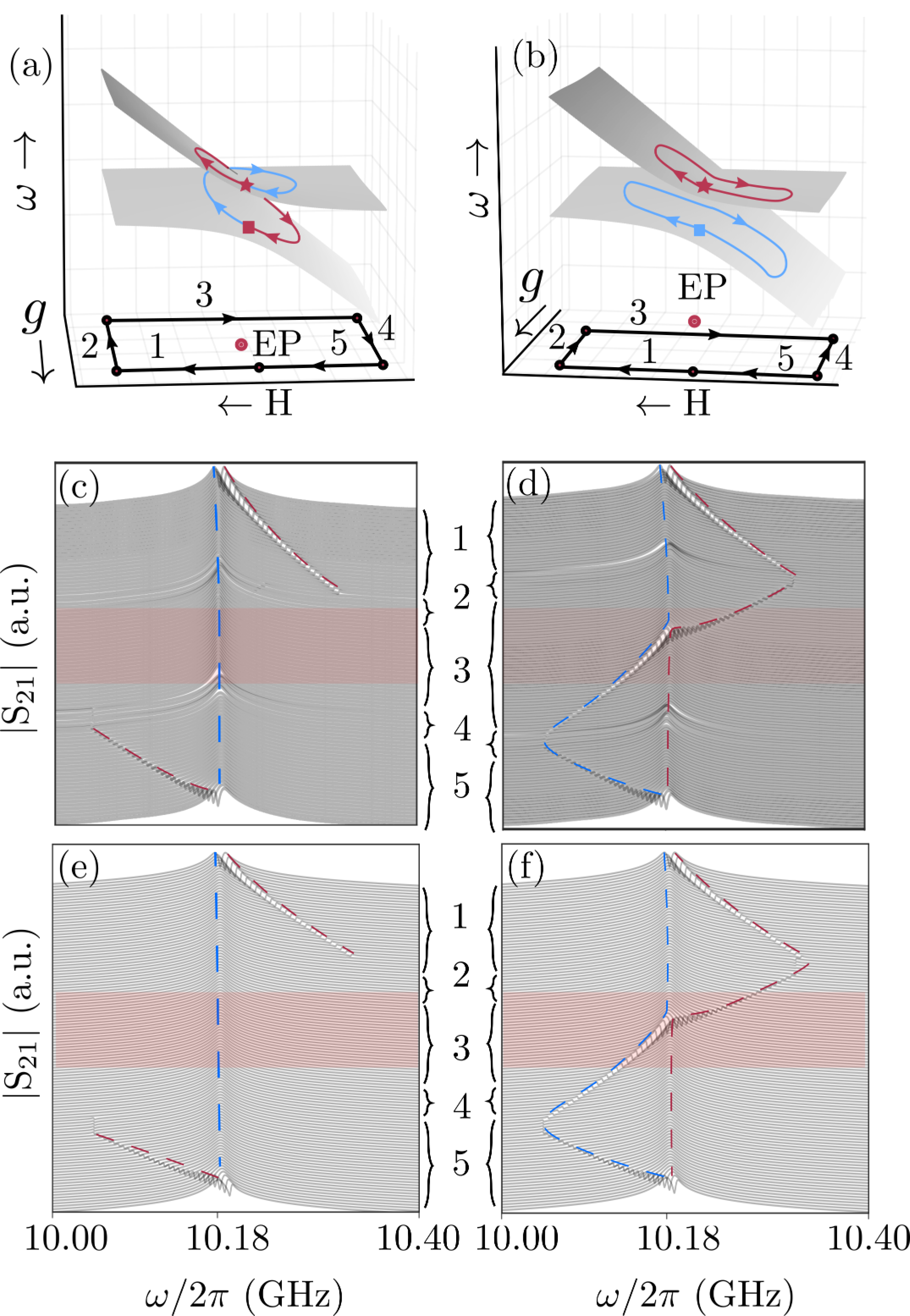}
\caption[Mode switching due to encircling the exceptional point]{(a) Path taken to encircle the EP.  A square (star) denotes the starting (ending) positions, while the red and blue curves indicates the path taken by mode 1 ($\omega_+$) and mode 2 ($\omega_-$) respectively.  When the EP is circled the two paths are connected.  (b) Without encircling the EP each mode behaves independently as shown by the red (mode 1, $\omega_+$) and blue (mode 2, $\omega_-$) curves.  (c) Transmission spectra observed while encircling the EP.  From top to bottom the system is tuned along paths $1 \to 5$.  One mode stays constant at the cavity frequency while the other mode shifts to high frequencies, disappears, and reappears at low frequencies.  Therefore the modes have switched positions.  (d) When the EP is not encircled an anticrossing occurs during path 3 and therefore the modes maintain their relative orientations.  This key difference along path 3 is highlighted with red shading.  (e) The theoretical spectra calculated according to Eq. \eqref{eq:simpDisp} when the EP is encircled and (d) when it is not.  The blue and red dashed curves in (c) - (f) are a guide for the eyes.  A modified version of this figure was originally published in Ref. \cite{Harder2017}.}
\label{fig:modeSwitching}
\end{figure}

The EP is also relevant in the magnon-photon system due to its impact on the global eigenspectrum. Fig. \ref{fig:modeSwitching} highlights the impact of controlling the $g-H$ parameter space.  In panel (a) a closed $g-H$ path, which starts and ends at the same location and encloses the EP, is shown.  The red and blue curves placed on the Riemann sheet show the dynamic behaviour of each eigenmode along this path.  Starting from the red square, the initially lower eigenmode follows the blue spiral upwards, transitioning to the upper sheet while the upper eigenmode follows the red path, transitioning to the lower sheet.  Therefore upon returning to the same starting $g-H$ coordinates, the initially lower eigenmode is now located at the upper sheet and vice versa, i.e. the eigenvalues have switched.\footnote{Actually, as the hybridized modes evolve they will not only interchange but also acquire a geometric phase \cite{Heiss1999, Dembowski2001, Rotter2010}.  However traditional magnon-photon experiments are not sensitive to this phase.}  This is in contrast to the behaviour shown in Fig. \ref{fig:modeSwitching} (b), where the path in the $g-H$ plane does not enclose the EP.  In this case after one continuous loop the lower mode, shown in blue, starts and ends at the blue square while the upper mode, shown in red, starts and ends at the red star.  Therefore the eigenmodes each move independently on their own Riemann sheet and return to their original locations when circling in the $g-H$ plane.  Experimental results demonstrating this effect are shown in Fig. \ref{fig:modeSwitching} (c) and (d).  In this figure the mode evolution is artificially highlighted by red and blue dashed curves.  Indeed when encircling the EP the modes are found to switch, as highlighted in panel (c), while in panel (d) the modes maintain their positions when the EP is avoided.  The calculated spectra according to the model of Eq. \eqref{eq:simpTrans} are shown in Fig. \ref{fig:modeSwitching} (e) and (f) for EP encircling and avoiding respectively, again demonstrating the switching behaviour.  This interesting behaviour is only one aspect of EP physics.  For example, the EP is related to coherent perfect absorption and PT symmetry breaking of the magnon-photon system \cite{Zhang2017a} and has been utilized in analogous systems to realize non-reciprocal waveguide transmission \cite{Doppler2016} and topological induced normal mode energy transfer \cite{Xu2016}. 

\subsubsection{Strong Coupling of Spin Waves} \label{subsec:spinWaves}

Hybridization can also be observed between microwaves and higher order spin wave modes, as has been demonstrated for standing spin waves (SSW) \cite{Bai2015, MaierFlaig2016}, forward volume modes \cite{Zhang2016} and backward volume modes \cite{Zhang2016} using both electrical detection and microwave transmission techniques.  The hybridized dispersion of the spin wave mode may be determined by expanding the 2 $\times$ 2 matrix of Eq. \eqref{eq:genMatrix} to a $3 \times 3$ matrix which includes both the FMR magnon, $m_f$ with resonance frequency $\omega_{rf}$, and the standing spin wave magnon, $m_s$ with resonance frequency $\omega_{rs}$ \cite{Bai2015},
\begin{equation}
\left(\begin{array}{ccc}
\omega - \omega_c + i \beta \omega_c & g_f & g_s \\
g_f & \omega - \omega_{rf} + i \alpha \omega_c & 0 \\
g_s & 0 & \omega - \omega_{rs} + i \alpha \omega_c \end{array}\right)\left(\begin{array}{c}
j \\
m_f \\
m_s \end{array}\right) = 0. \label{eq:sswDispersion}
\end{equation}
Here $g_f$ is the coupling strength between the cavity mode and the FMR, $g_s$ is the coupling strength between the spin wave and cavity modes, and the off diagonal zeros indicate that there is no direct coupling between the FMR and the spin wave.  Interestingly, in the case of SSW in the presence of spin-photon coupling it is possible that $\omega_{rf} > \omega_{rs}$ for $\omega < \omega_c$ and $\omega_{rf} < \omega_{rf}$ when $\omega > \omega_c$ \cite{Bai2015}.  This is in contrast to the uncoupled case where, due to the exchange interaction, the resonant frequency $\omega_{rs}$ of the SSW is always higher than $\omega_{rf}$ of the FMR. 

\begin{figure}[t!]
\centering
\includegraphics[width=10.5cm]{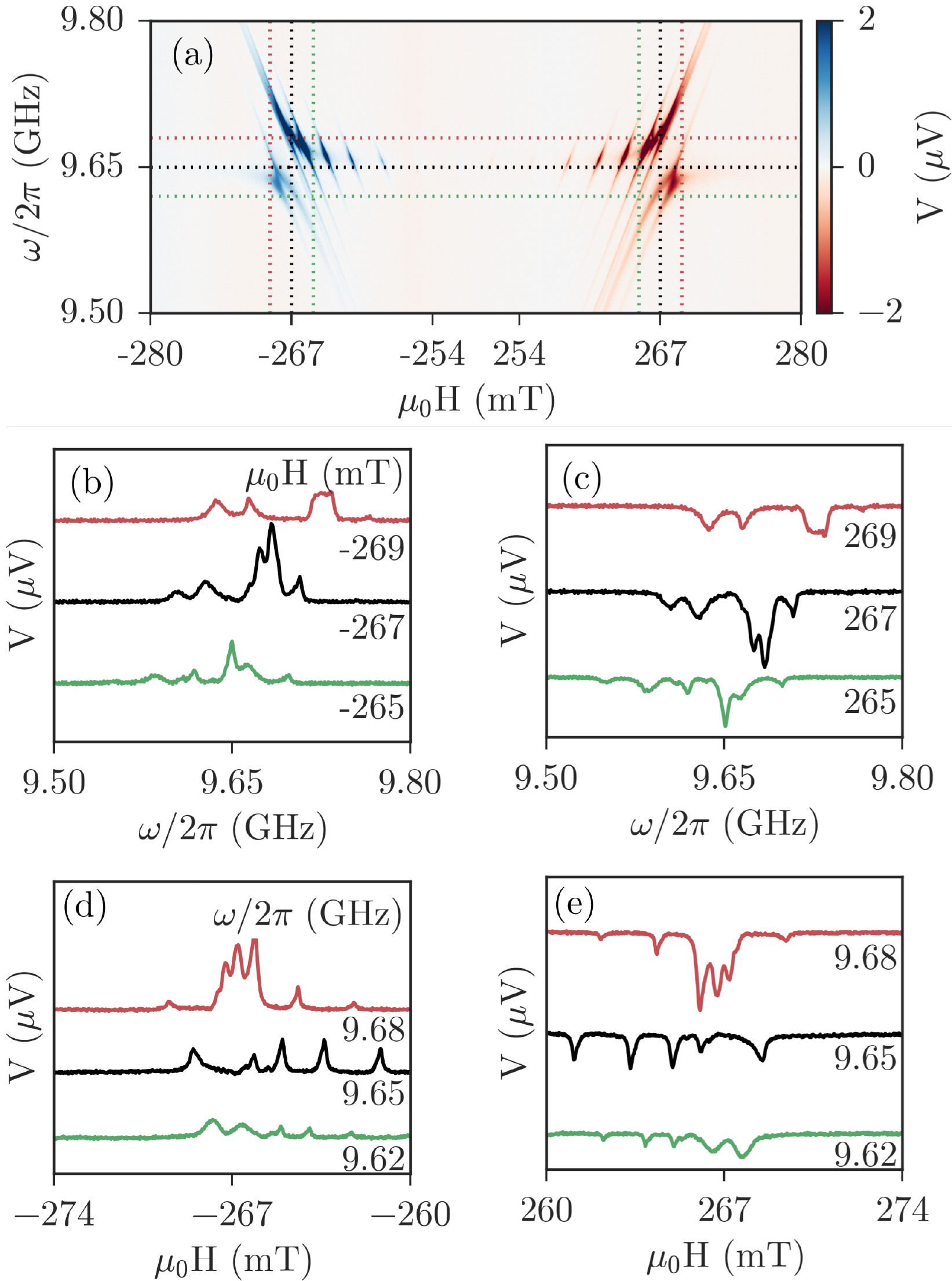}
\caption[Spin pumping voltage spectra of strongly coupled spin waves]{(a) Voltage mapping showing several strongly coupled spin wave modes in a YIG/Pt bilayer.  (b) |S$_{11}\left(\omega\right)$|$^2$ at several values of negative and (c) positive $\mu_0H$. (d) |S$_{11}\left(-|H|\right)$|$^2$ and (e)  |S$_{11}\left(|H|\right)$|$^2$ for several values of $\omega$.  The multiple spin waves are most clearly seen in the $\omega = \omega_c$ cut shown in black in panels (d) and (e).  The scale in panels (b) - (e) is the same as panel (a).  Note the $V\left(H\right) = - V\left(-H\right)$ symmetry which is evident in all panels.  Data used in this figure was originally published in Ref. \cite{MaierFlaig2016}.}
\label{fig:spinWaveVoltage}
\end{figure}

An example of the spin wave hybridization observed through electrical detection is shown in Fig. \ref{fig:spinWaveVoltage}.  These measurements were performed using a YIG(2.8 $\mu$m)/Pt(5 nm) bilayer with lateral dimensions $5$ mm $\times$ 2 mm in a commercial Bruker dielectric ring resonator \cite{MaierFlaig2016}.  A dominant SSW passes through the main FMR anticrossing.  Several additional SSW modes can be observed at the low field side of the main anticrossing, and stand out particularly well in voltage measurements (compared to microwave transmission) since a strong cavity mode background is not observed \cite{MaierFlaig2016}. 

\begin{figure}[t!]
\centering
\includegraphics[width=14cm]{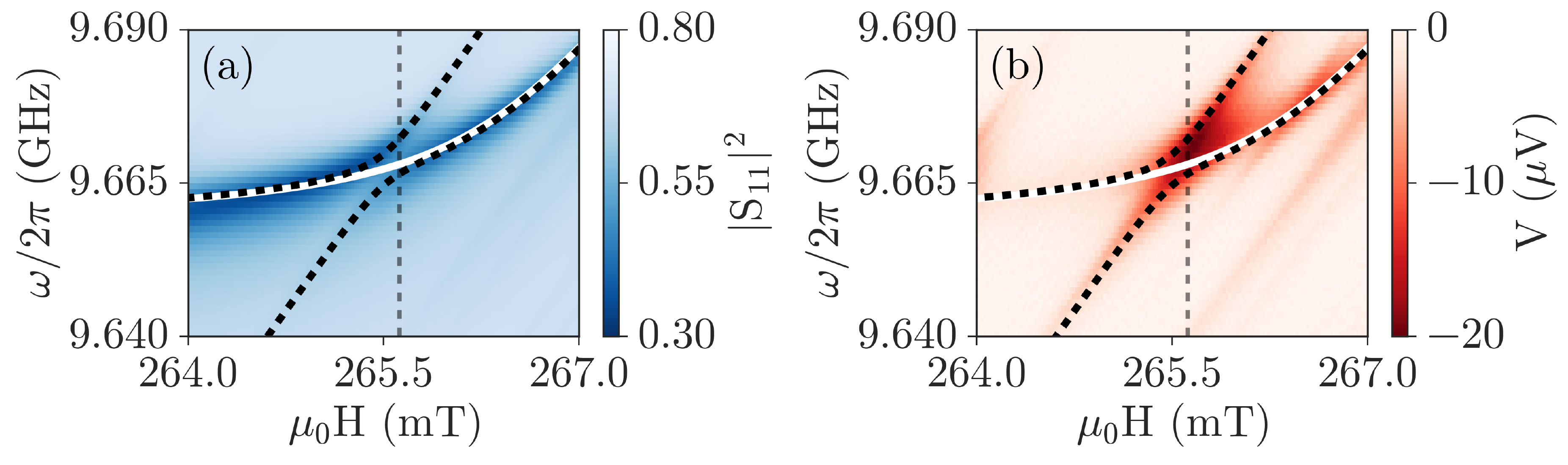}
\caption[Determination of coupling strength for strongly coupled spin wave modes]{(a) The reflection and (b) voltage mappings of typical SSW mode indicating strong hybridization even for higher order magnetic exciations (in this case an $n = 7$ SSW \cite{MaierFlaig2016}. The vertical dashed line indicates the crossing point.  (c) Reflection spectra at $\mu_0H = 265.62$ mT.  Data used in this figure was originally published in Ref. \cite{MaierFlaig2016}.}
\label{fig:spinWaveCouplingStrength}
\end{figure}

A spin wave anticrossing is further highlighted in Fig. \ref{fig:spinWaveCouplingStrength} using both microwave reflection in panel (a) and electrical detection in panel (b).  The solid white line indicates the main anticrossing (in the absence of spin wave modes), which is calculated using Eq. \eqref{eq:simpDisp} with a fitted coupling strength of $g/2\pi = 31.8$ MHz.\footnote{It is worth noting that, taking into account the number of spins in the sample, this corresponds to a single spin coupling rate of $g_0/2\pi = 0.1$ Hz \cite{MaierFlaig2016}, in agreement with experiments on paramagnetic systems \cite{Abe2011}.}  Although the spin wave dispersion can be calculated using the expanded three mode model of Eq. \eqref{eq:sswDispersion}, which would produce a single solution for the entire spectrum and also include the main anticrossing, this approach can become cumbersome when many spin wave modes are present, since, for example, the case of $n$-modes would require finding the roots of an $n^{th}$ order polynomial (the roots of the corresponding $n \times n$ matrix determinant).  A simpler approach, if there is no direct spin wave coupling, is to allow the hybridized FMR-cavity mode to act as a ``new cavity mode" which couples to the spin wave, that is, to describe the spin wave mode coupling $\omega_c$ in the $2 \times 2$ matrix of Eq. \eqref{eq:genMatrix} can be replaced by $\omega_+$ due to the FMR-cavity coupling (the white curve in Fig. \ref{fig:spinWaveCouplingStrength} (a) and (b)).  The result obtained through this approach, using a spin wave coupling strength of 3 MHz are shown as the black dashed curve in Fig. \ref{fig:spinWaveCouplingStrength} (a) and (b).  

Despite the strong hybridization of the spin waves, mode assignment is also possible \cite{MaierFlaig2016}.  
The most important result of such work has been the experimental confirmation of the predicted coupling strength dependence $g_n \propto 1/n$, where $n$ is the SSW index \cite{Cao2014, MaierFlaig2016}, highlighting the systematically variable coupling strengths that can be achieved through spin wave/cavity hybridization.
%
%
\subsubsection{Antiresonance Behaviour} \label{subsec:antiresonance}

In addition to the dispersion anticrossing, line width evolution and spin wave strong coupling, another fundamental feature of spin-photon hybridization is the presence of an antiresonance in the microwave spectra.  In contrast to a resonance, which is a maximum in the microwave transmission, an antiresonance is a position of minimum transmission.  In the language of harmonic oscillators, an antiresonance occurs when the driving force on $O_1$ destructively interferes with the back action force due to $O_2$, resulting in a stationary state of $O_1$.  In the case of spin-photon coupling an antiresonance will always occur at a frequency $\omega_\text{anti}$ in between the two hybridized modes, $\omega_- < \omega_\text{anti} < \omega_+$.  This feature is highlighted in Fig. \ref{fig:antiresonanceSpectra} (a) where experimental transmission spectra $\text{S}_{21}\left(\omega\right)$ are shown using blue circles for several $H$ fields \cite{Harder2016}.  The corresponding transmission phase behaviour, $\phi_{h}$, is shown in panel (b), where each spectra displays three $\pi$ phase jumps which occur at the CMP locations, $\omega = \omega_\pm$ and at the antiresonance, $\omega = \omega_\text{anti}$.  In both panels the solid black curves are calculated according to Eq. \eqref{eq:simpTrans}.  

\begin{figure}[b!]
\centering
\includegraphics[width=10.5cm]{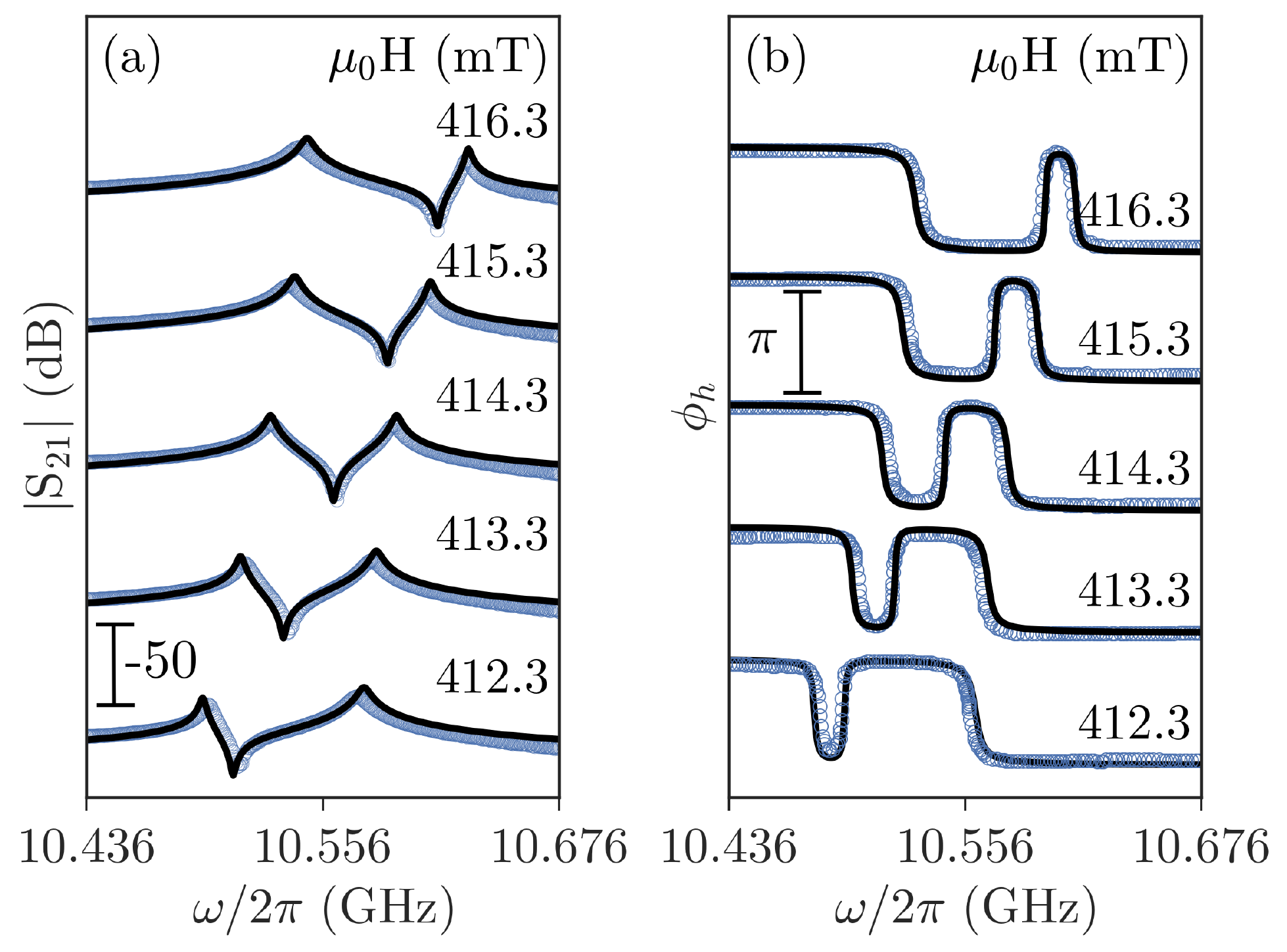}
\caption[Observation of an antiresonance in the CMP transmission spectra]{(a) Transmission spectra measured at different $\mu_0H$ and displayed on a logarithmic scale, highlighting the presence of an antiresonance between the two CMP modes.  Blue circles are experimental data and the solid black curves are calculated according to Eq. \eqref{eq:simpTrans}.  (b) The corresponding transmission phase $\phi_{h}$, which displays a $\pi$ phase jump at each of the CMP resonances as well as at the antiresonance position.  A modified version of this figure was originally published in Ref. \cite{Harder2016}.}
\label{fig:antiresonanceSpectra}
\end{figure} 

\begin{figure}[t!]
\centering
\includegraphics[width=10cm]{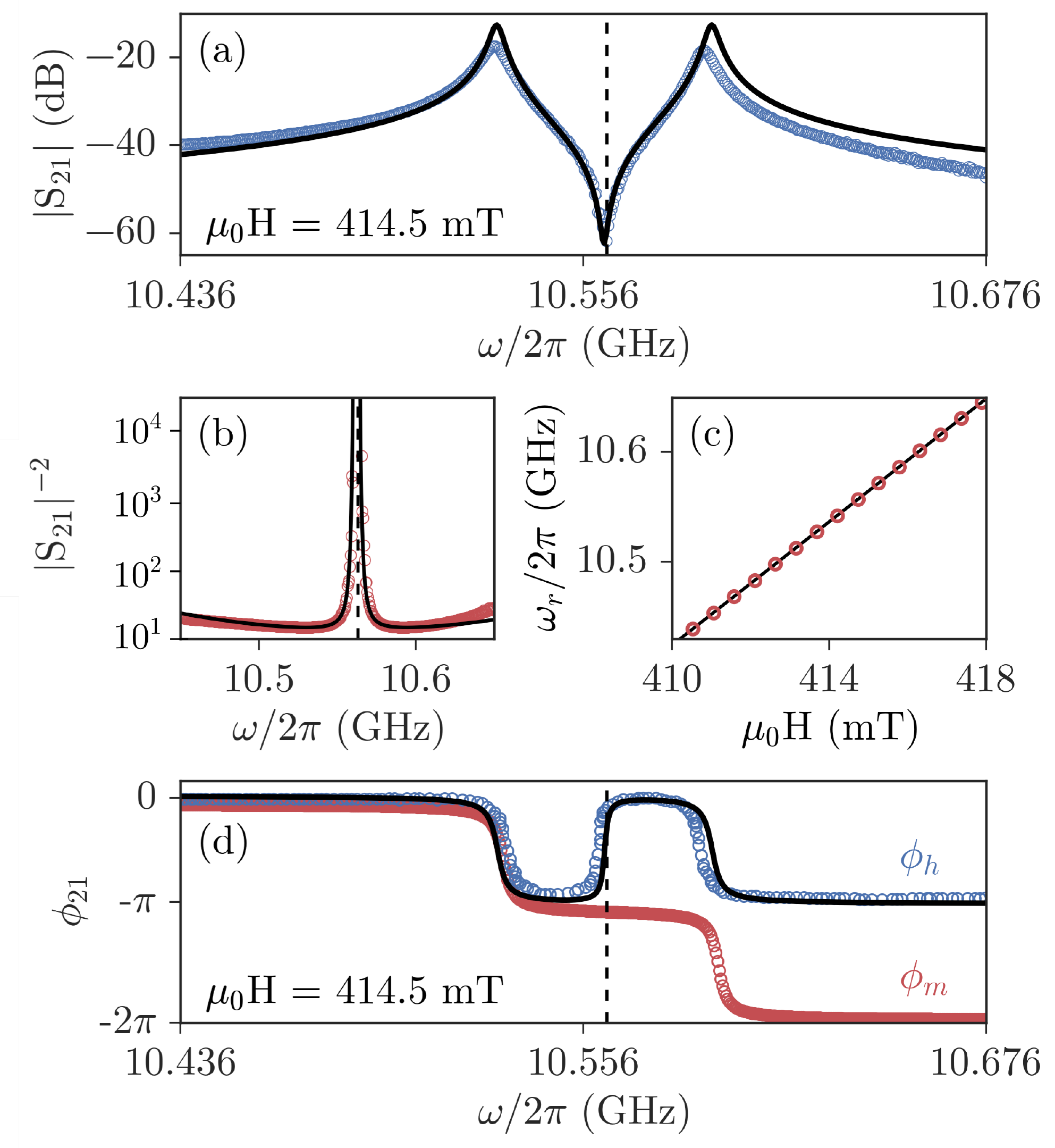}
\caption[Characterization of spin dispersion and phase based on antiresonance analysis]{Phase and dispersion analysis based on the antiresonance.  (a) Transmission spectra at the crossing point where $\omega_c = \omega_r$.  Blue circles are experimental data and the black solid curve is calculated according to Eq. \eqref{eq:simpTrans}.  The vertical dashed curve indicates the antiresonance position.  (b) The inverted transmission spectra which further highlights the antiresonance.  (c) A comparison between the antiresonance locations and the FMR dispersion.  Red circles are the experimental antiresonance positions and the solid black line is the calculated FMR dispersion.  (d) The transmission phase information.  The blue circles are the experimentally measured transmission phase while the red circles are the calculated magnetization phase using Eq. \eqref{eq:mPhase}.  The solid black curve is calculated according to Eq. \eqref{eq:simpTrans} and the vertical dashed line indicates the antiresonance position.  A modified version of this figure was originally published in Ref. \cite{Harder2016}.}
\label{fig:antiresonanceSpectra}
\end{figure}

The antiresonance can be observed by inverting Eq. \eqref{eq:simpTrans},
\begin{equation}
\frac{1}{\text{S}_{21}} \propto \frac{1}{\omega - \tilde{\omega}_r},
\end{equation}
which illustrates that $\text{S}_{21}^{-1}$ will be maximized ($\text{S}_{21}$ will be minimized) at $\omega = \omega_r$.  The fact that the antiresonance occurs at $\omega_r$ can be exploited in a useful way.  In general strong hybridization effects mean that the uncoupled system properties cannot be directly characterized from the transmission spectra, as would be done in traditional cavity based FMR experiments.  However by instead using the antiresonance, direct information about the spin subsystem can be determined, even in a strongly coupled regime.  This type of analysis is illustrated in Fig. \ref{fig:antiresonanceSpectra} using data from Ref. \cite{Harder2016}.  The transmission spectra at $\omega_r = \omega_c$ is shown in Fig. \ref{fig:antiresonanceSpectra} (a).  Experimental data are shown as blue circles and the black curve is a calculation according to Eq. \eqref{eq:simpTrans}.  The antiresonance location is marked by the vertical dashed line and lies directly between the two hybridized modes.  This is because this spectra has been measured at the crossing point.  By inverting the spectra of panel (a), as shown in Fig. \ref{fig:antiresonanceSpectra} (b), the influence of the antiresonance is clear.  The experimental antiresonance positions are plotted as symbols in Fig. \ref{fig:antiresonanceSpectra} (c) and the solid curve is a calculation of the FMR dispersion.  As anticipated the antiresonance positions follow the uncoupled FMR dispersion.  This means that in a strongly coupled system where, for example, the gyromagnetic ratio and Gilbert damping are not characterized, this information can still be directly examined utilizing the antiresonance. 

The antiresonance can also be used to understand the phase behaviour of the hybridized system.  It is well known that an oscillating system will display a phase jump upon passing through a resonance \cite{LandauBookMechanics}.  This can be observed in Fig. \ref{fig:antiresonanceSpectra} (d) which displays two $\pi$ phase jumps at the two CMP resonances.  Here blue symbols are experimental data and the solid black curve is calculated using Eq. \eqref{eq:simpTrans}.  However in addition to the resonance phase jumps there is an additional $\pi$ phase shift which occurs at a frequency between $\omega_+$ and $\omega_-$.  This shift exactly corresponds to the antiresonance location, as denoted by the vertical dashed line.  Therefore all of the phase shifts which are observed through either resonance or antiresonance phenomena are accounted for \cite{Harder2016}.  

Although experimentally it is only possible to access the phase information of the microwave transmission, which in the CMP model corresponds to the phase of $h$, $\phi_h$ the magnetization phase $\phi_m$ can be calculated using the experimentally defined $\phi_h$.  From Eq. \eqref{eq:genMatrix} it follows that
\begin{equation}
\phi_m = \phi_h + \text{arccot}\left(\frac{\omega - \omega_r}{\alpha \omega_c}\right). \label{eq:mPhase}
\end{equation}
Therefore given the transmission phase $\phi_h$ it is also possible to determine the magnetization phase behaviour \cite{Harder2016}.  Such a $\phi_m$ calculation is shown in Fig. \ref{fig:antiresonanceSpectra} (d) as red circles, using the blue $\phi_h$ data.  Although $\phi_h$ undergoes a phase shift at the antiresonance this phase shift does not influence $\phi_m$ --- mathematically it is compensated by the second term in Eq. \eqref{eq:mPhase} --- resulting in two, not three, phase shifts observed in the $\phi_m$ spectra. 
%
\subsubsection{Phase Analysis of Magnon-Dark Modes} \label{subsec:phase}
An intriguing application of strongly coupled spin-photon systems was proposed by Zhang et al. \cite{Zhang2015g}.  Using multiple YIG spheres in a microwave cavity they developed a gradient memory architecture based on magnon dark modes.  By applying a field gradient between the YIG samples, they were able to generate ``bright" magnon modes, which would couple to the cavity, and ``dark" magnon modes, which did not interact with the cavity.  As the dark modes do not couple to the cavity, they are longer lived and can be used to enhance storage times in memory applications.  A key characteristic of this multiple YIG system is the phase behaviour which can be understood through an antiresonance analysis \cite{Harder2016}.  

\begin{figure}[b!]
\centering
\includegraphics[width=12cm]{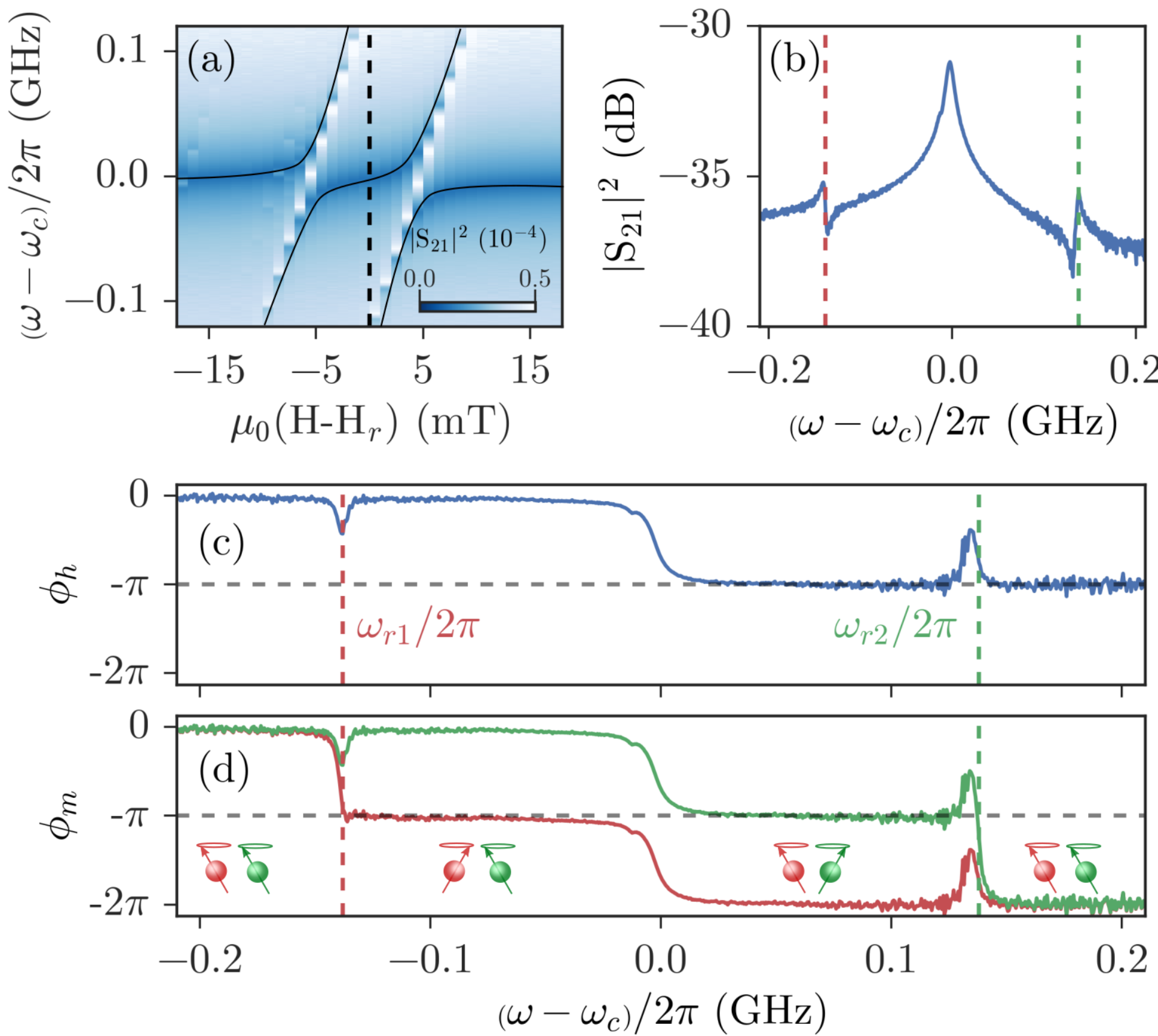}
\caption[Phase analysis of magnon-dark modes]{Phase analysis of a two YIG system.  (a) A microwave transmission mapping, $|\text{S}_{21}|^2$, as a function of microwave frequency and magnetic field for a microwave cavity coupled to two YIG spheres.  The microwave transmission spectra as a function of frequency is plotted in (b), at the magnetic field indicated by the vertical dashed line, showing two antiresonance frequencies.  (c) The measured microwave transmission phase has two opposite phase jumps due to the two antiresonances.  (d) The phases of the dynamic magnetization for each YIG FMR are calculated.  The inset sketch shows that the two FMRs are out of phase between the two antiresonance frequencies ($\omega_{r1}$ and $\omega_{r2}$).  A modified version of this figure was originally published in Ref. \cite{Harder2016}.}
\label{fig:antiresonanceDarkModes}
\end{figure}

The transmission spectra of a two YIG system is shown in Fig. \ref{fig:antiresonanceDarkModes} (a) using data from Ref. \cite{Harder2016}.  The three mode system consisting of two YIG spheres, $m_1$ and $m_2$, coupled to a single cavity mode, is described by the matrix \cite{Harder2016}, 
\begin{equation}
\left(\hspace{-0.1cm}\begin{array}{ccc}
\omega - \tilde{\omega}_c  & g_1 & g_2 \\
g_1 & \omega - \tilde{\omega}_{r1} & 0 \\
g_2 & 0 & \omega - \tilde{\omega}_{r2}  \end{array}\hspace{-0.1cm}\right) \left(\hspace{-0.1cm}\begin{array}{c}
h \\ m_1 \\ m_2 \end{array}\hspace{-0.1cm}\right) = \left(\hspace{-0.1cm}\begin{array}{c} \omega_c h_0 \\0 \\0 \end{array}\hspace{-0.1cm}\right) \label{eq:3x3Dark}
\end{equation}
and therefore the transmission spectra is given by,
\begin{equation}
\text{S}_{21} \propto \frac{\left(\omega - \tilde{\omega}_{r1} \right)\left(\omega - \tilde{\omega}_{r2} \right)\omega_c}{\left(\omega - \tilde{\omega}_c \right)\left(\omega - \tilde{\omega}_{r1} \right)\left(\omega - \tilde{\omega}_{r2}\right) - g_1^2\left(\omega - \tilde{\omega}_{r2} \right) - g_2^2 \left(\omega - \tilde{\omega}_{r1} \right)}. \label{eq:darkModeDisp}
\end{equation}
Here $\tilde{\omega}_{r1} = \omega_{r1} - i \alpha \omega_c$ and $\tilde{\omega}_{r2} = \omega_{r2} - i \alpha \omega_c$.  The dispersion of the three mode system is determined by calculating the roots of the $\text{S}_{21}$ denominator.  This yields the black curves in Fig. \ref{fig:antiresonanceDarkModes} (a).  The transmission spectra as a function of frequency at the magnetic field indicated by the vertical dashed line in panel (a), is plotted in Fig. \ref{fig:antiresonanceDarkModes} (b).  In this spectra three hybridized modes and two antiresonance locations are observed, corresponding to the FMR positions of the two YIG spheres.  The antiresonance positions are indicated by the red and green vertical dashed lines.  Here the presence of two antiresonances can be confirmed by two positive phase jumps which are observed in the microwave transmission phase $\phi_h$, as clearly shown in Fig. \ref{fig:antiresonanceDarkModes} (c).  Three phase delays due to the three hybridized resonances are also seen.  In analogy with the phase analysis of the preceeding section, the two magnetization phases, $\phi_{m1}$ and $\phi_{m2}$, can be related to the microwave transmission phase as,
\begin{align}
\phi_{m1}  & = \phi_h + \text{arccot}\left(\frac{\omega - \omega_{r1}}{\alpha \omega_c}\right), \label{eq:phasem1}\\
\phi_{m2} & = \phi_h + \text{arccot}\left(\frac{\omega - \omega_{r2}}{\alpha \omega_c}\right). \label{eq:phasem2}
\end{align}     
Using Eqs. \eqref{eq:phasem1} and \eqref{eq:phasem2} both magnetization phases were calculated and are plotted in Fig. \ref{fig:antiresonanceDarkModes} (d), where the relative phase of the two FMRs is shown by the insets.  Before the first antiresonance frequency, $\omega_{r1}$, the magnetizations of both FMRs are in phase with the microwave magnetic field.  In between $\omega_{r1}$ and $\omega_{r2}$ the magnetizations of the two YIG spheres are out of phase by $\pi$ with each other, forming a magnon dark mode (which was visible in this particular experiment since the anisotropy fields were distinct).  However, even within this range, while remaining out of phase with one another, the YIG magnetizations both experience a $\pi$-phase shift due to a hybridized mode near the cavity frequency $\omega_c$.  Finally, after the second antiresonance frequency $\omega_{r2}$, the magnetizations from both FMRs are still in phase with each other but out of phase by $\pi$ with the microwave magnetic field.  Therefore a combination of antiresonance and phase analysis directly reveals the in-phase/out-of-phase properties which enable the formation of dark modes, notably the fact that both systems undergo a simultaneous phase shift while maintaining a phase difference of $\pi$ in the dark mode region.


\subsection{Controlling Hybridization} \label{subsec:control}

The key parameter that controls the hybridization behaviour is the cooperativity, $C$, which determines whether or not mode splitting can be resolved at the crossing point.  When the cooperativity is large, $C \gg 1$, then $\omega_\text{gap}$ is much larger than the line width of either mode and both peaks can be clearly resolved.  On the other hand, when $C \ll1$ only a single peak will be observed at the crossing point, even though it is possible to have $g_\eta \ne 0$.\footnote{Actually it is possible to have $C > 1$ (although very close to 1) and still find $\omega_\text{gap} = 0$ near an exceptional point.  However in general $C = 1$ can be taken as the boundary between strong and weak coupling \cite{Harder2017}.}  Therefore the cooperativity, and hence the hybridization, can be controlled by tuning either of the damping rates $\alpha$ and $\beta$, or by tuning the coupling strength $g$ directly.  Tuning the damping of the spin system in a controllable way is challenging.  It would be possible to either use different magnetic samples or to exploit the temperature dependence of the Gilbert damping in certain materials.  However such methods would only produce small cooperativity changes and are cumbersome from an experimental perspective.  More successful hybridization control has been achieved by manipulating either the cavity damping or the coupling strength directly.

\subsubsection{Control of Cavity Damping} \label{sec:dampingControl}

\begin{figure}[t!]
\centering
\includegraphics[width=9cm]{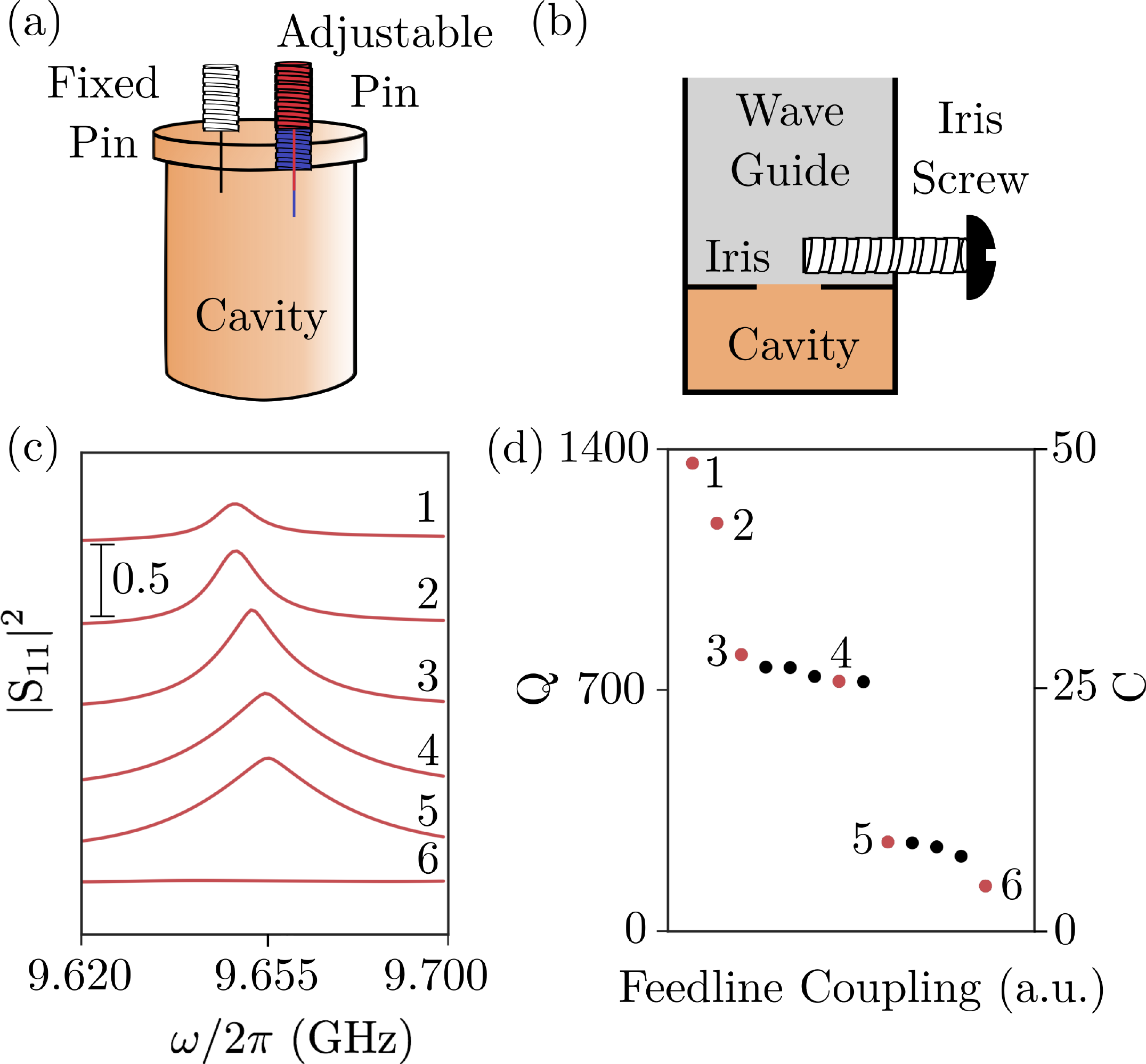}
\caption[Tuning cavity quality through external coupling]{(a) The quality of a 3D microwave cavity can be controlled by adjusting the pin length of the microwave input.  (b) Alternatively, in a dielectric ring resonator the size of the iris, which controls the feedline coupling between the waveguide and the cavity, can be adjusted to control the cavity damping.  (c) Example of reflection spectra at several iris diameters, showing the broadening of the cavity mode as the quality is reduced.  The quality factors of the numbered curves are shown by red circles in panel (d).  In this example the cavity quality can be controlled within the range $0 < Q < 1400$.}
\label{fig:tuneQSpectra}
\end{figure}

To understand the easiest way to control the cavity damping, recall that the loss rate of a loaded cavity has two contributions, $\beta_L = \beta_\text{int} + \beta_\text{ext}$.\footnote{In describing the electrodynamic phase correlation model the general case of two different loss rates at each port, $\beta_1^\text{ext}$ and $\beta_2^\text{ext}$ was considered.  For simplicity it can be assumed that $\beta_\text{ext} = \beta_1^\text{ext} + \beta_2^\text{ext}$.  The important point is that the total extrinsic loss rate can be controlled.  Whether this is done at port 1 or 2 is irrelevant.}  The intrinsic loss rate, $\beta_\text{int}$, is primarily determined by the cavity conductance and can therefore be increased by polishing or the use of superconducting materials, but is not easily tuneable.  However the total extrinsic loss rate, $\beta_\text{ext}$, is determined by the coupling of external microwaves from the feedlines into the cavity and is therefore controllable.  The loaded quality factor, $Q = 1/2\beta_L$, actually depends on the loaded cavity loss rate, and can therefore be tuned by controlling $\beta_\text{ext}$.  Fig. \ref{fig:tuneQSpectra} illustrates two examples of damping control in (a) a 3D cavity and (b) a commercial dielectric ring resonator.
Examples of the reflection spectra at different $Q$ and the quality factors that can be achieved in a dielectric ring resonator are shown in Fig. \ref{fig:tuneQSpectra} (c) and (d) respectively.  In this example $50 < Q < 1400$ with critical coupling at $Q = 700$.\footnote{The ideal scenario from a quality perspective would be to realize $\beta_\text{ext} = 0$, in which case $\beta_L$ would be minimized.  However if this happened no microwaves would be coupled into the cavity and the amplitude of the microwave transmission would be zero.  The ideal balance between signal amplitude and quality factor occurs when $\beta_\text{int} = \beta_\text{ext}$ \cite{PozarBook}.  This condition is known as critical coupling and is generally ideal for microwave transmission measurements.}  The corresponding cooperativity for $g \sim 60$ MHz, $\omega_c \sim 10$ GHz and $\alpha \sim 4 \times 10^{-4}$ is indicated on the right scale.  Traditional cavity based FMR experiments would correspond to very low cooperativities (due to the low sample to cavity volume filling factors), whereas cavity spintronics experiments operate in the $C \gg 1$ regime.

\begin{figure}[t!]
\centering
\includegraphics[width=14cm]{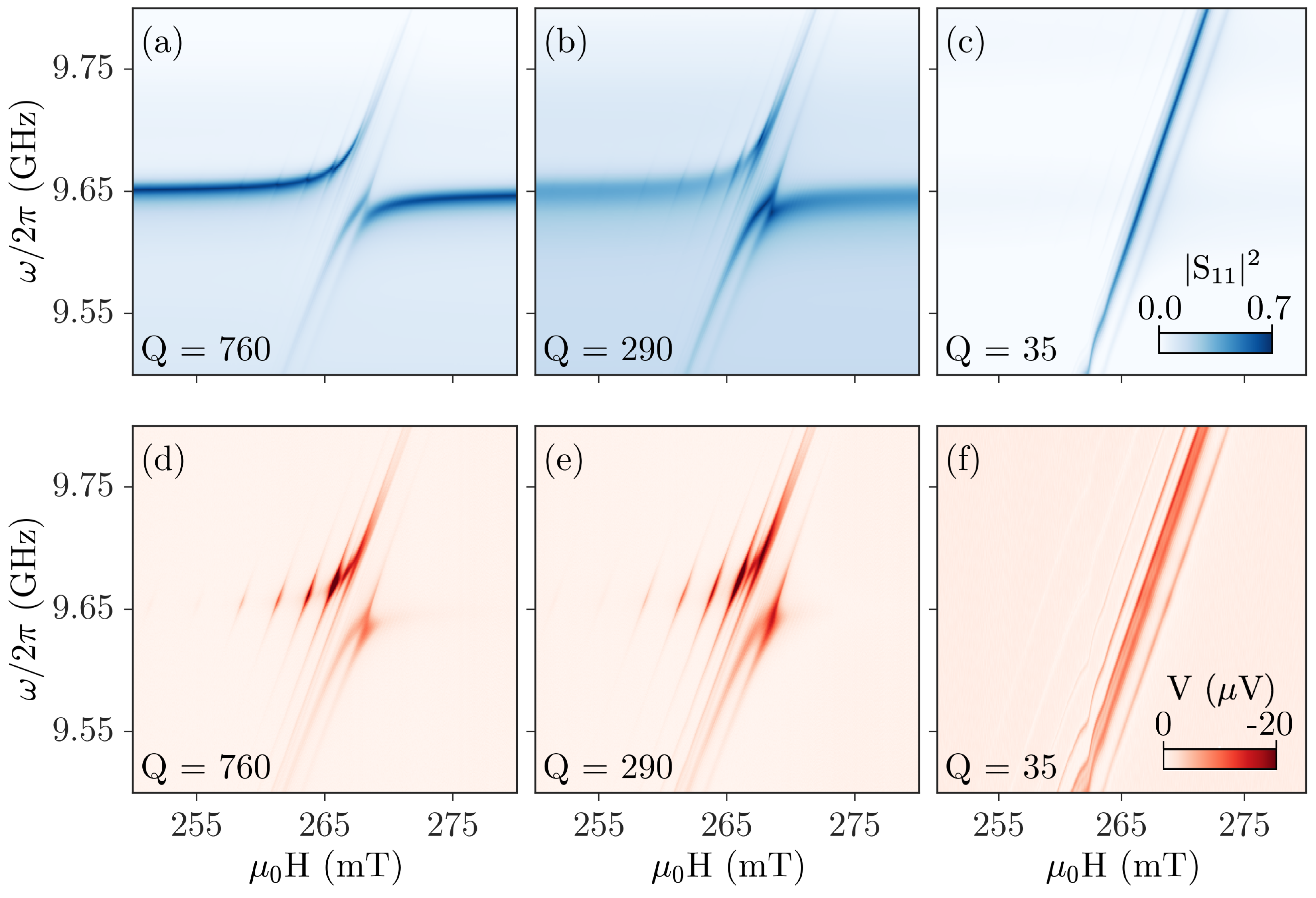}
\caption[Reflection spectra at various cooperativities]{Strong to weak coupling transition demonstrated by tuning the cavity quality.  Such a transition is marked by moving from a large anticrossing at the critically coupled value of $Q$ to the near destruction of the cavity mode and the resulting mode crossing at low $Q$.  Top panels are microwave reflection spectra and bottom panels are simultaneously measured voltage spectra.  A modified version of this figure was originally published in Ref. \cite{MaierFlaig2016}.}
\label{fig:tuneQMapping}
\end{figure}

Examples of the cavity quality influence on the hybridized spectra are shown in Fig. \ref{fig:tuneQMapping}.  The top panels show reflection spectra from Ref. \cite{MaierFlaig2016} with panel (a) - (c) at $Q = 760$, $Q = 260$, and $Q = 35$ respectively.  In general as $Q$ decreases the cavity mode broadens and a decreased amplitude is observed.  $\omega_\text{gap}$ also appears to decrease as the hybridized peaks merge.  Although in the weakly coupled regime $\omega_\text{gap}$ does depend on damping, while strongly coupled such a decrease is mainly an ``optical illusion" due to the broadening of the hybridized modes.  At extremely low $Q$ the cavity mode is essentially destroyed and no hybridization can be observed.  This is the case of conventional cavity based FMR measurements.  The multiple diagonal lines in Fig. \ref{fig:tuneQMapping} (c) correspond to higher order spin wave modes.  The slope of these lines has been used to determine the effective gyromagnetic ratio and saturation magnetization of such modes \cite{MaierFlaig2016}.

The behaviour observed in Fig. \ref{fig:tuneQMapping} (a) - (c) highlights the role of the microwave cavity as a filter.  When the cavity mode is narrow, as in panel (a), the bandwidth of microwaves which excite the magnetization is narrow.  As a result FMR absorption is only observed roughly in a narrow frequency range.  However, as the quality of the cavity is decreased, the bandwidth of the cavity increases and therefore the FMR resonance can be observed over a much wider frequency \cite{MaierFlaig2016}.  At ultra low $Q$ the microwave reflection is essentially flat over the observed frequency range except at the FMR and spin wave resonances.  The analogous behaviour observed in the spin pumping voltage is shown in Fig. \ref{fig:tuneQMapping} (f) - (h). 

\begin{figure}[t!]
\centering
\includegraphics[width=9.5cm]{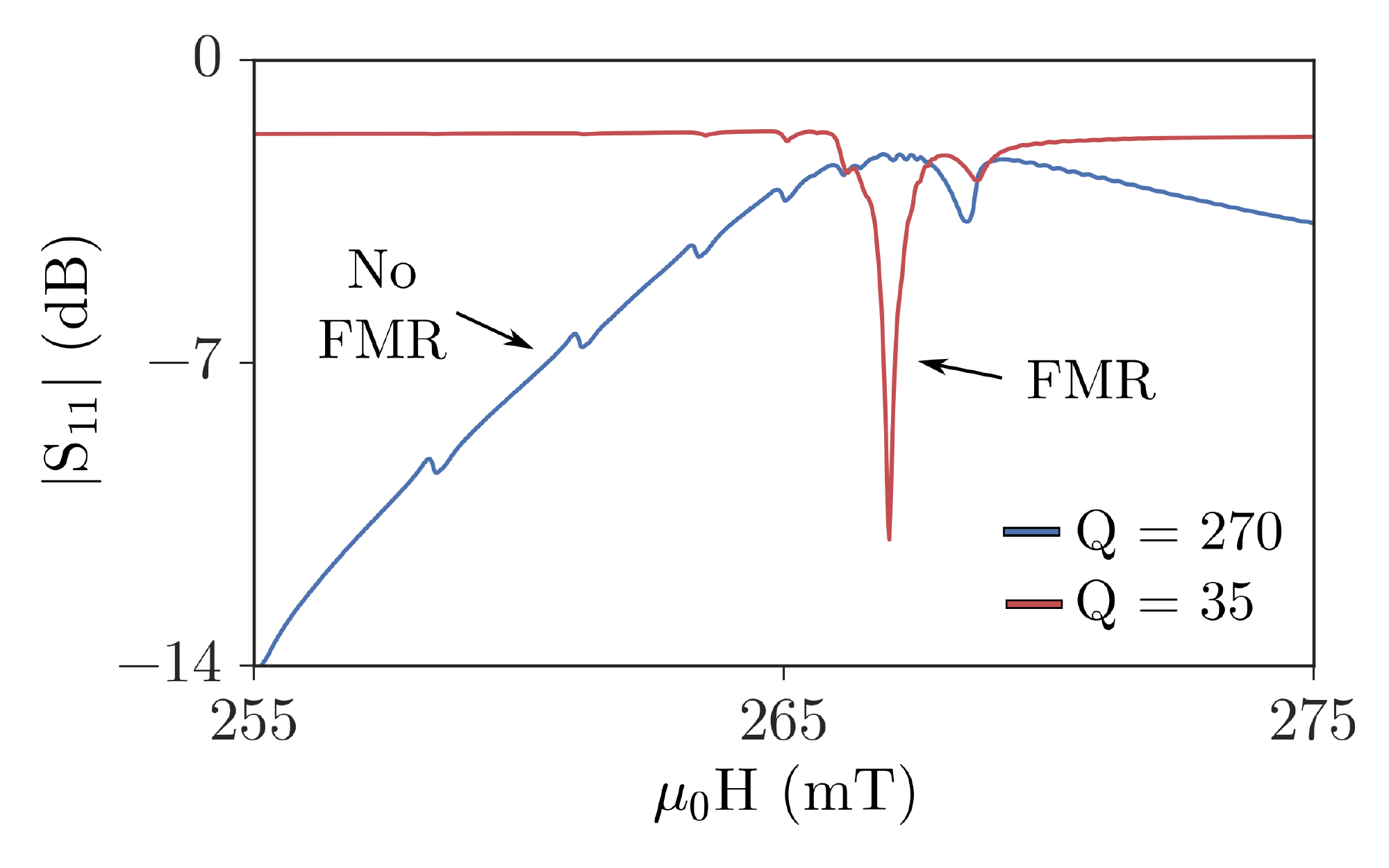}
\caption[Reflection spectra for strong and weak coupling]{Reflection spectra for a strongly coupled system (blue) and a weakly coupled system (red).  In the case of weak coupling the true FMR and spin wave properties can be characterized.  However when the spin-photon system is strongly coupled, the FMR peak is shifted due to hybridization and the spin subsystem cannot be directly accessed using conventional resonance properties.  A modified version of this figure was originally published in Ref. \cite{MaierFlaig2016}.}
\label{fig:tuneQMappingCut}
\end{figure}

A tuneable cavity quality highlights the importance of characterizing strong coupling before performing cavity FMR experiments.  Fig. \ref{fig:tuneQMappingCut} shows the reflection spectra, $|\text{S}_{11}|$, at $\omega = \omega_c$.  In the case of weak coupling (red curve) the spin-photon hybridization does not distort the dispersion and accurate FMR and spin wave properties can be extracted.  However when strong coupling is present (blue curve) no mode is seen at the uncoupled FMR position.  In this case the only signature of the original mode is the broad slope which leads to the hybridized peak.  This behaviour is consistent with observations of FMR line width enhancement due to resonant coupling \cite{Bai2015} and can be observed at easily realized sample volumes, indicating the importance of considering coupling effects when attempting to characterize magnetic systems \cite{MaierFlaig2016}.  It is worth noting that these effects are observed at low powers and are not the result of nonlinear behaviour.\footnote{Typically experiments performed in 3D microwave cavities deliver $\sim$ mW of power from the VNA and CST simulations confirm that the actual power delivered to the cavity is typically much less than 1\% of the applied microwave power \cite{Harder2016b}.  However by increasing the local power density at the sample non linear effects have been observed in other configurations \cite{Wang2016a, Wang2018}.}  

\begin{figure}[t!]
\centering
\includegraphics[width=11cm]{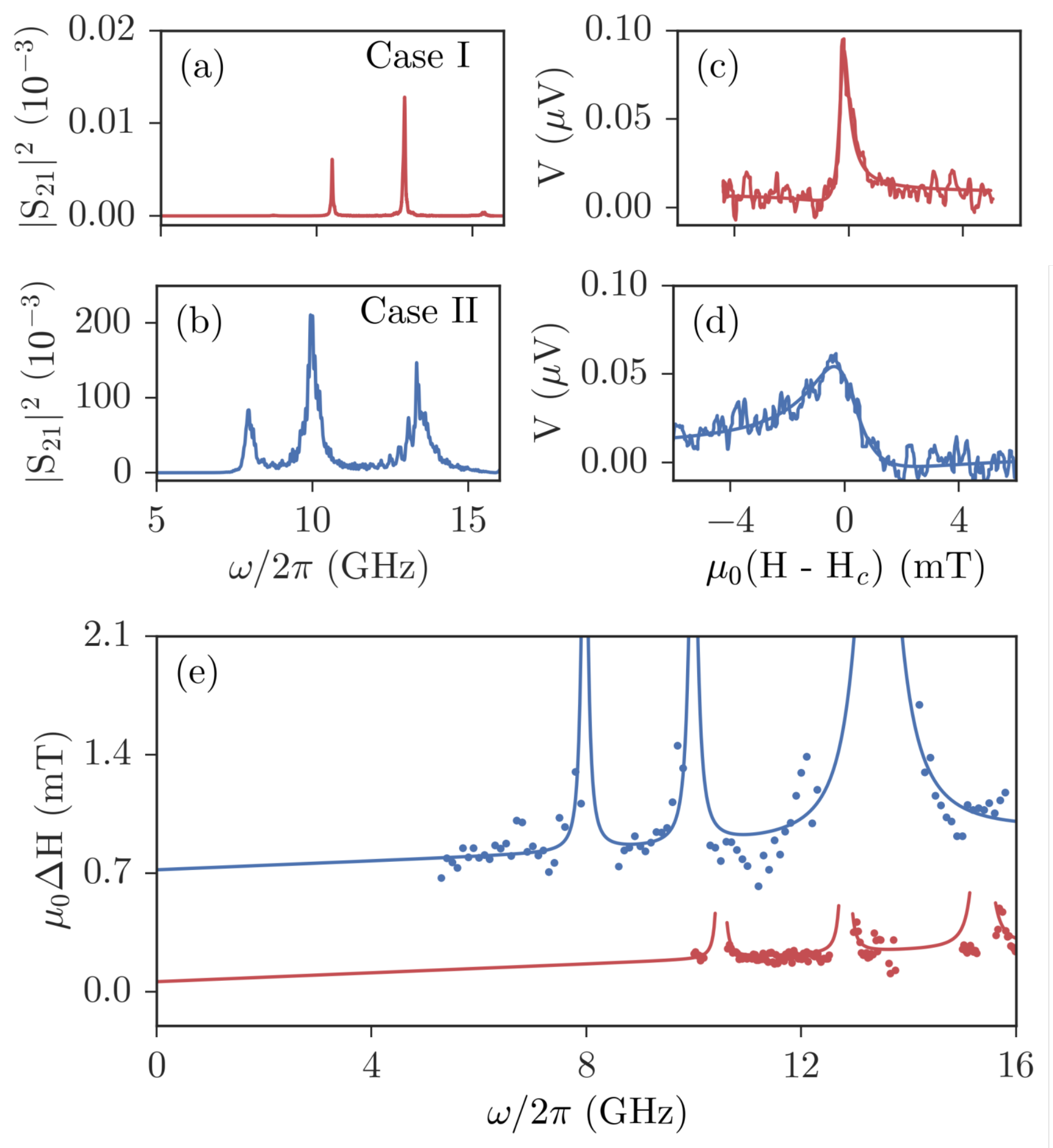}
\caption[Non-resonant line width broadening]{(a) and (b) Transmission spectra $\text{S}_{21}$ measured for different quality factors, labelled as Case I and Case II respectively.  The voltage spectra measured at $\omega/2\pi = 10$ GHz (c) for Case I and (d) Case II.  A significant change in the voltage line shape was observed.  The smooth solid curves are fits according to an asymmetric Lorentz line shape.  These fits produce $\mu_0 \Delta H = 0.20$ mT and $\mu_0 \Delta H = 1.03$ mT for case I and II respectively.  (e) Comparison of the frequency dependent FMR line width measured before and after tuning the cavity quality.  The solid curves are fits according to Eq. \eqref{eq:dhnr2}, which yield $\mu_0 \Delta H_0^I = 0.06$ mT and $\mu_0 \Delta H_0^{II} = 0.72$ mT.  A modified version of this figure was originally published in Ref. \cite{Bai2015}.}
\label{fig:nonresonantLW}
\end{figure}

The cavity quality will also influence the non-resonant line width enhancement as highlighted in Fig. \ref{fig:nonresonantLW}.  To illustrate this effect two different pin configurations were used in Ref. \cite{Bai2015}.  In case I, shown in Fig. \ref{fig:nonresonantLW} (a), three cavity modes of $\left(\omega_c/2\pi ~(\text{GHz}), Q\right)$ = (10.502, 420), (12.822, 420), and (15.362, 140) were excited in the cavity.  In case II, shown in Fig. \ref{fig:nonresonantLW} (b), three modes of $\left(\omega_c/2\pi ~(\text{GHz}), Q\right)$ = (7.969, 63), (9.990, 57) and (13.414, 21) were observed.  As shown in Fig. \ref{fig:nonresonantLW} (c) and (d), the line shape of the spin pumping spectrum $V(H)$ measured at $\omega/2\pi$ = 10 GHz was significantly tuned by the strongly coupled microwaves.  Without the influence of strong coupling the voltage spectra would have a Lorentz line shape.  However the presence of the cavity and the behaviour of the cavity modes influences this behaviour and instead an asymmetric line shape was found.\footnote{Such asymmetric voltage line shapes have previously been studied in detail in the context of spin rectification.  See e.g. \cite{Azevedo2011, Harder2011a}.}  The broad band behaviour of $\Delta H\left(\omega\right)$, which includes non-resonant regions between the cavity modes is shown in Fig. \ref{fig:nonresonantLW} (e) with experimental data as symbols.  In the case of multiple cavity modes Eq. \eqref{eq:dHNonResonant} can be generalized as 
\begin{equation}
\Delta H \left(\omega\right) = \Delta H_0 + \frac{\alpha \omega}{\gamma} + \frac{\omega^2 \omega_m}{\gamma} \sum_l K_l^2 \text{Im}\left(S_{c,l}\right), \label{eq:dhnr2}
\end{equation}
where the last term describes the coupling enhanced FMR line width near each cavity mode with $l$ summing over all cavity modes.  In the presence of hybridization the frequency slope which determines $\alpha$ is constant, independent of the cavity properties and the spin-photon coupling.  However the inhomogeneous broadening has been found to depend on the coupling properties, differing between the two cases highlighted above and from the value before loading the sample into the cavity \cite{Bai2015}.  This effect is due to a difference in the microwave density of states in the cavity \cite{Bai2015, Lagendijk1996} as highlighted by the difference in the transmission amplitude and line width of the cavity modes observed in Fig. \ref{fig:nonresonantLW} (a) and (b).  As a result, the measured $\Delta H_0$ is increased not only near each cavity mode, but also in the non-resonant regions between modes.  It is even possible for this extrinsic damping caused by the coherent spin-photon coupling to be larger than the intrinsic Gilbert damping of YIG, indicating the significant effect of FMR broadening due to non-resonant spin-photon coupling \cite{Bai2015}.  Analogous effects have recently been studied in spin systems driven by microwave striplines, where radiation damping plays an important and confounding role in the line width characterization, see, for example, Refs. \cite{Schoen2015, Klingler2017}.  This again highlights the importance of careful system characterization in order to disentangle the influence of strong coupling from traditional FMR behaviour. 


\subsubsection{Angular Control of Coupling Strength} \label{sec:angleControl}

Within the phase correlation model the magnetization dynamics are driven by a microwave magnetic field using the LLG equation.  The result of the LLG equation is that the magnetization dynamics are determined by the linear response function $\chi^+$, so that for the elliptically polarized modes $m^+ = \chi^+ h^+$.  The detailed nature of the coupling strength, such as its temperature dependence, relationship to the filling factor and spin density and its geometric details, are therefore determined by $\chi^+$.  In this sense the coupling strength generally depends on some set of parameters, $\textbf{p}$, sot that $g = g\left(\textbf{p}\right)$, and typically $g$ is simply measured for a given set of experimental conditions.  However, it is also possible to identify different contributions to the parameter set $\textbf{p}$ and therefore to control the coupling strenth directly.  For example, $g$ depends on the number of spins, $N_s$, $g \propto \sqrt{N_s}$ and therefore can be controlled by using different sizes of magnetic samples \cite{Tabuchi2014a} or by exploiting the strongly temperature dependent magnetization of compensated garnets \cite{MaierFlaig2017}.  

\begin{figure}[b!]
\centering
\includegraphics[width=11cm]{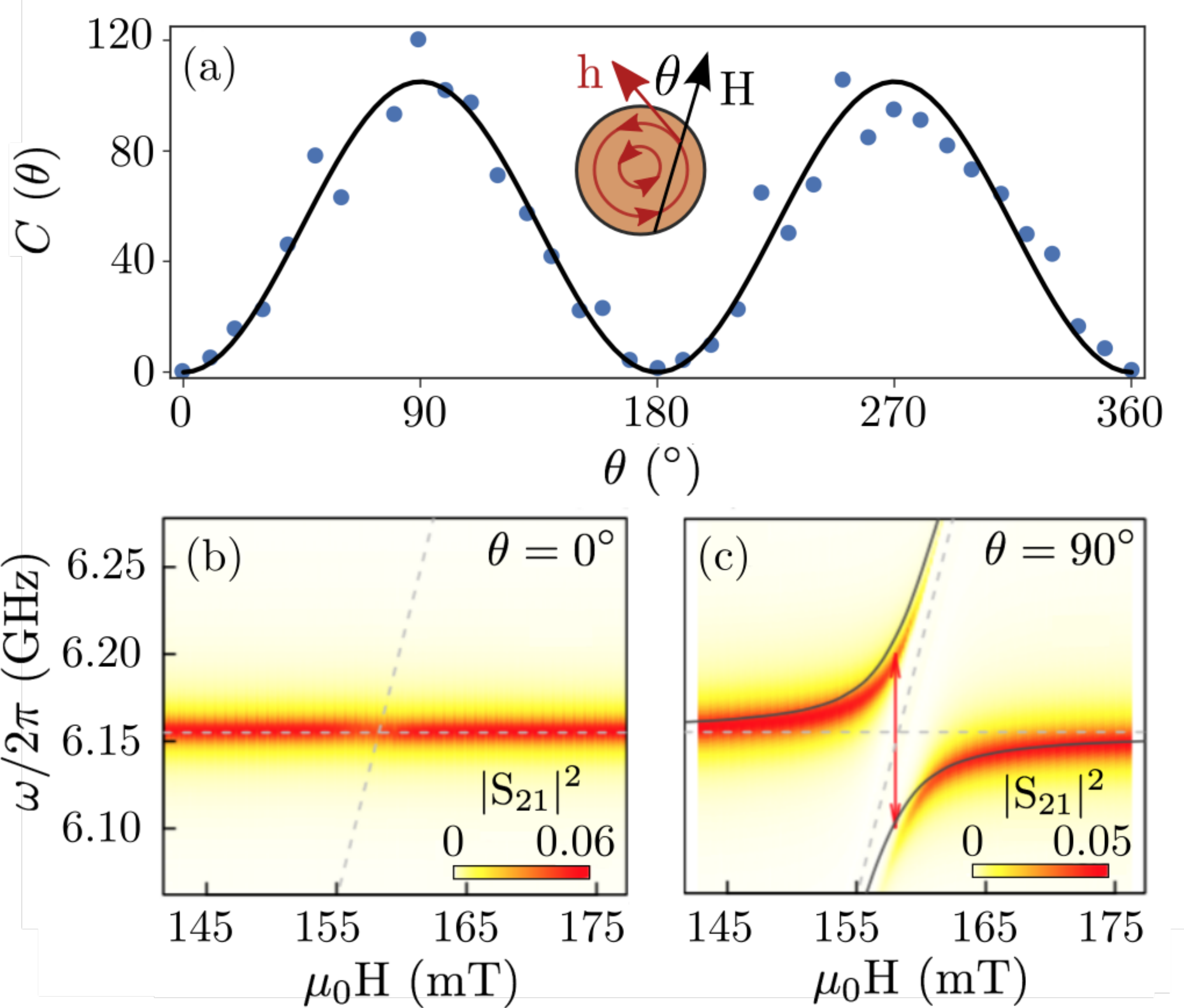}
\caption[Dependence of cooperativity on bias field orientation]{Control of the coupling strength by tuning the bias field angle.  (a) The inset shows the circular profile of a TM$_{011}$ magnetic field $h$.  $\theta$ indicates the local angle between $h$ and a bias field $H$ at the sample location.  Symbols indicate the angular dependent cooperativity while the black curve is a fit according to Eq. \eqref{eq:angleCoupling}. (b) Transmission spectra at $\theta = 0^\circ$ and (c) $\theta = 90^\circ$ demonstrating the minimized coupling strength at $\theta = 0^\circ$ and maximized coupling strength at $\theta = 90^\circ$.  Horizontal and vertical dashed lines indicate the uncoupled cavity and FMR dispersions respectively while the solid curves in (d) are a fit according to Eq. \eqref{eq:simpDisp}.  A modified version of this figure was originally published in Ref. \cite{Bai2017}.}
\label{fig:angularCooperativity}
\end{figure}

Since $\textbf{p}$ also depends on the local microwave field, it is also possible to control the coupling strength by manipulating the field profile.  For example, since this precession is due to a magnetic torque of the form $\textbf{M} \times \textbf{h}$, only the component of the microwave magnetic field which is perpendicular to the total magnetization $M$, and hence perpendicular to the static field $H$, will drive the precession.  Using the angle $\theta$ between the rf microwave field $h^+$ and the static bias field $H$, defined in the inset of Fig. \ref{fig:angularCooperativity} (a), the relevant component of the microwave field is $h^+|\sin\left(\theta\right)|$ and therefore $m^+ = \chi^+ h^+ |\sin\left(\theta\right)|$.  This geometric dependence can be included explicitly in the coupling strength so that
\begin{equation}
g = g_0 |\sin\left(\theta\right)|, \label{eq:angleCoupling}
\end{equation}
where $g_0$ is the coupling strength at $\theta = 90^\circ$.  Physically Eq. \eqref{eq:angleCoupling} means that when the microwave and static magnetic fields are aligned the magnetization dynamics are not driven efficiently and therefore the coupling is minimized.  On the other hand, when the microwave and static magnetic fields are perpendicular the coupling is at a maximum.  The angle $\theta$ therefore describes how effectively the magnetization dynamics can be driven by the cavity field, or in other words how efficiently the spin and photon systems are coupled.  

An example of coupling strength tuning via angular control is shown in Fig. \ref{fig:angularCooperativity} \cite{Bai2016, Bai2017} where a YIG($2.6~ \mu$m)/Pt(10 nm) bilayer with lateral dimensions of 10 mm $\times$ 7 mm was coupled to a circular TM$_{011}$ mode in a cylindrical microwave cavity.  The circular mode profile and in-plane bias field enabled straightforward angular control.  In this experiment the angle $\theta$ defines the orientation between the static magnetic field and the local microwave field at the sample location and the change in cooperativity as $\theta$ was tuned is shown in Fig. \ref{fig:angularCooperativity} (a), where symbols are experimental data and the solid black line is a fit according to Eq. \eqref{eq:angleCoupling}.  At $\theta = 0^\circ$ the cooperativity drops nearly to 0, resulting in a weakly coupled system, while the cooperativity is maximized at $\theta = 90^\circ$, resulting in strong spin-photon coupling.  This behaviour was confirmed by the transmission measurements shown in Fig. \ref{fig:angularCooperativity} (b) and (c).  In panel (b), at $\theta = 0^\circ$, the microwaves do not couple strongly to the spin system and therefore microwave transmission measurements only detect the cavity mode.  On the other hand in panel (c), at $\theta = 90^\circ$, a large anticrossing indicates the presence of strong spin-photon hybridization.

\begin{figure}[t!]
\centering
\includegraphics[width=9cm]{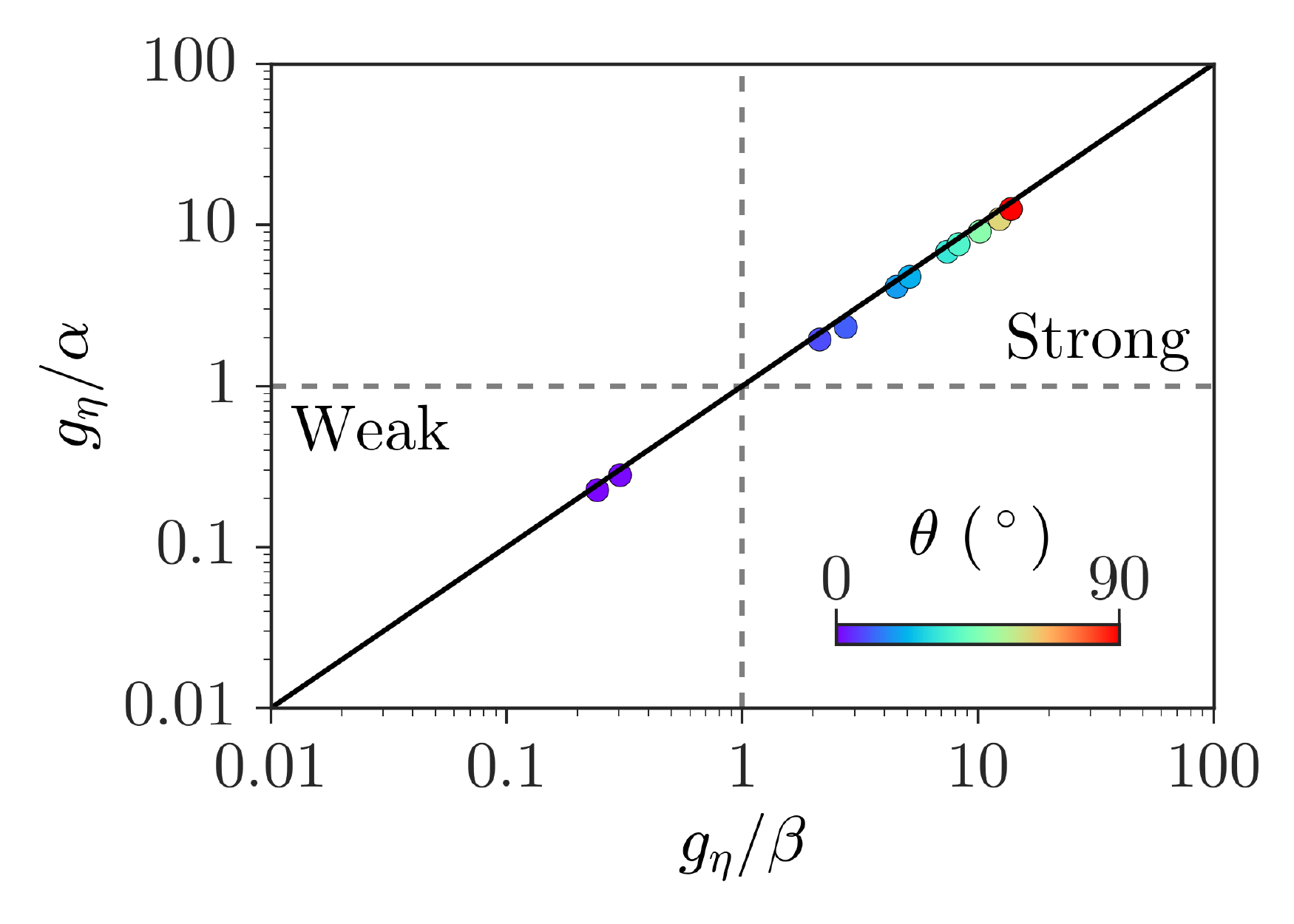}
\caption[From weak to strong coupling using bias field orientation]{The distribution of coupling strengths observed when $\theta$ is tuned.  This method enables systematic tuning between weak and strong coupling regimes.  For fixed $\alpha$ and $\beta$ the dissipation normalized coupling strengths which can be achieved in this method will always lie along a line of constant slope, as indicated by the black line.  A modified version of this figure was originally published in Ref. \cite{Bai2016}.}
\label{fig:angularWeakStrong}
\end{figure}

Using the angular control technique, systematic tuning between weak and strong coupling regimes was demonstrated in Ref. \cite{Bai2016}, as highlighted in Fig. \ref{fig:angularWeakStrong}.  In such experiments, as the angle is tuned from $0^\circ$ to $90^\circ$ the system moves along a diagonal line in the $g_\eta/\alpha - g_\eta/\beta$ parameter space, indicated by the solid line.  The circles indicate experimental measurements with the colour indicating the angle at which they were measured.  Such results highlight the tunability of cavity spintronic systems and their relevance for future device development.  

\subsubsection{Non-local Manipulation of Spin Current} \label{sec:nonlocal}

One interesting early application of cavity spintronics has been to realize non-local spin current manipulation.  In general the generation and manipulation of spin current is the kernel of spintronic development, and a variety of techniques exist which allow local spin current manipulation.  For example, Datta and Das \cite{Datta1990} proposed the spin field-effect transistor, in which spin current is manipulated by a gate voltage via a local spin-orbit interaction in a semiconductor channel \cite{Nitta1997}.  Alternatively, the exchange interaction is also commonly used to manipulate spin current, for example, spin current may drive magnetization dynamics through spin transfer torque \cite{Silsbee1979, Slonczewski1996, Berger1996}.  However since spin-orbit and exchange interactions have characteristic length scales of $\sim$ nm, the control offered by these interactions is inherently short ranged.  Actually, since spin current injection is a diffusive process,\footnote{Which is allowed because spin current, unlike charge current, need not be conserved.} devices which utilize spin-orbit or exchange interaction for spin current control are limited by the $\sim~\mu$m spin diffusion length,\footnote{This is another advantage of YIG.  Since it has a very small damping, the corresponding spin diffusion length is large.} which is determined by the rate at which the angular momentum carried by the spin current can be dissipated, for example by transferring to the lattice.  While this length scale is an improvement over the fundamental interaction length, realization of a long distance $(\gg \mu$m) spin current manipulation would be beneficial for spintronic applications.  

Since the spin current is proportional to $|m|^2$ the hybridized spin current can be determined by solving for $m$ in Eq. \eqref{eq:genMatrix}.  In doing so it is convenient to define the detuning parameters,
\begin{align}
\Delta_c &= \frac{\omega - \omega_c}{\beta \omega_c}, \\
\Delta_r &= \frac{\omega - \omega_r}{\alpha \omega_c},
\end{align}
and the spin-photon cooperativity $C_1 = g_1^2/\alpha \beta \omega_c^2$.  Here the subscript $1$ is a sample label.  In terms of the detunings, the real part of the transmission denominator in Eq. \eqref{eq:simpTrans}, which determines the hybridized dispersion, can be written as \cite{Bai2017}
\begin{equation}
\Delta_c \Delta_r = 1 + C_1. \label{eq:dcdrCond}
\end{equation}
  The $maximum$ spin current amplitude, $I_{s1}$, will occur on-shell, that is when the hybridized quasiparticles satisfy Eq. \eqref{eq:dcdrCond}, and therefore, evaluating $|m|^2$ from Eq. \eqref{eq:genMatrix} and imposing the constraint of Eq. \eqref{eq:dcdrCond}, the maximum spin current amplitude can be written as,
\begin{equation}
I_{s1} \propto \frac{C_1}{\left(\Delta_c + \Delta_r\right)^2}. \label{eq:singleCurrent}
\end{equation}
Eq. \eqref{eq:singleCurrent} describes how the spin current amplitude can be locally controlled by tuning the cooperativity in a cavity spintronic system.  This of course makes physical sense --- the spin current is influenced by hybridization and therefore if the coupling strength is changed the spin current should be modified.  Since cavity spintronic techniques allow for systematic control of the coupling strength it is therefore possible to systematically control the spin current of a single spin ensemble.  

However, a more intriguing effect can be observed when two identical spin systems (which therefore have the same spin resonance properties) are coupled to a single cavity mode.\footnote{The constraint that the spin systems be identical makes the theoretical formalism more transparent, highlighting the key physics involved.  However this effect could be realized even with distinct spin devices.}  This three mode system is described by the generalized $3 \times 3$ matrix of Eq. \eqref{eq:3x3Dark}, under the assumption that the two magnetic samples couple to the cavity mode with coupling strengths $g_1$ and $g_2$ respectively, but do not directly couple to each other.  Therefore, in analogy with the case of a single spin system, the real part of the dispersion produces the constraint
\begin{equation}
\Delta_c \Delta_r = 1 + C_1\left(\theta\right) + C_2, \label{eq:constraint2Spin}
\end{equation}
where $C_1$ and $C_2$ are the cooperativities of the two spin ensembles respectively.  Analogously the maximum spin current of each sample, $I_{s1}$ and $I_{s2}$ is
\begin{align}
I_{s1} \propto \frac{C_1\left(\theta\right)}{\left(\Delta_c + \Delta_r\right)^2} \label{eq:sC12Spin}\\
I_{s2} \propto \frac{C_2}{\left(\Delta_c + \Delta_r\right)^2}. \label{eq:sC22Spin}
\end{align}
In general this pattern continues and for $n$ identical samples the real part of the dispersion can be written as,
\begin{equation}
\Delta_c \Delta_r = 1 + \sum_{i = 1}^n C_i
\end{equation}
with the maximum spin current produced in the $i^{th}$ spin ensemble \cite{Bai2017}
\begin{equation}
I_{si} \propto \frac{C_i}{\left(\Delta_c + \Delta_r\right)^2}.
\end{equation}

An intriguing effect is now observed by carefully examining Eqs. \eqref{eq:constraint2Spin}, \eqref{eq:sC12Spin} and \eqref{eq:sC22Spin}.  Eq. \eqref{eq:constraint2Spin} describes a constraint on the detunings which depends on the cooperativities of both samples.  Therefore, even though the spin current of each device only depends directly on the local cooperativity (the numerator of Eqs. \eqref{eq:sC12Spin} and \eqref{eq:sC22Spin}) it will be influenced by the global properties of the system (the denominator of Eqs. \eqref{eq:sC12Spin} and \eqref{eq:sC22Spin}).  These global properties depend on the constraint of Eq. \eqref{eq:constraint2Spin} and therefore by locally tuning one sample a second sample, which is not directly tuned in any way, will also be manipulated.  

\begin{figure}[b!]
\centering
\includegraphics[width=15cm]{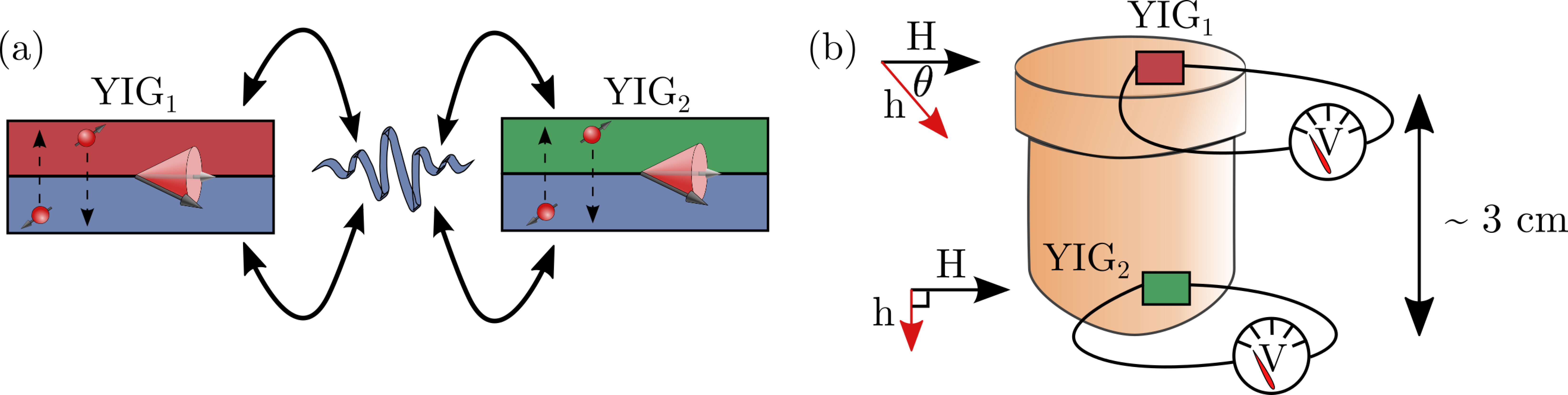}
\caption[Schematic illustration of photon mediated non-local spin current coupling]{Experimental setup used to demonstrate non-local spin current manipulation.  (a) The general idea is to use the cavity photon, which is coupled to both spin systems, as a bridge to carry information from YIG$_1$ to YIG$_2$.  Therefore by locally tuning YIG$_1$, an influence on YIG$_2$ can be observed.  (b) Experimentally this idea was realized by placing YIG$_1$ on the lid of a cylindrical microwave cavity which was rotated within a static magnetic field, $H$.  YIG$_2$, at the bottom of the cavity, had a fixed orientation with respect to $H$ \cite{Bai2017}.}
\label{fig:nonLocalSetup}
\end{figure}

The experimental setup that was used to demonstrate this idea is shown in Fig. \ref{fig:nonLocalSetup}.  Panel (a) indicates the schematic idea, which was to use the cavity photon as a bridge to carry information from the first sample, denoted as YIG$_1$, to the second sample, denoted as YIG$_2$.  Therefore by locally tuning YIG$_1$ the influence on a well separated YIG$_2$ could be observed.  In the experimental setup shown in Fig. \ref{fig:nonLocalSetup} (b), a YIG/Pt bilayer (YIG$_1$) was placed on the lid of a microwave cavity and wired out for electrical detection.  Meanwhile an identical YIG/Pt bilayer (YIG$_2$) was placed at the bottom of the cavity and also wired out for electrical detection.  As indicated in the figure, rotation of the cavity lid enables tuning of the angle $\theta$ between the local microwave and static field at the YIG$_1$ location.  The position of YIG$_2$ was fixed so that the microwave and static magnetic fields were perpendicular and maximum hybridization would be observed between the cavity and YIG$_2$. 

\begin{figure}[t!]
\centering
\includegraphics[width=11cm]{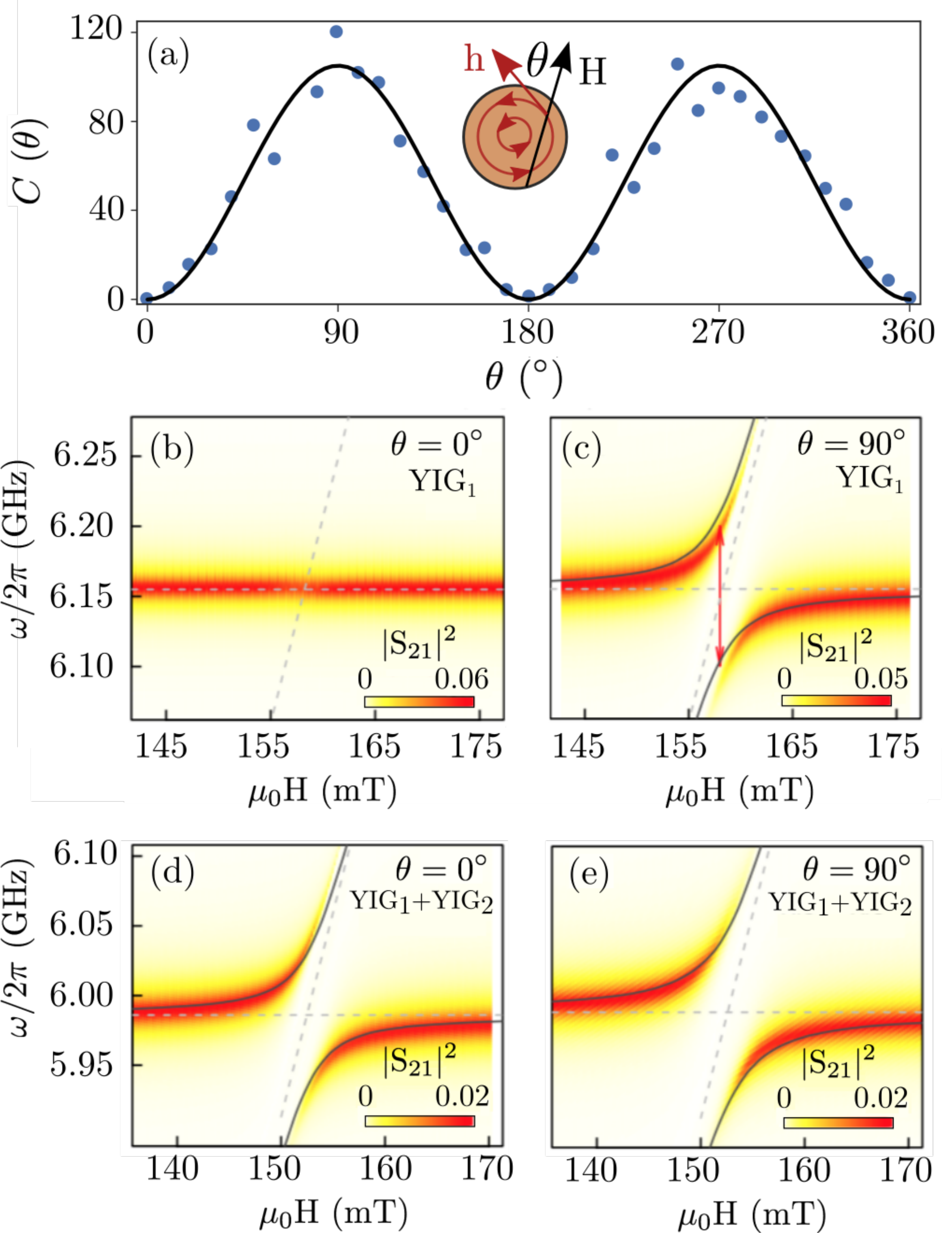}
\caption[Global control of transmission properties in a two YIG system]{Global properties of the single and two spin system.  (a) Angular dependence of the cooperativity for a single YIG/Pt bilayer.  (b) Microwave transmission with a single YIG/Pt sample placed on the lid of the cavity measured at (b) $\theta = 0^\circ$, where the minimum coupling is observed, and (c) $\theta = 90^\circ$, where the maximum coupling is observed. (d) Microwave transmission with both YIG/Pt bilayers inside the cavity at $\theta = 0^\circ$ and (e) $\theta = 90^\circ$.  In this experiment only the angle of the YIG sample at the top of the cavity was tuned, while the YIG placed at the bottom of the cavity was strongly coupled at all angles.  A modified version of this figure was originally published in Ref. \cite{Bai2017}.}
\label{fig:globalControl}
\end{figure}  

Fig. \ref{fig:globalControl} shows the global properties of the single and two spin systems, which are determined by the constraint of Eq. \eqref{eq:constraint2Spin}.  Panels (a) - (c) are the same as Fig. \ref{fig:angularCooperativity} and are included here again for easy comparison to panels (d) and (e), which show the transmission properties of the two spin system at (d) $\theta = 0^\circ$ and (e) $\theta = 90^\circ$.  While a noticeable increase in the dispersion gap was observed between $\theta = 0^\circ$ and $\theta = 90^\circ$ in panels (d) and (e), this difference is not as striking as the case of a single YIG sample.  This is because when both spin systems were in the cavity, YIG$_2$ was not tuned and always maximally coupled to the cavity mode.  $\omega_\text{gap}$ was therefore a superposition of both YIG samples coupled to the cavity and never went to zero, even when the coupling of YIG$_1$ was zero.  Nevertheless, the change to the global properties of the three mode system is evident. 

\begin{figure}[t!]
\centering
\includegraphics[width=14cm]{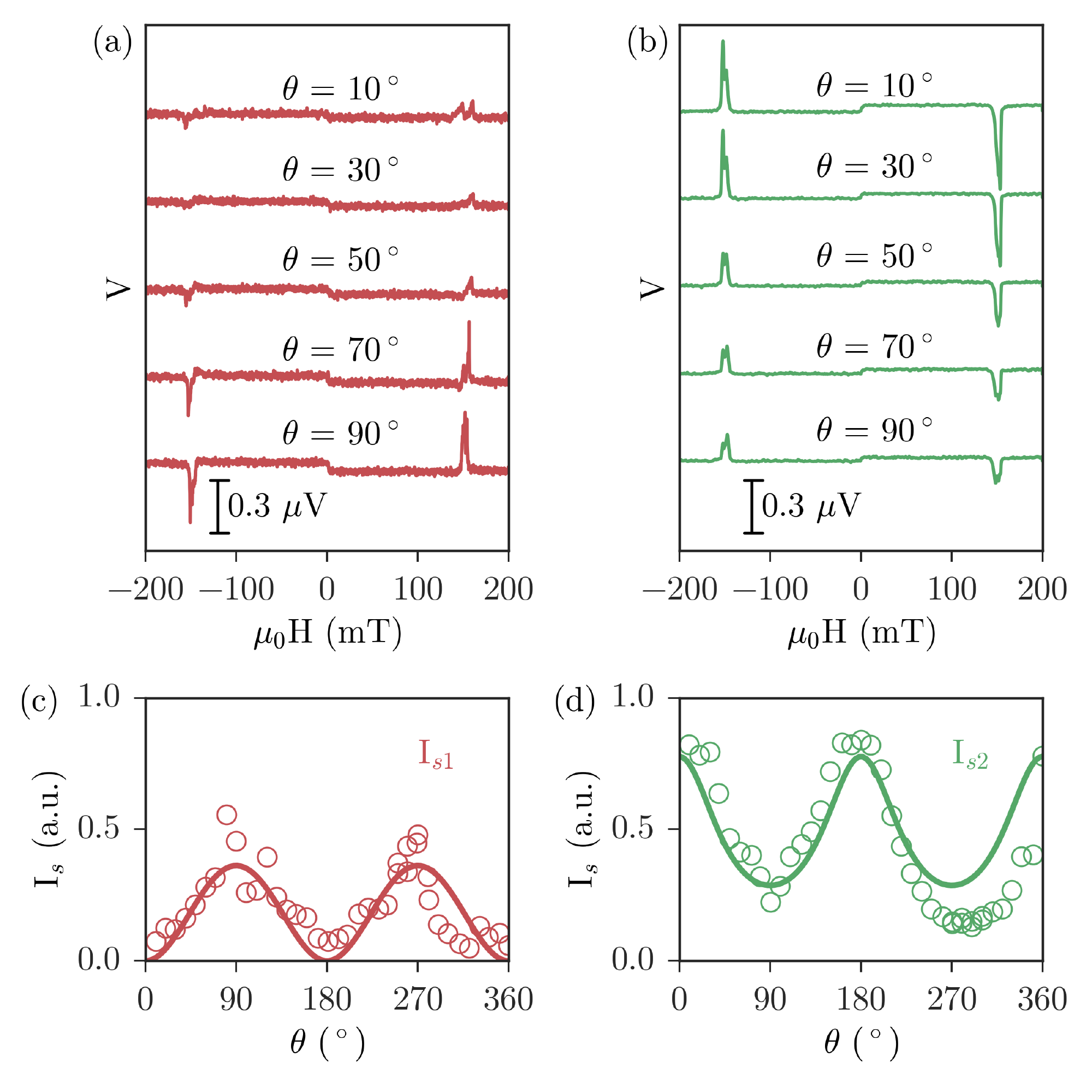}
\caption[Experimental demonstration of non-local spin current manipulation]{(a) The voltage generated in YIG$_1$ was locally tuned by controlling the cooperativity, while (b) the YIG$_2$ voltage signal was simultaneously controlled nonlocally, with no direct manipulation.  Converting the voltage signal into the spin current amplitude the angular dependence of (c) $I_{s1}$ and (d) $I_{s2}$ can be observed.  A modified version of this figure was originally published in Ref. \cite{Bai2017}.}
\label{fig:nonlocal}
\end{figure}

To demonstrate both local and non-local control of the spin current the voltage generated due to spin pumping in each YIG/Pt bilayer was measured.  As shown in Fig. \ref{fig:nonlocal} (a), when $\theta$ was tuned from $0^\circ$ to $90^\circ$, the voltage, and hence the spin current, that was generated in YIG$_1$ increased as the field torque on the magnetization increased.  Such a measurement demonstrates the direct control of the spin current in YIG$_1$ due to hybridization.  The simultaneously measured voltage output from YIG$_2$ is shown in Fig. \ref{fig:nonlocal}.  Here the inverse effect is observed.  As $\theta$ is increased the amplitude of the spin current in YIG$_2$, which is spatially separated from YIG$_1$ and not directly tuned in any way, decreased.  The maximum spin current amplitude at each angle, calculated using the relevant spin Hall angle \cite{Bai2015}, is shown in Fig. \ref{fig:nonlocal} (c) and (d) using open circles for YIG$_1$ and YIG$_2$ respectively.  Since the external magnetic field was not rotated during the measurement the spin currents from both samples maintain the same sign \cite{Bai2017}.  Both local and non-local control of the spin current was observed.  The solid curve for YIG$_1$ (YIG$_2$) is plotted according to Eq. \eqref{eq:sC12Spin} (Eq. \eqref{eq:sC22Spin}) using the constraint of Eq. \eqref{eq:constraint2Spin}.  
  
\begin{figure}[t!]
\centering
\includegraphics[width=12cm]{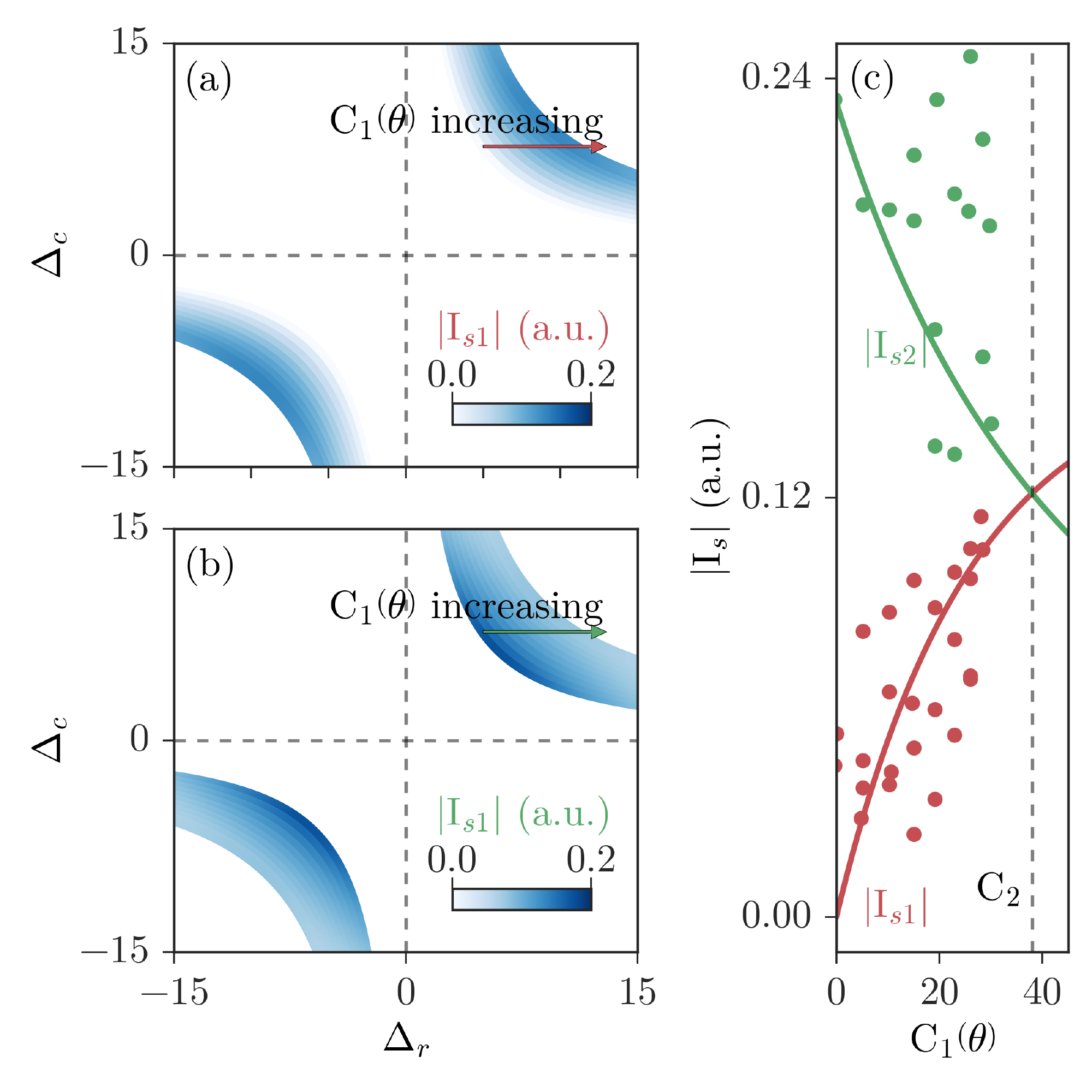}
\caption[Cooperativity dependence of spin current]{The cavity-FMR detuning dispersion calculated using Eq. \eqref{eq:constraint2Spin} for (a) YIG$_1$ and (b) YIG$_2$.  The colour scale indicates the spin current amplitudes, $I_{s1}$ and $I_{s2}$, which are calculated using Eq. \eqref{eq:sC12Spin} and Eq. \eqref{eq:sC22Spin} respectively for different cooperativities $C_1\left(\theta\right)$.  The change in $C_1\left(\theta\right)$ is indicated by the red and green arrows for $I_{s1}$ and $I_{s2}$ respectively.  (c) Along the arrows, at a fixed cavity frequency detuning of $\Delta_c = 7.7$, the amplitudes of both spin currents are plotted as a function of $C_1\left(\theta\right)$.  Symbols indicate experimental data and the solid curves are calculations using Eqs. \eqref{eq:sC12Spin} and \eqref{eq:sC22Spin}.  The vertical dashed line indicates the fixed cooperativity of YIG$_2$.  A modified version of this figure was originally published in Ref. \cite{Bai2017}.}
\label{fig:nonlocalCooperativity}
\end{figure}

The non-local spin current control originates from the influence of the global hybridization on local electrical detection measurements.  To highlight how this originates based on the cavity and FMR detunings, a $\Delta_c - \Delta_r$ dispersion following from Eq. \eqref{eq:constraint2Spin} is plotted in Fig. \ref{fig:nonlocalCooperativity} (a) and (b) for different values of the cooperativity $C_1\left(\theta\right)$.  The arrows in each figure indicate the direction of increasing $C_1\left(\theta\right)$.  Panel (a) is plotted with a colour scale that indicates the amplitude of $I_{s1}$, calculated from Eq. \eqref{eq:sC12Spin}, with the colour scale of panel (b) indicating the amplitude of $I_{s2}$ according to Eq. \eqref{eq:sC22Spin}.  The two hybridized modes are only excited when the cavity and FMR detunings are either both positive or both negative, reflecting the fact that both cooperativities are positive.  Based on Fig. \ref{fig:nonlocalCooperativity} (a) and (b) the spin current features of this three mode coupled system as: (i) The hybridized modes of the coupled system rely on the sum of cooperativities of all magnetic samples with the coupling strength increasing when more magnetic samples are added; (ii) The spin current pumped by each magnetic sample depends on both the global properties of the normal mode detunings and the local cooperativity with the cavity mode; and (iii) The amplitude of $I_{s1}$ (spin current in the directly tuned sample) increases as $C_1$ is increased, while $I_{s2}$ (spin current of the distant sample) has the opposite behaviour, decreasing as $C_1$ is increased.  In general, since the spin current in the directly controlled sample is proportional to its cooperativity, while the spin current in the non local sample is inversely proportional, the two spin currents will always change in opposite directions.  However if, for example, three devices were used, it may be possible to locally tune two samples in such a way that the spin current in a third remains unchanged.   

In this example the spin current was measured at a fixed cavity mode frequency detuning of $\Delta_c = 7.7$.  This position is indicated by the arrows in Fig. \ref{fig:nonlocalCooperativity} (a) and (b).  Along these arrows the amplitudes of both spin currents are plotted as a function of $C_1$ in Fig. \ref{fig:nonlocalCooperativity} (c).  The solid curves are $I_{s1}$ and $I_{s2}$, calculated using Eqs. \eqref{eq:sC12Spin} and \eqref{eq:sC22Spin} respectively.  The data plotted in this manner is effectively ``third order" compared to the raw data of Fig. \ref{fig:nonlocal} (a) and (b), meaning that the spin currents were first calculated from the raw voltage measurement, and then the cooperativity was calculated for each field orientation.  As a result several sources of additional uncertainty are introduced in this method.  For example, the spin Hall angle is notorious for its large uncertainties \cite{Liu2011a}.  Additionally, the determination of the coupling strength, which is needed for the cooperativity calculation, is difficult for angles where the coupling strength, and hence amplitude, is small.  For this reason there are large uncertainties present in Fig. \ref{fig:nonlocalCooperativity} (c), which make a quantitative comparison between theory and experiment using this method difficult.  Nevertheless, the qualitative agreement between the model and the experimental data indicates that the non-local manipulation of spin current in YIG$_2$, which can be detected locally through spin pumping, is due to the local cooperativity control of YIG$_1$.  This point may perhaps be better emphasized by the angular control demonstrated in Fig. \ref{fig:nonlocalCooperativity} (d).  This experiment therefore demonstrated one way in which cavity spintronics, and strong spin-photon coupling, breaks the limits of conventional spintronics by utilizing the emergent properties of the cavity-magnon-polariton. 

\section{Summary}

In summary, advances in magnetism, spintronics, and light-matter interactions have created a new frontier of condensed matter research studying Cavity Spintronics. Via the quantum physics of spin-photon entanglement on one hand, and via classical electrodynamic coupling on the other, this frontier merges progress in spintronics with advances in cavity QED and cavity polaritons. 

Intended to give a useful introduction for new researchers entering this emerging field, we have provided in this article different perspectives for understanding the basic physics of spin-photon coupling, by reviewing different theoretical models established based on the intuitive, classical, and quantum pictures. With our focus on potential spintronics applications, we have summarized some of our own experiments investigating the physics and applications of spin-photon coupling. Over the past few years, this emerging field has witnessed a rapid expansion. New avenues beyond the scope of this review, such as studying quantum magnonics \cite{Tabuchi2015, Zhang2015b, Osada2015, Kusminskiy2016, Liu2016}, and cavity magnomechanics \cite{Zhang2015f} by using spin-photon coupling, have been created. It is expected that many more exciting applications will be demonstrated due to the rapid pace of development in this new frontier of research.

\section{Acknolowdgements}

This early review article has been based on the thesis research of one of us (M. H.) performed in the Dynamics Spintronics Group at the University of Manitoba, Canada. It involves many collaborative studies as identified in the reference section. In particular, we would like to thank the DSG members and alumni L. H. Bai, Y. S. Gui, P. Hyde, S. Kaur, J. W. Rao, and B. M. Yao for their contributions. We would also like to thank the groups of C. L. Chien, S. T. B. Goennenwein, H. Guo, H. Huebl, W. Lu, K.-P. Marzlin, M. Weiler, J. Q. Xiao, D. S. Xue, and J. Q. You for their collaborations. This work was supported by NSERC and the National Natural Science Foundation of China (Oversea Scholar Collaborative Research Grant No. 11429401).

\newpage

\begin{appendices}

\section{Simplification of Harmonic Oscillator Dispersion} \label{sec:appOscDispSimp}

For the frequency dispersion, analytic expressions for the real and imaginary components of the eigenvalues are easily determined using de Moivre's theorem:
\begin{align}
\omega_\pm &= \frac{1}{2}\left[\omega_c + \omega_r \pm \frac{1}{\sqrt{2}} \left[|z| + \left(\omega_c - \omega_r\right)^2 - \omega_c^2\left(\beta - \alpha\right)^2 + \frac{1}{4}\kappa^4 \omega_c^2\right]^{1/2}\right], \label{eq:redeMoivre}\\
\Delta \omega_\pm &= \frac{1}{2} \left[\omega_c \left(\alpha + \beta\right) \pm \frac{1}{\sqrt{2}}\frac{2\omega_c \left(\beta - \alpha\right)\left(\omega_c - \omega_r\right)}{\left[|z| + \left(\omega_c-\omega_r\right)^2 - \omega_c^2\left(\beta - \alpha\right)^2 + \frac{1}{4}\kappa^4 \omega_c^2\right]^{1/2}}\right], \label{eq:imdeMoivre}
\end{align}
where
\begin{equation}
|z| = \sqrt{\left[\left(\omega_c - \omega_r\right)^2 - \omega_c^2 \left(\beta - \alpha\right)^2 + \frac{1}{4}\kappa^4 \omega_c^2\right]^2 + 4 \omega_c^2 \left(\beta - \alpha\right)^2 \left(\omega_c - \omega_r\right)^2}.
\end{equation}
Here $\text{Re}\left(\tilde{\omega}_\pm\right) = \omega_\pm$ and $\text{Im}\left(\tilde{\omega}_\pm\right) = \Delta \omega_\pm$.


\section{Simplification of Harmonic Oscillator Transmission Spectra} \label{sec:appOscTransSimp}
$\text{S}_{21}$ can be written in terms of $\omega_\pm$ as,
\begin{equation}
\text{S}_{21} \propto \frac{\omega - \tilde{\omega_r}}{\left(\omega - \tilde{\omega}_+ \right)\left(\omega - \tilde{\omega}_-\right)}.
\end{equation}
The frequency can be expanded near the eigenmodes as $\omega = \omega_\pm + \epsilon = \omega_\mp \pm \delta \omega + \epsilon$, where $\delta \omega = \omega_+ - \omega_-$ and $\epsilon$ is small compared to $\omega$.  Assuming the two eigenmodes are well separated, meaning the system is sufficeintly detuned from the crossing point or strongly coupled, $\delta \omega \gg \epsilon$, and $\delta \omega \gg \Delta\omega_\pm$.  If in addition the transmission is measured not too close to the FMR mode, so that $\omega_\pm - \omega_r \gg \epsilon$, then
\begin{equation}
|\text{S}_{21}|_\pm^2 \propto \frac{\left(\omega_\pm - \omega_r\right)^2 + 2 \left(\omega - \omega_\pm\right) \left(\omega_\pm - \omega_r\right)}{\delta\omega^2 \left[\left(\omega - \omega_\pm\right)^2 + \Delta\omega_\pm\right]^2},
\end{equation}  
where $|\text{S}_{21}|^2_\pm = |\text{S}_{21}\left(\omega \sim \omega_\pm\right)|^2$.  Therefore the transmission spectra near either eigenmode may be written as
\begin{equation}
|\text{S}_{21}|^2_\pm \propto L + \left(\omega_\pm - \omega_r\right)^{-1} D
\end{equation}
as in Eq. \ref{eq:s21LandD}.

For the field swept case the transmission spectra may be simplified by expanding near $\tilde{\omega}_e$.  In that case,
\begin{equation}
\text{S}_{21} \propto \frac{\omega - \tilde{\omega}_r}{\omega_r - \tilde{\omega}_e}
\end{equation}
and provided the spectra is probed several line widths away from the FMR frequency, $\omega - \omega_r \gg \alpha \omega_c$, $\text{S}_{21}$ can be written as
\begin{equation}
|\text{S}_{21}|^2 \propto \frac{\left(q\Delta \omega_e + \omega_r - \omega_e\right)^2}{\left(\omega_r - \omega_e\right)^2 + \Delta \omega_e^2}
\end{equation}
as found in Eq. \eqref{eq:fano}.


\section{Normal Modes of Harmonic Oscillator} \label{sec:appOscNormalModes}
Using the classical rotating wave approximation and in the absence of a driving force, the equations of motion given by Eq. \ref{eq:omegaMatrix} become
\begin{equation}
\left(\begin{array}{cc}
\omega - \tilde{\omega}_c & -\frac{\kappa^2 \omega_c}{2} \\
-\frac{\kappa^2 \omega_c}{2} & \omega - \tilde{\omega}_r
\end{array}\right) \left(\begin{array}{c}
x_1 \\
x_2 \end{array} \right) = 0.
\end{equation}
The normalized eigenvectors of the CMP are therefore,
\begin{equation}
X_\pm = \frac{1}{\sqrt{2 \Omega}} \left(\begin{array}{c}
\pm \sqrt{\Omega \mp \Delta} \\
\sqrt{\Omega \pm \Delta} \end{array}\right) \label{eq:eigenvectors}
\end{equation}
where $\Omega = \sqrt{\left(\tilde{\omega}_r - \tilde{\omega}_c\right)^2 + \kappa^4 \omega_c^2}$ and $\Delta = \tilde{\omega}_r - \tilde{\omega}_c$, which defines the transformation between hybridized and spin/photon states in Eq. \eqref{eq:hopfieldTransform}.


\section{Details of Input-Output Formalism} \label{sec:inoutDetails}
Here the relationships between input and output fields and the bath photons are derived.  The Heisenberg equations of motion for the two baths are
\begin{align}
\dot{c}_q &= -\frac{i}{\hbar} \left[c_q, H\right] = - i \omega_q c_q + \lambda_c a, \label{eq:cEOM} \\
\dot{d}_q &= -\frac{i}{\hbar} \left[d_q, H\right] = - i \omega_q d_q + \lambda_d a, \label{eq:dEOM}
\end{align}
which can be solved in terms of the bath states at either $t_i < t$ or $t_f > t$, corresponding to initial and final states before and after interaction with the cavity,
\begin{align}
c_q \left(t\right) &= e^{-i \omega_q \left(t-t_i\right)} c_q\left(t_i\right) + \lambda_c \int_{t_i}^t d\tau e^{-i \omega_q \left(t-\tau\right)}a\left(\tau\right), \nonumber \\
c_q \left(t\right) &= e^{-i \omega_q \left(t-t_f\right)} c_q\left(t_i\right) - \lambda_c \int_{t}^{t_f} d\tau e^{-i \omega_q \left(t-\tau\right)}a\left(\tau\right)
\end{align}
with analogous expressions for $d_q$.  Using these solutions a sum over cavity modes becomes, 
\begin{align}
\lambda_c \sum_q c_q &= \lambda_c \sum_q e^{-i \omega_q \left(t-t_i\right)} c_q\left(t_i\right) + \lambda_c^2 \sum_q \int_{t_i}^t d\tau e^{-i\left(\omega_q - \omega_c\right)\left(t - \tau\right)} \left[e^{-i \omega_c \left(\tau - t\right)} a\left(\tau\right)\right], \nonumber \\
\lambda_c \sum_q c_q &= \lambda_c \sum_q e^{-i \omega_q \left(t-t_f\right)} c_q\left(t_f\right) - \lambda_c^2 \sum_q \int_{t}^{t_f} d\tau e^{-i\left(\omega_q - \omega_c\right)\left(t - \tau\right)} \left[e^{-i \omega_c \left(\tau - t\right)} a\left(\tau\right)\right]
\end{align}
with similar expressions for $d_q$.  Fermi's golden rule can be used to define the external coupling rate for the cavity mode to the bath as,
\begin{equation}
2\kappa_c \left(\omega_c\right) = 2 \pi \lambda_c^2 \rho = 2\pi \lambda_c^2 \sum_q \rho\left(\omega_q\right),
\end{equation}
where $\rho\left(\omega_q\right)$ is the density of states of the photon bath for mode $q$ and $\rho = \sum_q \left(\omega_q\right)$ is the full density of states.  For a single mode of the cavity $\rho\left(\omega_q\right) = \delta \left(\omega_c - \omega_q\right)$ and therefore write
\begin{equation}
2 \int_{-\infty}^\infty d\nu \kappa_c \left(\omega_c + \nu\right) e^{-i\nu\left(t-\tau\right)} = 2\pi \lambda_c^2 \sum_q \int_{-\infty}^\infty d\nu e^{-i\nu\left(t-\tau\right)} \delta \left(\omega_c+\nu-\omega_q\right).
\end{equation}
Under a Markov approximation $\kappa_c$ is approximately constant near $\omega_c$ and therefore
\begin{equation}
2\kappa_c \delta\left(t-\tau\right) = \lambda_c^2 \sum_q e^{-i \left(\omega_q - \omega_c\right) \left(t-\tau\right)}
\end{equation}
where the delta function is defined as $\delta\left(\omega\right) = \left(2\pi\right)^{-2} \int e^{-i\omega t} dt$.  Therefore
\begin{align}
\lambda_c \sum_q c_q\left(t\right) &= \lambda_c \sum_q e^{-i\omega_q \left(t-t_i\right)} c_q\left(t_i\right) + 2 \int_{t_i}^t d\tau k_c \delta \left(t-\tau\right)e^{-i\omega_c\left(\tau-t\right)}a\left(\tau\right), \nonumber \\
\lambda_c \sum_q c_q\left(t\right) &= \lambda_c e^{-i\omega_q\left(t-t_f\right)}c_q\left(t_f\right) - 2 \int_t^{t_f}d\tau k_c \delta \left(t-\tau\right)e^{-i\omega_c\left(\tau-t\right)} a\left(\tau\right).
\end{align}
Assuming that $a\left(\tau\right) e^{-i\omega_c \left(\tau - t\right)}$ goes smoothly to 0 as $t\to \pm \infty$,
\begin{equation}
\int_{t_i}^t d\tau \delta \left(t-\tau\right) e^{i\omega_c\left(\tau-t\right)}a\left(\tau\right) = \frac{1}{2} a\left(t\right),
\end{equation}
and thus defining the input and output modes as the wave packets
\begin{align}
c_\text{in} \left(t\right) &= -\frac{1}{\sqrt{2\pi \rho}} \sum_q e^{-i\omega_q \left(t - t_i\right)} c_q\left(t_i\right),\nonumber \\
c_\text{out} \left(t\right) &= \frac{1}{\sqrt{2\pi \rho}} \sum_q e^{-i\omega_q \left(t - t_f\right)} c_q\left(t_f\right),
\end{align}
it follows that
\begin{align}
\lambda_c \sum_q c_q\left(t\right) &= -\sqrt{2\kappa_c} c_\text{in}\left(t\right) + \kappa_c a\left(t\right), \\
\lambda_c \sum_q c_q\left(t\right) &= \sqrt{2\kappa_c} c_\text{out}\left(t\right) - \kappa_c a\left(t\right).
\end{align}
Therefore 
\begin{align}
c_\text{in}\left(t\right) + c_\text{out}\left(t\right) &= \sqrt{2\kappa_c}a\left(t\right), \nonumber \\
d_\text{in}\left(t\right) + d_\text{out}\left(t\right) &= \sqrt{2\kappa_d}a\left(t\right). \label{eq:cincoutApp}
\end{align}

\end{appendices}

\newpage
\section*{References}
\bibliography{PRCMP.bbl}

\end{document}